\documentclass[12pt,a4paper]{article}

\pdfoutput=1
\usepackage{jheppub,psfrag,slashed,cancel,lscape,caption,array,graphicx,subcaption}
\usepackage[utf8]{inputenc}
\usepackage{amsmath}
\usepackage{mathtools}
\usepackage{mathrsfs}
\usepackage{multirow}
\usepackage{booktabs} 
\usepackage{chngcntr}

\usepackage{tikz}
\usetikzlibrary{decorations.pathmorphing}
\usetikzlibrary{decorations.markings}
\usetikzlibrary{positioning, shapes, snakes, arrows}

\tikzset{
	graviton/.style={line width=.8pt, -latex,decorate, decoration={snake, segment length=4pt,amplitude=1.8pt, pre length=.1cm, post length=.25cm}},
	worldline/.style={gray, line width=1pt},
	worldlineBold/.style={black, line width=.6pt},
	zUndirected/.style={line width=1pt},
	zParticle/.style={line width=1pt,postaction={decorate},decoration={markings,mark=at position .6 with {\arrow[#1]{latex}}}},
	zParticleF/.style={line width=1pt,postaction={decorate}},
	cscalar/.style={line width=1pt,postaction={decorate},decoration={markings,mark=at position .6 with {\arrow[#1]{latex}}}},
	cscalar2/.style={line width=1pt,postaction={decorate},decoration={markings,mark=at position .8 with {\arrow[#1]{latex}}}},
	photon/.style={line width =.8pt, decorate, decoration={snake, segment length=4pt, amplitude=1.8pt,  pre length=.1cm, post length=.1cm}}
}

\DeclareFontFamily{OT1}{pzc}{}
\DeclareFontShape{OT1}{pzc}{m}{it}{<-> s * [1.350] pzcmi7t}{}
\DeclareMathAlphabet{\mathpzc}{OT1}{pzc}{m}{it}

\def\cO{\mathcal{O}}
\def\cD{\mathcal{D}}

\def\cM{\mathcal{M}}

\def\eps{\epsilon}

\def\d{\mathrm{d}}

\def\dd{\delta\!\!\!{}^-\!}

\def\d{\mathrm{d}}
\def\eps{\epsilon}

\def\braket#1{\langle #1 \rangle}

\def\nn{\nonumber}

\def\ie{i.e. }
\def\eg{e.g. }

\def\eqn#1{eq.~\eqref{#1}}
\def\eqns#1#2{eqs.~\eqref{#1} and~\eqref{#2}}

\def\Fig#1{Figure~{\ref{#1}}}

\def\Sec#1{Section~{\ref{#1}}}
\def\Secs#1#2{Sections~{\ref{#1}} and~{\ref{#2}}}

\def\App#1{Appendix~{\ref{#1}}}
\def\rcite#1{ref.~\cite{#1}}
\def\rcites#1{refs.~\cite{#1}}

\newcommand{\vev}[1]{\langle #1\rangle}
\newcommand{\Vev}[1]{\Bigl \langle #1 \Bigr\rangle}
\newcommand{\dx}{{\Delta x}}
\newcommand{\xdot}{{\dot x}}

\newcommand{\be}{\begin{equation}}
\newcommand{\ee}{\end{equation}}
\newcommand{\ba}{\begin{align}}
\newcommand{\ea}{\end{align}}
\ifx\genfrac\sdflkaj\else\fi
\newcommand{\sfrac}[2]{{\textstyle\frac{#1}{#2}}}
\newcommand{\WQFTbraket}[1]{ \langle #1 \rangle_{\rm WQFT}}

\newcommand{\ga}{{\mathfrak a}}
\newcommand{\gb}{{\mathfrak b}}
\newcommand{\gc}{{\mathfrak c}}

\begin{document}

\begin{flushright}
\begingroup\footnotesize\ttfamily
	UUITP-37/20 \\ HU-EP-20/22-RTG
\endgroup
\end{flushright}

\vspace{15mm}

\begin{center}
{\LARGE\bfseries 
	Classical black hole scattering from a worldline quantum field theory
\par}

\vspace{15mm}

\begingroup\scshape\large 
	Gustav~Mogull,${}^{1,2,3}$ Jan~Plefka${}^{2}$ and Jan~Steinhoff${}^{3}$  
\endgroup
\vspace{5mm}
					
\textit{${}^{1}$Department of Physics and Astronomy, Uppsala University, Box 516,\\
 75108 Uppsala, Sweden	}\\[0.25cm]		
\textit{${}^{2}$Institut f\"ur Physik und IRIS Adlershof, Humboldt-Universit\"at zu Berlin, \phantom{${}^2$}\\
  Zum Gro{\ss}en Windkanal 6, D-12489 Berlin, Germany} \\[0.25cm]
\textit{${}^{3}$Max-Planck-Institut f\"ur Gravitationsphysik
(Albert-Einstein-Institut)\\
M\"uhlenberg 1, D-14476 Potsdam, Germany } \\[0.25cm]

\bigskip
  
\texttt{\small\{gustav.mogull@aei.mpg.de,
jan.plefka@hu-berlin.de, jan.steinhoff@aei.mpg.de\}} 

\vspace{10mm}

\emph{In memory of Sissi Plefka and J\"urgen Steinhoff} \vspace{10mm}

\textbf{Abstract}\vspace{5mm}\par
\begin{minipage}{14.7cm}
A precise link is derived between scalar-graviton S-matrix elements
and expectation values of
operators in a worldline quantum field theory (WQFT),
both used to describe classical scattering of black holes.
The link is formally provided by a worldline path integral representation
of the graviton-dressed scalar propagator,
which may be inserted into a traditional definition of the S-matrix
in terms of time-ordered correlators.
To calculate expectation values in the WQFT a new set of
Feynman rules is introduced which treats the gravitational field $h_{\mu\nu}(x)$
and position $x_i^\mu(\tau_i)$ of each black hole on equal footing.
Using these both the 3PM three-body gravitational radiation
$\braket{h^{\mu\nu}(k)}$ and 2PM two-body deflection $\Delta p_i^\mu$
from classical black hole scattering events are obtained.
The latter can also be obtained from the eikonal phase of a 
$2\to2$ scalar S-matrix,
which we show corresponds to the free energy of the WQFT.
\end{minipage}\par

\end{center}
\setcounter{page}{0}
\thispagestyle{empty}
\newpage

\tableofcontents

\section{Introduction}

Black holes are fascinating objects intimately tied to the fundamental
properties of space, time and matter. Rightly they have
been referred to as ``the most perfect macroscopic objects in the universe'' \cite{Chandrasekhar:1979qb}. Their internal state is completely 
determined by their mass, charge and spin;
in this respect they strongly resemble elementary particles,
the equally fascinating constituents of matter and fundamental forces. 
These microscopic cousins of black holes are described using quantum field
theory; their observables (such as cross sections)
are derived from scattering amplitudes,
which in turn have been called
``the most perfect microscopic structures in the universe'' \cite{Dixon:2011xs}.

With the advent of gravitational wave astronomy able to observe
the binary inspirals and mergers of black holes and neutron stars
\cite{Abbott:2016blz,TheLIGOScientific:2017qsa,LIGOScientific:2018mvr},
the need for high-precision theoretical predictions of
their classical potentials and emerging gravitational radiation has arisen \cite{Purrer:2019jcp}.
This is similar to the need for high-precision predictions of scattering
cross section of elementary particles ---
a highly developed subject in quantum field theory.
A number of complementary classical theoretical approaches
to this central problem in general relativity
have been established over recent years
\cite{Blanchet:2013haa,Schafer:2018kuf,Futamase:2007zz,Pati:2000vt,Bel:1981be,Westpfahl:1985tsl}.
Taking up the parallelism with elementary particles,
quantum field theoretical methods of perturbative quantum
gravity have proven themselves very efficient for determining the classical
gravitational interactions of black holes.

The gravitational two-body problem has traditionally been approached in a
perturbative post-Newtonian (PN) weak-field and low-velocity approximation, where one simultaneously
expands in powers of Newton's constant $G$ and in the relative velocity of the two bodies
$v/c$ that are linked by the virial theorem for a bound system ($\frac{v^{2}}{c^{2}}\sim \frac{Gm}{c^2r}$).
The non-relativistic general relativity formalism (NRGR)
\cite{Goldberger:2004jt,Goldberger:2006bd,Goldberger:2009qd}
uses an effective field theory (EFT) to model the massive bodies as 
point-like massive particles coupled to the gravitational field,
and is valid for widely separated massive objects. 
Integrating out the  suitably non-relativistically decomposed
graviton field $h_{\mu\nu}(x)$ \cite{Kol:2007bc} in the path integral yields
a Feynman diagrammatic expansion for the classical effective potential of
the black holes and associated gravitational radiation ---
see \eg \rcites{Goldberger:2007hy,Foffa:2013qca,Rothstein:2014sra,Porto:2016pyg,Levi:2018nxp}. 
The suitably non-relativistically decomposed graviton field
$h_{\mu\nu}(x)$ is integrated out in 
the path integral while the worldline
trajectories of the black holes $x_i^{\mu}(\tau_{i})$ are kept as classical background sources. The state of the art is 4PN level for the potential 
\cite{Damour:2014jta,Damour:2016abl,Bernard:2016wrg,Foffa:2016rgu,Damour:2017ced,Foffa:2019rdf,Foffa:2019yfl,Blumlein:2020pog,Porto:2017dgs,Marchand:2017pir,Galley:2015kus},
parts of 5PN \cite{Foffa:2019hrb,Blumlein:2019zku,Bini:2019nra,Blumlein:2020pyo} and 6PN \cite{Blumlein:2020znm,Cheung:2020gyp,Bini:2020nsb,Bini:2020wpo,Bini:2020hmy,Bini:2020uiq},
and 3PN \cite{Blanchet:2001aw,Blanchet:2004ek,Blanchet:2008je} for the gravitational radiation
emitted from a quasi-circular inspiral (see also \rcite{Leibovich:2019cxo}).
Spin effects may also be taken into account \cite{Blanchet:2013haa,Schafer:2018kuf,Porto:2005ac,Levi:2015msa}, in the conservative dynamics \cite{Levi:2020kvb,Antonelli:2020aeb,Levi:2016ofk} (see refs.~\cite{Porto:2006bt,Porto:2008tb,Levi:2008nh,Porto:2008jj,Porto:2010tr,Perrodin:2010dy,Levi:2010zu} for important early EFT work) and radiation \cite{Mishra:2016whh,Buonanno:2012rv,Porto:2010zg,Porto:2012as,Maia:2017gxn,Maia:2017yok}.

Inspired by the progress made calculating scattering amplitudes,
approaches involving a post-Minkowskian (PM) expansion in Newton's constant,
which re-sum the entire PN expansion in velocity,
have recently been gaining prominence.
For instance, the worldline EFT may also be deployed in a PM weak-field 
scenario as one may naturally use perturbative quantum gravity to represent
gravitons as a metric fluctuation about flat Minkowskian space-time.
This is also the right approximation for
black hole scattering events or $N$-body interaction scenarios.
A worldline EFT formalism for the PM expansion was recently established in
\rcite{Kalin:2020mvi} for conservative binary dynamics (including tidal effects, see also ref.~\cite{Bini:2020flp,Cheung:2020sdj,Kalin:2020lmz,Haddad:2020que}),
and has now been successfully applied to order 3PM ($\cO(G^3)$)
\cite{Kalin:2020fhe}.
Earlier worldline-based PM calculations can be found in refs.~\cite{Westpfahl:1985tsl,Bel:1981be,Ledvinka:2008tk,Damour:2016gwp,Blanchet:2018yvb} for the conservative sector, in refs.~\cite{Kovacs:1977,Kovacs:1978} for radiation, and in refs.~\cite{Bini:2017xzy,Vines:2017hyw,Bini:2018ywr} for spin effects.

A fruitful alternative approach to capture the
classical interactions of massive bodies in gravity
has also been explored through a more direct examination of 
scattering amplitudes in perturbatively quantized gravity.
While there are early works on the subject  \cite{Iwasaki:1971vb,Duff:1973zz,Holstein:2004dn}, this
approach has blossomed in recent years upon employing modern on-shell methods for scattering amplitudes 
\cite{Neill:2013wsa,Bjerrum-Bohr:2013bxa,Luna:2017dtq,Bjerrum-Bohr:2018xdl,Kosower:2018adc}. These works
have  led us to 
the 2PM \cite{Bjerrum-Bohr:2018xdl,Cheung:2018wkq,Cristofoli:2019neg} and 3PM \cite{Bern:2019nnu,Bern:2019crd,Cheung:2020gyp} results for the effective gravitational potential, as well as early results including spin effects \cite{Vaidya:2014kza,Guevara:2018wpp,Guevara:2019fsj,Maybee:2019jus,Bern:2020buy,Damgaard:2019lfh,Aoude:2020onz}. The computational method established so far
is somewhat intricate: starting from the scattering amplitude of two massive flavored scalar particles minimally
coupled to gravity, and taking a subtle classical limit \cite{Kosower:2018adc,Damour:2019lcq}, one matches the amplitudes obtained
to those of an EFT of non-relativistic scalar particles in order to determine its conservative two-body potential \cite{Cheung:2018wkq,Damour:2017zjx}.
The so-obtained effective potential is then used to compute observables such as
the scattering angle or the periastron advance in the bound system \cite{Kalin:2020mvi,Kalin:2019rwq,Kalin:2019inp}.

Both approaches --- involving the worldline EFT and modern scattering amplitudes --- 
agree on the final results for observables
and conservative potentials in the PM expansion;
the question of efficiency is a matter of debate (and taste). 
What has remained unclear, however, is whether there is a more
direct connection between the amplitude and worldline EFT approaches. 
The present work fills this gap.

Our key observation is that the Feynman-Schwinger or worldline
representation of the  graviton-dressed scalar propagator
\cite{Bastianelli:2000pt,Bastianelli:2002fv} provides this link. 
Inserting it into a time-ordered correlation function of scalars and gravitons
yields a precise map to expectation values of operators
in a worldline quantum field theory (WQFT). 
This WQFT is the same worldline EFT discussed above
\cite{Goldberger:2004jt,Goldberger:2006bd,Goldberger:2009qd,Kalin:2020mvi},
but with the important additional ingredient that the worldline trajectories
are also quantized. We write 
\be
x^\mu(\tau)=b^\mu+\tau v^\mu+z^\mu(\tau)\, ,
\label{eq:bvzdef}
\ee
where $z^\mu$ describes the  perturbation of a black hole from its
original straight-line trajectory in a binary scattering process, and integrate
out $z^\mu$ in the WQFT path integral together with the graviton $h_{\mu\nu}$.
So both the worldline and the graviton field are treated on an equal
footing in our approach.

Previous results for expressing observables of the black holes
(such as their deflections and radiation)
encountered through scattering amplitudes derived in \rcite{Kosower:2018adc}
follow elegantly from correlators in our WQFT.
The tedious procedure in the traditional worldline EFT approach 
of first finding the effective potential by integrating out the graviton ---
and thereafter solving the
resulting equations of motion in terms of a perturbative ansatz for the $z^{\mu}$ of \eqn{eq:bvzdef} --- is 
streamlined through WQFT Feynman rules,
which provide a fast track to the integrands yielding the observables.
The classical eikonal of the scattering of two massive particles, encoding the classical part of the 4-point amplitude, can be calculated directly from the WQFT.\footnote{Interestingly, a similar connection was indeed studied a long time ago in 
\rcite{Fabbrichesi:1993kz} recovering the eikonal result \cite{Amati:1988tn,Amati:1990xe}
for the ultra-relativistic limit of a string scattering computation.}
So we expect our new formalism to not only be of foundational interest in clarifying
the connection between scattering amplitudes and the worldline theory, but also to be of calculational advantage for 
precision calculations in the classical gravity two-body problem.
We demonstrate this by establishing
the sub-leading corrections to the deflection and radiation, the latter having not appeared in the
literature.

The rest of our paper is organized as follows.
In \Sec{sec:2} we introduce the Feynman-Schwinger representation
of the gravitationally dressed scalar propagator,
and demonstrate how it may be inserted into time-ordered correlation functions.
Then in \Sec{sec:3} we explain how to move from correlators to S-matrices,
by cutting the propagators of external legs.
We also begin our discussion of the eikonal phase of scalar scattering,
demonstrating that it corresponds precisely to the free energy of the WQFT.
In \Sec{sec:feynmanRules} we introduce Feynman rules for the WQFT,
which we can use to conveniently calculate expectation values in Fourier space.
Using these, in \Secs{sec:radiation}{sec:deflections} we
compute the 2PM radiation $k^2\braket{h^{\mu\nu}(k)}$ integrand
and deflection $\Delta p_1^\mu$
from an inelastic scattering of two black holes respectively,
drawing a close comparison with the equivalent amplitudes-based calculations.
In \Sec{sec:radiation} we also compute the 3PM
radiation integrand from an inelastic scattering of three black holes.
Finally, in \Sec{sec:eikonal} we revisit the eikonal phase and demonstrate how useful
observables, including the deflection and scattering angle,
can be obtained from it.
In \Sec{sec:discussion} we conclude.

\section{Worldline actions versus S-matrices}
\label{sec:2}

In this section we show how expectation values of operators in a worldline theory,
corresponding to gravitational observables,
can be directly obtained from S-matrices in the classical limit.
The link is formally provided by a worldline representation of the
massive scalar propagator in a fixed gravitational background,
which we refer to as the Feynman-Schwinger form.
First we rewrite the worldline action.

\subsection{Worldline action}

We seek to describe the scattering of two
(or more) unbound black holes. The spinless black holes
may be described in an effective field theory (EFT)
framework~\cite{Goldberger:2004jt} as relativistic massive particles
moving along their worldlines and coupled to gravity:
\begin{equation}
\label{eq:Sinit}
S=S_{\rm EH}+S_{\rm gf}+\sum_{i}S^{(i)}_{\rm pm}\,.
\end{equation}
$S_{\rm EH}$ is the usual Einstein-Hilbert action
(working in $D$ dimensions):
\begin{equation}\label{eq:einsteinHilbert}
S_{\rm EH}=-2m_{\rm Pl}^{D-2}\int\!\d^Dx\sqrt{-g}R\,.
\end{equation}
Using the weak field approximation we expand
\be
\label{eq:gravitondef}
g_{\mu\nu}=\eta_{\mu\nu}+\kappa h_{\mu\nu}\, ,
\ee
where $\kappa=m_{\rm Pl}^{1-D/2}$,
thereafter raising and lowering indices with the ``mostly minus'' Minkowski
metric $\eta_{\mu\nu}={\rm diag}(+1,-1,-1,-1)$.
Our gauge-fixing term $S_{\rm gf}$ is
\begin{equation}
S_{\rm gf}=\int\!\d^Dx\,
\big(\partial_\nu h^{\mu\nu}-\frac12\partial^\mu{h^\nu}_\nu\big)^2\,,
\end{equation}
which imposes the usual de Donder gauge condition
$\partial_\nu h^{\mu\nu}=\frac12\partial^\mu{h^\nu}_\nu$.

The point mass action for a single extended object (such as a black hole) moving along a worldline $x^{\mu}(\tau)$ and with proper time
$\d\tau = \sqrt{g_{\mu\nu} dx^{\mu}dx^{\nu}}$ reads
\begin{equation}\label{eq:pmAction2ndO}
S_{\rm pm}=- m \int\!\d\tau + c_{R} \int\!\d\tau R(x) + c_{V}
\int\!\d\tau R_{\mu\nu}(x) \dot{x}^{\mu}\dot{x}^{\nu} + \ldots \, .
\end{equation}
The first term induces geodesic motion with respect to the metric $g_{\mu\nu}$.
In addition, we allow for non-minimal couplings of the point mass
to the gravitational field parametrized by a priori unknown
Wilson coefficients $c_{R/V}$. There is an infinite number of
terms beyond these two organized in higher powers
of the curvature tensor and derivatives. These terms account for the internal structure
of the extended object to be described. 
The first two leading terms above do not contribute to physical observables as they may may be removed by a (singular) field redefinition of $h_{\mu\nu}$,
as was argued \rcite{Goldberger:2004jt} and subsequently 
demonstrated at 4PN in \rcite{Foffa:2019yfl}
(see also \rcite{Porto:2016pyg}).
We shall drop them for the time
being, yet the $c_{R}$ term will have a role to play shortly.

The point-mass action can also be written in a Polyakov form:
\begin{equation}\label{eq:pmAction}
	S_{\rm pm}=-\frac{m}2\int_{-\infty}^\infty\!\d\tau
	\left(e^{-1}g_{\mu\nu}\dot{x}^\mu\dot{x}^\nu+e\right)\,,
\end{equation}
where $e(\tau)$ is the einbein with equation of motion
$e^2=g_{\mu\nu}\dot{x}^\mu\dot{x}^\nu$.
Solving for $e(\tau)$ we recover the point-mass action in \eqn{eq:pmAction2ndO};
alternatively, we can gauge fix $e=1$ to enforce 
$g_{\mu\nu}\dot{x}^\mu\dot{x}^\nu=1$,
which identifies $\tau$ as the proper time.
In the latter case the point-mass action becomes
\begin{equation}\label{eq:Spm3}
S_{\rm pm}=-\frac{m}2\int_{-\infty}^{\infty}\!\d\tau\left(
g_{\mu\nu}\dot{x}^\mu \dot{x}^\nu+1\right)\,.
\end{equation}
This form of the particle action is superior to
the initial one \eqn{eq:pmAction2ndO} as it does not involve any square roots and only
displays a linear coupling of the worldline to the graviton field $h_{\mu\nu}$ in the weak-field expansion.

\subsection{Dressed propagators in the Feynman-Schwinger representation}

Next we consider a massive complex scalar field $\phi(x)$
coupled to Einstein gravity as the QFT avatar of a single black hole.
For a binary system one simply generalizes to
two differently flavored massive scalars $\phi_i(x)$.
The relevant action reads
\begin{align}\label{eq:Sprime}
\begin{aligned}
S' &= S_{\rm EH}+ S_{\rm gf} + \sum_{i=1}^{2} S_{i}\,,\\
\text{with}\quad  S_{i} &= \int\!\d^Dx\, \sqrt{-g} \left ( g^{\mu\nu}\, \partial_{\mu}\phi_{i}^{\dagger}
\partial_{\nu}\phi_{i} - m_i^{2}\, \phi_{i}^{\dagger}\phi_{i} 
- \xi\, R\, \phi_{i}^{\dagger} \phi_{i}\right )\,,
\end{aligned}
\end{align}
where we allow for a non-minimal coupling of the scalar field
to the background curvature
controlled by the dimensionless parameter $\xi$.
In a \emph{fixed gravitational background}
the associated Green's function $G(x,x')$ of the scalar field obeys 
the partial differential equation
\be
\left ( \nabla_{\mu}\nabla^{\mu}+ m^{2}+ \xi \, R\right )\, G(x,x') =
\sqrt{-g}\, \delta^{(D)}(x-x')\,,
\ee
where $\nabla_{\mu}$ denotes the gravitational covariant derivative,
\ie $ \nabla_{\mu}
\nabla^{\mu} G= 
 \partial_{\mu}\partial^{\mu}G + \Gamma^{\mu}{}_{\mu\nu}\partial^{\nu}G$.
There exists a worldline path integral representation
for $G(x,x')$ that we shall now review. 
 
Let us first consider the analogous situation in scalar QED. The
Green's function $G_{A}(x,x')=\langle \Omega | T\{\phi(x) \phi^{\dagger}(x')\}|\Omega\rangle$ for a massive charged scalar propagating in an electromagnetic background $A_{\mu}(x)$ 
 obeys
\be
\left ( {D}_{\mu}{D}^{\mu}+ m^{2} \right )\, G_{A}(x,x') =  \delta^{(D)}(x-x')
\ee
with $D_{\mu}=\partial_{\mu}+ie A_{\mu}$. It was first proposed by Feynman \cite{Feynman:1950ir} 
in the birth phase  of QED that this Green's function has
a worldline path integral representation\footnote{The derivation of this classic result is nicely reviewed in chapter 33 of \rcite{Schwartz:2013pla}.}
\be
\label{eq:GA}
G_{A}(x,x') = \int_{0}^{\infty}\!\d s\,  e^{-ism^{2}}\, \int_{x(0)=x}^{x(s)=x'} D [x]
\exp\Bigl[ -i\int_{0}^s\!\d \sigma\, \Big(\sfrac{1}{4} \eta_{\mu\nu}\frac{\d x^{\mu}}{\d\sigma}
\frac{\d x^{\nu}}{\d\sigma}
+ e A_{\mu} \, \dot{x}^{\mu}\Big) \Bigr ]
\, ,
\ee
which reduces to the Schwinger proper time representation of the propagator in the free ($e=0$) case.
Notice that $\sigma$ (and therefore $s$) has dimensions of $m^{-2}$,
so we distinguish it from the proper time $\tau$ with dimensions of $m^{-1}$.

This
worldline representation of the photon-dressed propagator --- which we refer to as the Feynman-Schwinger
representation --- is very efficient for computing effective actions at one-loop order, e.g.~to compute the
Euler-Heisenberg action in the case of constant electromagnetic field strengths.
The generalization to the non-abelian case is straightforward: simply insert a trace over 
color states in the path integrand and replace the gauge field by $A_{\mu}^{a}\, T^{a}$ with
the generators $T^{a}$ in the representation of the scalar. This representation of the
Green's function has been used for efficient calculations of one-loop amplitudes
and effective actions in gauge theories \cite{Strassler:1992zr},
and it also arises through the point particle limit of open strings \cite{Bern:1991aq} ---
see \rcites{Schubert:2001he,Edwards:2019eby} for comprehensive reviews.
 
In gravity the problem is more intricate and subject to a longer
discussion in the literature. Naively one
would expect to simply generalize \eqn{eq:GA} to a curved background upon promoting
$\eta_{\mu\nu}$ to $g_{\mu\nu}$, plus including possible curvature couplings:
\be
G(x,x') \sim   \int_{0}^{\infty}\!\d s\, e^{-ism^{2}}\int_{x(0)=x}^{x(s)=x'}\mathcal{D} [x] \, 
\exp\Bigl[ -i\int_{0}^s\!\d \sigma\, \left (\sfrac{1}{4} g_{\mu\nu} \frac{\d x^{\mu}}{\d\sigma}
\frac{\d x^{\nu}}{\d\sigma}
+ \tilde\xi\, R(x) \right )\Bigr ]
\, .
\ee
The first claim
of such a representation of the massive scalar Green's
function $G(x,x')$ in a \emph{gravitational} background as a worldline
path integral goes back to De Witt \cite{DeWitt:1957at} and Parker \cite{Parker:1979mf,Bekenstein:1981xe}.\footnote{They wrongly claimed
this result with $\tilde \xi = \xi -\frac{1}{3}$, with $\xi$ the non-minimal
coupling of \eqn{eq:Sprime}.}   One issue is that the path integral measure
becomes metric dependent, i.e.~schematically one has 
\be
\mathcal D [x] = D[x] \prod_{0\leq\sigma\leq s}\sqrt{-\text{det} g_{\mu\nu}[x(\sigma)] }\, ,
\ee
where $D[x]=\prod_{\sigma}\d^{D}x(\sigma)$ is the standard flat space path integral
measure. This metric dependence may be conveniently controlled through bosonic $\ga^{\mu}$
and fermionic $\gb^{\mu}, \gc^{\mu}$ ``Lee-Yang'' ghosts \cite{Bastianelli:1992ct}:\footnote{The fermionic path integral yields a factor of $[-\det g]$ while the bosonic one contributes $[-\det g]^{-1/2}$ yielding the desired total $[-\det g]^{1/2}$ .}
\be
\prod_{0\leq\sigma\leq s}\sqrt{-\text{det} g_{\mu\nu}[x(\sigma)]} = 
\int D[\ga,\gb,\gc]\, \exp\Bigl[ -i\int_{0}^s\!\d \sigma\, \left (\sfrac{1}{4} g_{\mu\nu} 
(\ga^{\mu}\ga^{\nu} + \gb^{\mu}\gc^{\nu} \right ) \Bigr ] \, .
\ee
With these ghosts included all divergences in the worldline QFT have been shown to cancel, yet a finite counter term  $\frac{1}{4} R(x)$ remains \cite{Bastianelli:2000pt}.\footnote{In non-covariant regularization schemes, such as mode regularization, additional terms proportional to Christoffel symbols appear, $-\frac{1}{12}g^{\mu\nu}g^{\rho\kappa}g_{\sigma\eta}
\Gamma^{\sigma}_{\mu\rho}\Gamma^{\eta}_{\nu\kappa}$.}

The upshot is the following representation of the scalar Green's function in a gravitational background that generalizes
\eqn{eq:GA} to the gravitational case \cite{Bastianelli:2000pt,Bastianelli:2002fv}:
\begin{align}\label{eq:BekensteinParker}
G(x,x') =&  \int_{0}^{\infty}\!\d s\,e^{-ism^{2}}\, \int_{x(0)=x}^{x(s)=x'}D [x] \, \int D[\ga,\gb,\gc] \\ &
\exp\Bigl[ -i\int_{0}^s\!\d \sigma\, \left (\sfrac{1}{4} g_{\mu\nu} \left (\frac{\d x^{\mu}}{\d\sigma}\frac{\d x^{\nu}}{\d\sigma}+ \ga^{\mu}\ga^{\nu} + \gb^{\mu}\gc^{\nu} \right)
+ (\xi-\sfrac{1}{4}) R(x) \right )\Bigr ]\nn
\, .
\end{align}
Writing $\sigma=\frac{\tau}{2m}$
(where $\tau$ is the proper time) and $s=\frac{T}{2m}$
yields an expression excitingly close to the worldline action
$S_{\rm pm}$ we obtained in \eqn{eq:Spm3}:
\begin{align}
\label{eq:BP2}
G(x,x') =& \int_{0}^{\infty}\frac{\d T}{2m} \, \int_{x(0)=x}^{x(T)=x'} D [x]
\exp\Bigl[ -i\int_{0}^T\!\d \tau\, \left (\frac{m}{2} g_{\mu\nu}\dot{x^{\mu}}\dot{x^{\nu}} + \frac{m}{2}
+ \sfrac{1}{2m}(\xi-\sfrac{1}{4}) R(x) \right )\Bigr ]  \nn \\
& \quad  \times\int D[\ga,\gb,\gc]\, \exp\Bigl[ -i\int_{0}^T\!\d \tau\, \left (\sfrac{1}{8m} g_{\mu\nu} 
(\ga^{\mu}\ga^{\nu} + \gb^{\mu}\gc^{\nu} \right ) \Bigr ] \, ,
\end{align}
that is if we ignore the ghosts and the non-minimal coupling to $R$. The ghosts are in fact non-propagating
and their purpose in life is to cancel divergences of coinciding worldline fields, 
i.e.~$\langle \xdot^{\mu}(\tau)\xdot^{\nu}(\tau)\rangle \sim \delta(0)$.
A graphical representation of the gravitationally dressed Green's function in the
weak-field approximation is given in \Fig{fig:SchwingerProperTime}.

\begin{figure}[t]
  \centering
$$
G(x,x')=
\raisebox{0.1cm}{\begin{tikzpicture}[baseline={(current bounding box.center)}]
\node (in) at (-0.75,0) {$x$};
\node (out) at (0.75,0) {\raisebox{0.1cm}{$x'$}};
\draw [worldlineBold] (in) -- (out);
\end{tikzpicture}}
+
\raisebox{0.55cm}{
\begin{tikzpicture}[baseline={(current bounding box.center)}]
\node (in) at (-1,0) {$x$};
\node (out) at (1,0) {\raisebox{0.1cm}{$x'$}};
\coordinate (x) at (0,0);
\node (k) at (0,1) {$h$};
\draw [fill] (x) circle (.08);
\draw [worldlineBold] (in) -- (x);
\draw [worldlineBold] (x) -- (out);
\draw [graviton] (x) -- (k);
\end{tikzpicture}}
+
\raisebox{0.55cm}{\begin{tikzpicture}[baseline={(current bounding box.center)}]
\node (in) at (-1,0) {$x$};
\node (out) at (1,0) {\raisebox{0.1cm}{$x'$}};
\coordinate (x1) at (-0.2,0);
\coordinate (x2) at (0.2,0);
\node (k1) at (-0.3,1) {$h$};
\node (k2) at (0.3,1) {$h$};
\draw [worldlineBold] (in) -- (x1);
\draw [worldlineBold] (x1) -- (x2);
\draw [worldlineBold] (x2) -- (out);
\draw [graviton] (x1) -- (k1);
\draw [graviton] (x2) -- (k2);
\draw [fill] (0,0) ellipse (0.45 and 0.1);
\end{tikzpicture}}
+
\raisebox{0.55cm}{\begin{tikzpicture}[baseline={(current bounding box.center)}]
\node (in) at (-1,0) {$x$};
\node (out) at (1,0) {\raisebox{0.1cm}{$x'$}};
\coordinate (x1) at (-0.3,0);
\coordinate (x2) at (0.3,0);
\coordinate (x3) at (0,0);
\node (k1) at (-0.4,1) {$h$};
\node (k2) at (0.4,1) {$h$};
\node (k3) at (0,1) {$h$};
\draw [worldlineBold] (in) -- (x1);
\draw [worldlineBold] (x1) -- (x3);
\draw [worldlineBold] (x3) -- (x2);
\draw [worldlineBold] (x2) -- (out);
\draw [graviton] (x1) -- (k1);
\draw [graviton] (x2) -- (k2);
\draw [graviton] (x3) -- (k3);
\draw [fill] (0,0) ellipse (0.55 and 0.1);
\end{tikzpicture}}
+ \ldots
$$
\caption{Graphical representation of the Green function $G(x,x')$ of \eqn{eq:BP2} for a massive scalar
moving in a  weak gravitational background $g_{\mu\nu}=\eta_{\mu\nu}+\kappa h_{\mu\nu}$,
excluding the $R\phi^2$ interaction in \eqn{eq:Sprime}.
A closed expression for the Green's function
 in momentum space may be found in \eqn{eq:Fromaggio}.
 }
\label{fig:SchwingerProperTime}
\end{figure}

\subsection{From the S-matrix to the worldline}

Using the gravitationally dressed Green's function $G$ we can now
write S-matrix elements as expectation values of operators in the worldline theory.
Assuming a fixed gravitational background we write $G$ as a two-point function via a genuine quantum field theoretical path integral:
\begin{equation}
G_i(x,x')=\mathcal{Z}_{i}^{-1}\int\!\cD[\phi_i]\,\phi_i(x)\phi_i^\dagger(x')\,
e^{iS_i}\,.
\end{equation}
For the black hole scattering we are interested in we require the S-matrix element
of two scalars with or without a final state graviton
$\phi_{1}\, \phi_{2}\to \phi_{1}\, \phi_{2}(+h)$
in the \emph{classical limit}, \ie~suppressing
virtual loops in the process. These processes may be computed
by inserting two gravitationally dressed
Green's functions $G_{i}$ with masses $m_{i}$ into the gravitational path integral. 
Consider the time-ordered correlator:
\begin{align}\label{eq:timeOrderedCorrelator}
\begin{aligned}
&\langle \Omega |T\{ h_{\mu\nu}(x)\phi_1(x_1)\phi^\dagger_1(x_1')\phi_2(x_2)\phi^\dagger_2(x_2')\} |\Omega
\rangle\\
&\qquad=\tilde{\mathcal{Z}}^{-1}\int\!\cD[h_{\mu\nu},\phi_1,\phi_2]\,
h_{\mu\nu}(x)\phi_1(x_1)\phi^\dagger_1(x_1')\phi_2(x_2)\phi^\dagger_2(x_2')\, e^{iS'}\\
&\qquad=\mathcal{Z}^{-1}\int\!\cD[h_{\mu\nu}]\, h_{\mu\nu}(x)\, G_1(x_1,x_1')\, G_2(x_2,x_2')\, e^{i(S_{\rm EH}+ S_{\rm gf})}\,.
\end{aligned}
\end{align}
In the last step of integrating out the scalars $\phi_{1}$ and $\phi_{2}$
we have neglected virtual scalar loops that are mediated via gravitons,
which is acceptable in the classical limit.
For a pure $2\to 2 $ scattering without a radiated
graviton simply drop $h_{\mu\nu}(x)$ above.

The S-matrix then follows via
LSZ reduction and Fourier transforming to momentum space:
\begin{align}\label{eq:AmpinG12}
\braket{\phi_1\phi_2(+h)|S|\phi_1\phi_2}
=& \mathcal{Z}^{-1} \int\!\d^D[x_{i},x_{i}',x]\, e^{ip_{i}\cdot x_{i}-ip_{i}'\cdot x_{i'}(-ik\cdot x)}\\ 
\int \mathcal{D}[h_{\mu\nu}]\,  & \left (\epsilon^{\mu\nu}(k) h_{\mu\nu}(x) \right )\, {G}_{1}(x_{1},x'_{1})\, 
{G}_{2}(x_{2},x'_{2})\, e^{i (S_{\rm EH}+ S_{\rm gf})}\, \Bigr |_{\stackrel{\text{amputated}}{\text{\tiny connected}}}\,.\nn
\end{align}
Note that in the path integral above pure scalar loops never appear, which is why this relation only holds in the $\hbar\to 0$ limit. The classical limit
on the right-hand side then additionally suppresses  virtual gravitons in the loops, as well as mixed loops
of gravitons and worldline fluctuations that we will describe shortly.
Now inserting the worldline path integral representation of the $G_{i}$ from \eqn{eq:BP2}
 on the
right-hand side of \eqn{eq:AmpinG12} we see that the emerging action in the exponent of the path integral ---
which should now be interpreted as a QFT on the worldline coupled to the gravitational path integral ---
is very close to the worldline expression we
arrived at in eqs.~(\ref{eq:Sinit}) and (\ref{eq:Spm3}). 
Yet, there are two decisive differences that we shall discuss in turn.
Firstly, the worldline action of \eqn{eq:Spm3} calls for an integral over infinite total
proper time $\tau\in[-\infty,\infty]$, whereas 
in \eqn{eq:BP2} we integrate over an ensemble of finite proper times  $\tau\in[0,T]$. 
Secondly, there is the coupling to the Ricci scalar along the worldline appearing 
in \eqn{eq:BP2}, which was in principle also allowed in \eqn{eq:pmAction2ndO}.

We shall deal with the first point in the following section as it requires a detailed analysis
of the LSZ reduction. 
Addressing the second point, we argue that the non-minimal gravitational $\xi$-coupling of scalars in 
the action \eqref{eq:Sprime} is irrelevant for the classical limit of the S-matrix 
$\phi_{1}\, \phi_{2}\to \phi_{1}\, \phi_{2}(+h)$. For this consider the leading Feynman vertex
originating from the interaction  term
$\xi\int\!\d^Dx \sqrt{-g}\,R \phi_{i}^{\dagger}\phi_{i}$ in  \eqn{eq:Sprime}:
\be
\label{eq:FDxi}
\begin{tikzpicture}[baseline={(current bounding box.center)}]
\coordinate (in) at (0,-1);
\coordinate (out) at (0,1);
\coordinate (x) at (0.5,0);
\node (k) at (2,0) {$\mu\nu$};
\draw (out) node [right] {$\phi_{i}$};
\draw (in) node [right] {$\phi_{i}$};
\draw [cscalar] (in) -- (x);
\draw [cscalar] (x) -- (out);
\draw [graviton] (x) -- (k) node [midway, below] {$q$};
\draw [fill] (0.4,-0.1) rectangle (0.6,0.1);
\end{tikzpicture}= i \xi\, \kappa
\left ( q^{2} \eta_{\mu\nu} - q_{\mu} q_{\nu} \right )\,.
\ee
The important point is that it couples quadratically to the transfered momentum $q$.
As was pointed out in \rcite{Kosower:2018adc} the classical limit of a $\phi_{1}\, \phi_{2}\to \phi_{1}\, \phi_{2}$ scattering process amounts to taking the momentum transfer to zero
($q= \hbar \bar{q} $ with $\hbar\to 0$). Hence, there is no contribution of this term to the classical limit of the amplitude.\footnote{Note that here it is important that $q$ appears quadratically:
The linear terms in $q$ in the numerators turn out to be the leading contributions as
the $q$-independent (``superclassical'') terms cancel out, see \cite{Kosower:2018adc}
and an explicit demonstration in section \ref{sect:5.1}.} So we may conveniently set $\xi=\sfrac{1}{4}$ in \eqn{eq:BP2} 
to remove it from the worldline action. This argument is in line with the arguments presented
in \rcite{Goldberger:2004jt} for disregarding the Ricci scalar coupling on a worldline quantum field theory
in the  classical limit. 

In summary: we have shown that there is a direct connection
between scalar-graviton S-matrices and the worldline QFT
in the classical limit via the path integral representation of the
gravitationally dressed scalar propagator given in \eqn{eq:BP2}. 

\section{Graviton-dressed propagator for a massive scalar field}
\label{sec:3}

In the previous section we showed how the Feynman-Schwinger representation
of a gravitationally dressed scalar propagator could be inserted
into a QFT correlator, yielding an expectation value in the worldline theory.
However, to study S-matrices we must still apply LSZ reduction.
This will convert correlators into S-matrices by cutting
the propagators on their external legs,
sending those states to the boundary where they interact weakly.
In this section we achieve this from the worldline perspective
by first deriving a momentum space representation
of the gravitationally dressed propagator.
The overall effect of putting the scalar legs on-shell 
is to switch from a worldline action integrated over a finite proper time domain
to one over an infinite domain $\tau\in[-\infty,\infty]$.
We will then compare with the expectation values
one would compute in a worldline QFT.
As our first example we examine the eikonal phase of a $2\to2$
S-matrix in the classical limit,
which corresponds to the free energy of the worldline theory.

\subsection{Momentum space representation}
\label{sec:momSpaceRep}

Let us now introduce a master formula for the gravitationally
dressed two-point function
of a massive scalar field coupled to $N$ external gravitons with all legs off-shell,
\ie the momentum space version of $G(x,x')$ in \Fig{fig:SchwingerProperTime}. 
We work in the \emph{non-minimally} coupled theory
with $\xi=1/4$ in \eqns{eq:Sprime}{eq:BekensteinParker}.
To our knowledge only the single-graviton $N=1$ case has been established so far
\cite{Ahmadiniaz:2019ppj}.

Starting from the position space propagator $G(x,x')$ in
\eqn{eq:BekensteinParker} we insert a weak gravitational background
of the form
\begin{equation}\label{eq:gravBackground}
h_{\mu\nu}=\sum_{l=1}^{N}\, \epsilon_{\mu\nu}^{(l)} e^{i\, k_{l}\cdot x(\sigma_{l})}
\end{equation}
into the path integral representing $N$ (off-shell) gravitons ---
we do not require $k_{l}^{2}=0$ or $k_{l}\cdot \epsilon_{l}=0$.
In order to deal with the boundary conditions of the $x^{\mu}(\sigma)$ path integral we perform a background field expansion
about straight line trajectories (which solve the flat space equations of motion):
\be
x^{\mu}(\sigma) = x^{\mu}+ \dx^{\mu} \frac{\sigma}{s}\ + q^{\mu}(\sigma)
\, , \quad \dx^{\mu}= x^{\prime\, \mu}-x^{\mu}\, .
\ee
Inserting this
and Fourier transforming \eqn{eq:BekensteinParker} in $x$ and $x'$ to the
momentum space variables $p$ and $p'$ for the scalar particles yields\footnote{We include the $i\epsilon$ prescription to make the $s$ integral well-defined. It leads to the bulk Feynman propagator in the final result.}
\begin{align}\label{eq:momProp}
D(p,p',\{\epsilon^{(l)},&k_{l}\})  = \left(\!-\frac{i\kappa}{4}\right)^N
\int_{0}^{\infty}\!\d s\, e^{-is(m^{2}-i\epsilon)}
\int\!\d^{D}x\int\!\d^{D}x' e^{i(p\cdot x - p'\cdot x')- \frac{i}{4s}\dx^{2}} \\
&
\Bigl \langle\prod_{l=1}^{N}\int_{0}^s\!\d \sigma_{l}\, \epsilon^{(l)}_{\mu\nu}
 \Bigl ( \xdot^{\mu}(\sigma_{l})\xdot^{\nu}(\sigma_{l})+ \ga^{\mu}(\sigma_{l})\ga^{\nu}(\sigma_{l})
+ \gb^{\mu}(\sigma_{l})\gc^{\nu}(\sigma_{l}) \Bigr ) e^{ik_{l}\cdot x(\sigma_{l})}\Bigr\rangle\,.\nn
\end{align}
We take $p$ as ingoing and $p'$ as outgoing.
The expectation value above is defined as an unnormalized path integral over the
fluctuations $q$ and the ghost fields:
\be
\langle \mathcal{O}(\ga,\gb,\gc,q) \rangle := \int D[q,\ga,\gb,\gc]\,  \mathcal{O}(\ga,\gb,\gc,q)\, e^{-i\int_{0}^s\!\d \sigma \frac{1}{4}({\dot q}^{2} +\ga^{2} + \gb\cdot \gc)}\, .
\ee
All fluctuating  fields now have vanishing boundary conditions.

Our task now is to evaluate the correlator in \eqn{eq:momProp},
and then take the Fourier transform.
For this we insert the relevant two-point functions on the worldline:
\begin{align}
\begin{aligned}
\vev{q^{\mu}(\sigma) q^{\nu}(\sigma')}&= 
	\phantom{+}2i \eta^{\mu\nu}\, \Delta(\sigma,\sigma') \,, \\
\vev{\ga^{\mu}(\sigma) \ga^{\nu}(\sigma')}&=
	-2i \eta^{\mu\nu}\, \delta(\sigma-\sigma')\,,\\
\vev{\gb^{\mu}(\sigma) \gc^{\nu}(\sigma')}&=
	\phantom{+}4i \eta^{\mu\nu}\, \delta(\sigma-\sigma')\,,
\end{aligned}
\end{align}
where the propagator on a worldline of finite length $s$
is (see \eg \rcite{Schubert:2001he})
\begin{align}
\label{eq:DeltaProps}
\Delta(\sigma,\sigma') &=
\sfrac{1}{2} | \sigma-\sigma'| + \frac{\sigma\sigma'}{s} - \frac{\sigma+\sigma'}{2} 
\, , \qquad \text{with } \sigma,\sigma'\in[0,s]\,.
\end{align}
It is a straightforward exercise to evaluate the path integrals,
though as the details are somewhat involved
a full discussion is relegated to \App{app:propDeriv}.
The final result is a compact master formula for the
gravitationally dressed scalar propagator:
\begin{align}
\label{eq:Fromaggio}
D(p,p',\{\epsilon^{(l)},&k_{l}\})  = \left(\!-\frac{i\kappa}{4}\right)^N\delta^{(D)}(p-p'+\sum_{l=1}^{N}k_{l})
\, \int_{0}^{\infty}\!\d s \,  e^{is(p^{\prime\, 2}-m^{2}+i\epsilon) }\nn\\
\prod_{l=1}^{N}& \int_{0}^s\!\d \sigma_{l}\, 
\epsilon^{(l), \, \mu\nu}\,
\Bigl [  \partial_{\epsilon_{l}^{\mu}}\partial_{\epsilon_{l}^{\nu}}   
+ \partial_{\alpha_{l}^{\mu}}\partial_{\alpha_{l}^{\nu}}+ \partial_{\beta_{l}^{\mu}}\partial_{\gamma_{l}^{\nu}}
\Bigr ]
 \exp \Biggl [- (p+p')\cdot \sum_{l=1}^{N}(ik_{l}\sigma_{l}+\epsilon_{l}) \nn\\
&-i \sum_{l,l'=1}^{N}\Bigl \{
\frac{|\sigma_{l}-\sigma_{l'}|}{2}k_{l}\cdot k_{l'} - i \, \text{sign}(\sigma_{l}-\sigma_{l'})\, \epsilon_{l}\cdot k_{l'} \\ & \qquad
+ \delta(\sigma_{l}-\sigma_{l'}) (\epsilon_{l}\cdot \epsilon_{l'} + \alpha_{l}\cdot \alpha_{l'}
-4 \gamma_{l}\cdot \beta_{l'})
\Bigr \}\Biggr ]\Biggr|_{\epsilon_{l}=\alpha_{l}=\beta_{l}=\gamma_{l}=0}\, .\nn
\end{align}
Here we have introduced fiducial ``polarization''
vectors $\epsilon_{l}^{\mu}$ and $\alpha_{l}^{\mu}$,
as well as anti-commuting vectors $\beta_{l}^{\mu}$ and  $\gamma_{l}^{\mu}$.
The expression is remarkably similar (in the double copy sense)
to the one obtained for the $N$-photon-dressed 
\cite{Daikouji:1995dz,Ahmadiniaz:2015kfq} propagator
: to insert
a photon leg in lieu of a graviton one simply takes
a single $\partial_{\epsilon_{l}^{\mu}}$ derivative there\footnote{
It would be interesting to work out the double copy relation to the
 $N$-gluon dressed propagator found in \cite{Ahmadiniaz:2015xoa} in detail.}.

To better understand this formula it is instructive to work
out the single graviton ($N=1$) case.
Noting $\text{sign}(0)=0$ and the cancellation of the $\delta(0)$
terms when all polarization derivatives hit the same leg $l=l'$, \ie
\be
\epsilon^{(l), \, \mu\nu}\,\Bigl [ \partial_{\epsilon_{l}^{\mu}}\partial_{\epsilon_{l}^{\nu}}   
+ \partial_{\alpha_{l}^{\mu}}\partial_{\alpha_{l}^{\nu}}+ \partial_{\beta_{l}^{\mu}}\partial_{\gamma_{l}^{\nu}}
\Bigr ]
(\epsilon_{l}\cdot \epsilon_{l} + \alpha_{l}\cdot \alpha_{l}
-4 \gamma_{l}\cdot \beta_{l}) = (2 + 2 - 4) \epsilon^{(l), \, \mu}{}_{\mu} =0\nn
\ee
one straightforwardly works out
\be
D(p,p',k;\epsilon) = \frac{i}{p^{2}-m^{2}+i\epsilon}\,\frac{i}{p^{\prime\, 2}-m^{2}+i\epsilon}\,(-\sfrac{i\kappa}{4}) \, \delta^{(D)}(p-p'+k)\,
(p+p')^{\mu}\, (p+p')^{\nu} \, \epsilon_{\mu\nu}\, ,
\ee
reproducing \rcite{Ahmadiniaz:2019ppj}.
Amputating the scalar legs and stripping off the momentum-conserving
$\delta^{(D)}(P)$ function and polarization tensor we obtain the three-point vertex
\be
-\frac{i\kappa}{4} \, (p+p')^{\mu}\, (p+p')^{\nu} \, .
\ee
Let us compare this result to the QFT three-point vertex of two scalars
and a graviton. 
For a general $\xi$ coupling there are two vertices: the 
three-point interaction vertex between two 
scalars $\varphi$ and a graviton $h_{\mu\nu}$ from the minimal coupling (all scalar
momenta ingoing):
\begin{align}
\begin{tikzpicture}[baseline={(current bounding box.center)}]
\coordinate (in) at (-1,0);
\coordinate (out) at (1,0);
\coordinate (x) at (0,0);
\node (k) at (1,-1) {$q$};
\draw (in) node [left] {$p$};
\draw (out) node [right] {$p'$};
\draw [fill] (x) circle (.08);
\draw [cscalar] (in) -- (x);
\draw [cscalar] (x) -- (out);
\draw [graviton] (x) -- (k);
\end{tikzpicture}=-i\kappa \left [
p^{(\mu}p^{\prime\nu)}-\frac{\eta^{\mu\nu}}2\left(p\cdot p'-m^2\right) \right ]
\end{align}
To this we need to  add the non-minimal $\xi$ coupling vertex of \eqn{eq:FDxi}
\begin{align}
\begin{tikzpicture}[baseline={(current bounding box.center)}]
\coordinate (in) at (0,-0.8);
\coordinate (out) at (0,0.8);
\coordinate (x) at (0.4,0);
\node (k) at (1.7,0) {${}_{\mu\nu}$};
\draw [cscalar] (in) -- (x) node [midway, right] {$p$};
\draw [cscalar] (x) -- (out) node [midway, right] {$p'$};
\draw [graviton] (x) -- (k) node [midway, below] {$q$};
\draw [fill] (x) circle (.1);
\end{tikzpicture}
+
\begin{tikzpicture}[baseline={(current bounding box.center)}]
\coordinate (in) at (0,-0.8);
\coordinate (out) at (0,0.8);
\coordinate (x) at (0.4,0);
\node (k) at (1.7,0) {${}_{\mu\nu}$};
\draw [cscalar] (in) -- (x) node [midway, right] {$p$};
\draw [cscalar] (x) -- (out) node [midway, right] {$p'$};
\draw [graviton] (x) -- (k) node [midway, below] {$q$};
\draw [fill] (0.3,-0.1) rectangle (0.5,0.1);
\end{tikzpicture}
=& -i\kappa \Bigl [
p^{(\mu}p^{\prime\nu)}-\frac{\eta^{\mu\nu}}2\left(p\cdot p'-m^2\right)-
 \xi\, \left ( q^{2} \eta_{\mu\nu} - q_{\mu} q_{\nu} \right ) \Bigr ] \nn\\
 =& -i\kappa \Bigl [ \sfrac 1 4 (p+p')^{\mu}(p+p')^{\nu} +(\sfrac 1 4 - \xi) \left ( q^{2} \eta_{\mu\nu} - q_{\mu} q_{\nu} \right )\Bigr ]
\end{align}
where --- crucially ---
in the last line we have used the on-shell condition $p^{2}=m^{2}=p^{\prime 2}$
on the \emph{scalar} legs.
We have a match for $\xi=\frac{1}{4}$,
but only if we put the scalar legs on-shell.\footnote{This might be in fact the simplest derivation of the $\tilde\xi=\xi-\frac{1}{4}$ relation in \eqn{eq:BekensteinParker}.}

It is a simple exercise to also include the $\tilde\xi R[x(\tau)]$ term in the
worldline action and perform the path integral for $N=1$ as well.
One quickly arrives at the above expression for the general $\xi$ case.

\subsection{Putting the scalar legs on shell}

Now that we have a momentum space representation of the gravitationally
dressed scalar propagator we can proceed to put the scalar legs on-shell.
As we have already seen, this is necessary in order to match to the
QFT expression which is then effectively a form factor
$F(p,p';\{k_{i},\epsilon_{i}\})$ with off-shell graviton legs:
\be
F(p,p'|\{k_{i},\epsilon_{i}\}) = 
\vev{-p'|\prod_{i=1}^{N} \epsilon_{i}\cdot h(k_{i})|p}=
\raisebox{0.75cm}{\begin{tikzpicture}[baseline={(current bounding box.center)}]
\node (in) at (-1.6,0) {$p$};
\node (out) at (3.0,0) {\raisebox{0.1cm}{$p'$}};
\coordinate (x1) at (-0.4,0);
\coordinate (x2) at (0.4,0);
\coordinate (x3) at (1.8,0);
\node (k1) at (-0.6,1.4) {$\epsilon_{1},k_{1}$};
\node (k2) at (0.4,1.4) {$\epsilon_{2},k_{2}$};
\node (kn) at (1.2,1.4) {$\ldots$};
\node (k3) at (2.0,1.4) {$\epsilon_{n},k_{n}$};
\draw [fill] (x1) circle (.08);
\draw [fill] (x2) circle (.08);
\draw [fill] (x3) circle (.08);
\draw [cscalar] (in) -- (-0.5,0);
\draw [worldline] (x1) -- (x2);
\draw [dashed] (x3) -- (x2);
\draw [cscalar] (2.0,0) -- (out);
\draw [graviton] (x1) -- (k1);
\draw [graviton] (x2) -- (k2);
\draw [graviton] (x3) -- (k3);
\draw [fill] (0.7,0) ellipse (1.3 and 0.2);
\end{tikzpicture}}\,.
\ee
Let us perform the LSZ reduction on
$D(p,p',\{\epsilon,k\})$ of \eqn{eq:Fromaggio} now.
First we put the outgoing $p'$ scalar leg on shell:
\be
-i(p^{\prime\, 2}-m^{2}+i\epsilon)\, D(p,p',\{\epsilon,k\})\Bigr |_{p^{\prime\, 2}=m^{2}-i\epsilon}\,.
\ee
Therefore we pull the inverse propagator into
the $s$ integral in \eqn{eq:Fromaggio} and use
\be
\label{eq:An1}
-i(p^{\prime\, 2}-m^{2}+i\epsilon)\, \, \int_{0}^{\infty}\!\d s \,  e^{is(p^{\prime\, 2}-m^{2}+i\epsilon) }  \Omega(s) = - \, \int_{0}^{\infty}\!\d s \, \frac{\d}{\d s}
\left ( e^{is(p^{\prime\, 2}-m^{2}+i\epsilon) } \right )\, \Omega(s)
\ee
where we have introduced
\begin{align}
\label{eq:defsigmas}
\Omega(s):=& \left(-\frac{i\kappa}{4}\right)^{N}\, \delta^{(D)}(p-p'+\sum_{l=1}^{N}k_{l})
\, \prod_{l=1}^{N} \int_{0}^s\!\d \sigma_{l}\, 
\epsilon^{(l), \, \mu\nu}\,
\Bigl [  \partial_{\epsilon_{l}^{\mu}}\partial_{\epsilon_{l}^{\nu}}   
+ \partial_{\alpha_{l}^{\mu}}\partial_{\alpha_{l}^{\nu}}+ \partial_{\beta_{l}^{\mu}}\partial_{\gamma_{l}^{\nu}}
\Bigr ] \nn\\ &
 \exp \Biggl [ -(p+p')\cdot \sum_{l=1}^{N}(ik_{l}\sigma_{l}+\epsilon_{l}) -i \sum_{l,l'=1}^{N}\Bigl \{
\frac{|\sigma_{l}-\sigma_{l'}|}{2}k_{l}\cdot k_{l'} - i \, \text{sign}(\sigma_{l}-\sigma_{l'})\, \epsilon_{l}\cdot k_{l'} \nn \\ & \qquad
+ \delta(\sigma_{l}-\sigma_{l'}) (\epsilon_{l}\cdot \epsilon_{l'} + \alpha_{l}\cdot \alpha_{l'}
-4 \gamma_{l}\cdot \beta_{l'})
\Bigr \}\Biggr ]\Biggr|_{\epsilon_{l}=\alpha_{l}=\beta_{l}=\gamma_{l}=0}\, .
\end{align}
Partially integrating \eqn{eq:An1} and using $\Omega(0)=0$ yields
\be
-i(p^{\prime\, 2}-m^{2}+i\epsilon)\, D(p,p',\{\epsilon,k\})\Bigr |_{p^{\prime\, 2}=m^{2}-i\epsilon}
= \Omega(\infty) \Bigr |_{p^{\prime\, 2}=m^{2}-i\epsilon} \, .
\ee
The overall effect is therefore to send $s\to\infty$.\footnote{See also the recent \cite{Bonocore:2020xuj} making the same argument}. 

It remains to put the incoming  $p$ scalar leg on shell.
For this we first go to ``center of mass'' proper time coordinates:
\be
\label{CoMPTV}
\tilde{\sigma}_{l}:=\sigma_{l}- \sigma_{+} \qquad \text{with} \quad
 \sigma_{+}= \frac{1}{N}\sum_{l=1}^{N}\sigma_{l}\,,
 \ee
 and we pick up the constraint $\sum_{l}\tilde{\sigma}_{l}=0$.
 The $N$-fold integral over the $\sigma_{l}$'s may then be rewritten as
 \be
 \prod_{l=1}^{N} \int_{0}^{\infty}\!\d \sigma_{l} \ldots =
 \prod_{l=1}^{N} \int_{-\infty}^{\infty}\!\d\tilde{\sigma}_{l} \, \int_0^\infty\!\d\sigma_{+}
 \, \delta\left (\sum_{l=1}^{N}\tilde{\sigma}_{l} \right)\ldots 
 \ee
 Note the change of the integration region to $\mathbb{R}$ in the new proper-time coordinates
 $\tilde\sigma_{l}$ matching the one performed in the worldline QFT.
 Moreover, as $\sigma_{l}-\sigma_{l'}=\tilde{\sigma}_{l}-\tilde{\sigma}_{l'}$ the variable $\sigma_{+}$ only couples
 to the $\exp[-i(p+p')\cdot\sum_{l}k_{l}\sigma_{l}]$ term in $\Omega(\infty)$ from \eqn{eq:defsigmas}.
 One then easily performs the $\sigma_{+}$ integral:
 $$
 \int_0^\infty\!\d\sigma_{+} e^{-i(p+p')\cdot \sum_{l=1}^{N}k_{l}\sigma_{l}}
 = 
e^{-i(p+p')\cdot \sum_{l=1}^{N}k_{l}\tilde{\sigma}_{l}} \int_0^\infty\!\d\sigma_{+} e^{i(p+p')\cdot (p-p') \sigma_{+}} = \frac{ie^{-i(p+p')\cdot \sum_{l=1}^{N}k_{l}\tilde{\sigma}_{l}}}{p^{2}-m^{2}+i\epsilon}\,,
$$
 where we have used total momentum conservation and the mass-shell condition for $p'$.
 But this precisely extracts the incoming scalar propagator!

Hence the net effect of
LSZ reducing the graviton dressed propagator of \eqn{eq:Fromaggio} 
to a form factor is very mild and can be done explicitly: drop
the overall $s$ integral, insert a total proper-time delta function and take the
proper time integrals to run over $\mathbb{R}$.\footnote{These steps to put the scalar legs on shell apply generically
to \emph{any} Feynman-Schwinger representation of a gluon, photon or graviton dressed propagator, and have to the best of our knowledge not been observed before.} The final result is (dropping the tildes on $\sigma$)
\begin{align}
\label{eq:OnShellFromaggio}
F(&p,p'|\{\epsilon^{(l)},k_{l}\})  = \left(-\frac{i\kappa}{4}\right)^{N} 
\delta^{(D)}(P)
\,  \prod_{l=1}^{N}\Bigl [ \int_{-\infty}^{\infty}\!\d\sigma_{l}\, 
\epsilon^{(l), \, \mu\nu}\,
\Bigl (  \partial_{\epsilon_{l}^{\mu}}\partial_{\epsilon_{l}^{\nu}}   
+ \partial_{\alpha_{l}^{\mu}}\partial_{\alpha_{l}^{\nu}}+ \partial_{\beta_{l}^{\mu}}\partial_{\gamma_{l}^{\nu}}
\Bigr ) \Bigr ]\nn\\ &
 \delta\left (\sum_{l=1}^{N}\sigma_{l} \right)\
 \exp \Biggl [- (p+p')\cdot \sum_{l=1}^{N}(ik_{l}\sigma_{l}+\epsilon_{l}) 
 -i \sum_{l,l'=1}^{N}\Bigl \{
\frac{|\sigma_{l}-\sigma_{l'}|}{2}k_{l}\cdot k_{l'}\\
&- i \, \text{sign}(\sigma_{l}-\sigma_{l'})\, \epsilon_{l}\cdot k_{l'} 
+ \delta(\sigma_{l}-\sigma_{l'}) (\epsilon_{l}\cdot \epsilon_{l'} + \alpha_{l}\cdot \alpha_{l'}
-4 \gamma_{l}\cdot \beta_{l'})
\Bigr \}\Biggr ]\Biggr|_{\footnotesize\begin{matrix}
\epsilon_{l}=\alpha_{l}=\beta_{l}=\gamma_{l}=0; \cr 
p^{2}=m^{2}-i\epsilon=p^{\prime\, 2}\end{matrix}}\nn
\end{align}
with $P=p-p'+ \sum_{l} k_{l}$.
This is a surprisingly compact
result for an $N$-graviton emission expression.

\subsection{Link to position space expression}

Let us see how this form factor
relates to the analogous expressions one
would compute in a worldline QFT (WQFT).
Here the starting point is that of \eqn{eq:BekensteinParker},
except with an integral over infinitely extended proper times:
\begin{align}\label{eq:sinftystart}
\Xi(b,v;\{\epsilon^{(l)},k_{l}\}) := \int D [x] \, \int D[\ga,\gb,\gc]
\exp\Bigl[ -i\int_{-\infty}^\infty\!\d \sigma\left (\sfrac{1}{4} g_{\mu\nu} \left (\xdot^\mu\xdot^\nu+ \ga^{\mu}\ga^{\nu} + \gb^{\mu}\gc^{\nu} \right)\right )\Bigr ],
\end{align}
where again we begin with a collection of plane waves for the graviton
with momenta $k_l$ and polarizations $\eps^{(l)}$: \eqn{eq:gravBackground}.
This is equivalent to
\begin{align}
\Xi(b,v;&\{\epsilon^{(l)},k_{l}\})= \\
&  \left (-\frac{i\kappa}4\right)^N
\int D [x] \, \int D[\ga,\gb,\gc] 
\prod_{l=1}^{\infty}\Bigl [ \int_{-\infty}^{\infty} d\sigma_{l}\, \epsilon^{\mu\nu}_{l}
\Bigl (\xdot^{\mu}(\sigma_{l}) \xdot^{\nu}(\sigma_{l})+
\ga^{\mu}(\sigma_{l}) \ga^{\nu}(\sigma_{l}) \nn\\
&+  \gb^{\mu}(\sigma_{l}) \gc^{\nu}(\sigma_{l}) \Bigr) e^{ik_{l}\cdot x(\sigma_{l})}   \Bigr ]
\,
\exp\Bigl[ -i\int_{-\infty}^{\infty}\!\d\sigma\, \left (\frac{1}{4} \left (\xdot^{2}(\sigma)+ \ga^{2}(\sigma) + \gb(\sigma)\cdot \gc(\sigma) \right)
\right )\Bigr ]
\, . \nn
\end{align}
We note from the action appearing in the last 
exponential that the momentum associated to $x^{\mu}$ is
$p_{\mu}= -\frac{1}{2}\xdot_{\mu}$, which is somewhat unconventional.
Inserting the proper time $\tau=2m\sigma$
as done above \eqn{eq:BP2} would yield the canonical relation.

We now consider the background field expansion for $x^{\mu}(\sigma)$:
\be
\label{eq:bvdef}
x^{\mu}(\sigma) = b^{\mu}+ v^{\mu}\, \sigma + z^{\mu} (\sigma)\, .
\ee
In order to integrate out the $z^{\mu}$ field and the ghosts we use generic 
translation-invariant propagators:
\begin{align}\label{eq:genericProps}
\begin{aligned}
\vev{z^{\mu}(\sigma) z^{\nu}(\sigma')} &= 2i \eta^{\mu\nu}\, \Delta(\sigma-\sigma')\,,\\
\vev{\ga^{\mu}(\sigma) \ga^{\nu}(\sigma')} &= -2i \eta^{\mu\nu}\,\delta(\sigma-\sigma')\,,\\
\vev{\gb^{\mu}(\sigma) \gc^{\nu}(\sigma')} &=+4i\eta^{\mu\nu}\,\delta(\sigma-\sigma')\,.
\end{aligned} 
\end{align}
Concerning $\Delta(\sigma)$ we shall at this point only assume that $\partial_{\sigma}\partial_{\sigma'}\Delta(\sigma-\sigma')=-\delta(\sigma-\sigma')$, which holds true for a time-symmetric as well as
retarded (or advanced) propagator on the infinitely extended worldline.
With this one straightforwardly finds (again going to ``center of mass'' proper time
coordinates as we did in  \eqn{CoMPTV}):
\begin{align}
\label{eq:SpaceTimeFromaggio}
&\Xi(b,v;\{\epsilon^{(l)},k_{l}\})=
\Xi_{0}\, \delta\left ( \sum_{l=1}^{N} k_{l}\cdot v \right )\,
e^{i\sum_{l=1}^{N} k_{l}\cdot b} \\
&\,\, \left(-\frac{i\kappa}{4}\right)^N\, 
\,  \prod_{l=1}^{N}\Bigl [ \int_{-\infty}^{\infty}\!\d\sigma_{l}\, 
\epsilon^{(l), \, \mu\nu}\,
\Bigl (  \partial_{\epsilon_{l}^{\mu}}\partial_{\epsilon_{l}^{\nu}}   
+ \partial_{\alpha_{l}^{\mu}}\partial_{\alpha_{l}^{\nu}}+ \partial_{\beta_{l}^{\mu}}\partial_{\gamma_{l}^{\nu}}
\Bigr ) \Bigr ]\nn\\ &\,\,
 \delta\!\left (\sum_{l=1}^{N}\sigma_{l} \right)\
 \exp \Biggl [ v\cdot \sum_{l=1}^{N}(ik_{l}\sigma_{l}+\epsilon_{l}) 
 -i \sum_{l,l'=1}^{N}\Bigl \{
\Delta(\sigma_{l}-\sigma_{l'})\, k_{l}\cdot k_{l'}\nn \\
&\,\,- i \, \partial_{\sigma_{l}}\Delta(\sigma_{l}-\sigma_{l'})\, \epsilon_{l}\cdot k_{l'} 
+ \delta(\sigma_{l}-\sigma_{l'}) (\epsilon_{l}\cdot \epsilon_{l'} + \alpha_{l}\cdot \alpha_{l'}
-4 \gamma_{l}\cdot \beta_{l'})
\Bigr \}\Biggr ]\Biggr|_{
\epsilon_{l}=\alpha_{l}=\beta_{l}=\gamma_{l}=0}\, . \nn
\end{align}
Here $\Xi_{0}$ is an overall measure factor
\begin{equation}
\Xi_{0}:=\lim_{T\to\infty}
\Bigl [ \frac{i}{(2\pi T)^{d/2}}e^{-iT v^{2}/2} \Bigr ]
\end{equation}
that we may drop as it falls out of normalized correlation functions. Now if we identify the
boundary conditions in terms of the momenta as (recall $p_{\mu}=-\frac{1}{2} \frac{dx_{\mu}}{d\sigma}$)
\be
\label{eq:pqvdefs}
p^{\mu}= -\frac{1}{2} \xdot(-\infty) = -\frac{v^{\mu}}{2} + \frac{q^{\mu}}{2}\, , \qquad
p^{\prime\, \mu}=-\frac{1}{2} \xdot(+\infty) = -\frac{v^{\mu}}{2}- \frac{q^{\mu}}{2}\, ,
\ee
where $q$ is the total momentum transfer of the scattered scalar particle, 
we see that \eqref{eq:SpaceTimeFromaggio} is dauntingly close to the form factor expression \eqref{eq:OnShellFromaggio} upon noting that $-v=p+p'$!
Concretely, if we pick the worldline propagator to be time-symmetric,
\be
\label{eq:feynprop}
\Delta(\sigma) = \frac{|\sigma|}{2}\,,
\ee
we arrive at our central relation linking the QFT form factor to the WQFT correlator:
\be
\label{centralrel}
\boxed{
\frac{\Xi(b,v;\{\epsilon^{(l)},k_{l}\})}{\Xi_{0}}
 =  \delta\left ( \sum_{l=1}^{N} k_{l}\cdot v \right )\,
e^{i\sum_{l=1}^{N} k_{l}\cdot b}\, F(p,p'|\{\epsilon^{(l)},k_{l}\}) \, ,
}
\ee
where the use of Feynman propagators is understood in the form factor.
Note the emergence of the total momentum transfer $q= \sum_{l=1}^{N} k_{l}$ in the above.

So that the significance of \eqn{centralrel} is properly understood,
let us briefly recap the steps that have led us here.
We started with the scalar Green's function $G(x,x')$
in a gravitational background \eqref{eq:BekensteinParker},
which can be inserted into time-ordered correlation functions containing pairs
of distinctly flavored scalars --- see \eqn{eq:timeOrderedCorrelator}.
Moving from time-ordered correlators to S-matrices
required us to obtain a momentum space representation of $G(x,x')$ ---
$D(p,p',\{\epsilon^{(l)},k_{l}\})$, given in \eqn{eq:Fromaggio} ---
and then cut into its external scalar legs,
yielding the form factor $F(p,p'|\{\epsilon^{(l)},k_{l}\})$.
What \eqn{centralrel} therefore tells us ---
provided the external legs are on shell ---
is that we can identify S-matrices with expectation values
in the WQFT using the classical $\hbar\to0$ limit.
The expectation values in the $n$-body case are\footnote{Factors of $\Xi_{0}$ are absorbed into the path integral measure $D[z_i]$.}
\begin{align}\label{eq:expValues}
\Bigl \langle \cO(h,\{x_i\})\Bigr \rangle_{\text{WQFT}}
&=\mathcal{Z}_{\text{WQFT}}^{-1}
\int D[h_{\mu\nu}]
\int \prod_{i=1}^{n} D [z_{i},\ga_{i},\gb_{i},\gc_{i}] \, \cO(h,\{x_i\}) e^{i (S_{\rm EH}+S_{\rm gf})}\nn\\ 
&\qquad\exp\Bigl[ -i\sum_{i=1}^{n}\int_{-\infty}^{\infty}\!\d\tau_{i}\, \frac{m_{i}}{2}
g_{\mu\nu}\left (\xdot_{i}^{\mu}\xdot_{i}^{\nu}+ \ga_{i}^{\mu}\ga_{i}^{\nu}+ \gb_{i}^{\mu} \gc_{i}^{\nu} \right )\Bigr ]\, ,
\end{align}
where $g_{\mu\nu}(x)=\eta_{\mu\nu}+\kappa h_{\mu\nu}(x)$ and
$x_{i}(\tau_{i})=b_{i}+v_{i}\tau_i +z_{i}(\tau_{i})$.
$\mathcal{Z}_{\text{WQFT}}$ is the partition function 
\begin{align}\label{ZWQFTdef}
\mathcal{Z}_{\text{WQFT}}
&:= \text{const} \times
\int D[h_{\mu\nu}]
\int \prod_{i=1}^{n} D [z_{i},\ga_{i},\gb_{i},\gc_{i}] \,  e^{i (S_{\rm EH}+S_{\rm gf})}
\nn\\ &\qquad
\exp\Bigl[ -i\sum_{i=1}^{n}\int_{-\infty}^{\infty}\!\d\tau_{i}\, \frac{m_{i}}{2}
g_{\mu\nu}\left (\xdot_{i}^{\mu}\xdot_{i}^{\nu}+ \ga_{i}^{\mu}\ga_{i}^{\nu}+ \gb_{i}^{\mu} \gc_{i}^{\nu} \right )\Bigr ]\, ,
\end{align}
and $\text{const}$ 
ensures that $\mathcal{Z}_{\text{WQFT}}=1$ in the non-interacting case ($\kappa=0$).

\subsection{Towards the eikonal phase}
\label{sec:toEikonal}

Equipped with \eqn{centralrel} we discover an intriguing relation between
the free energy of the WQFT and the eikonal phase
of a $2\to 2 $ scalar S-matrix in the classical limit.
The exponentiated eikonal phase is defined as a Fourier transform
of the S-matrix into impact parameter space
transverse to the $(D-2)$-dimensional scattering plane \cite{Amati:1987wq,Amati:1990xe}: 
\be \label{eq:eikonalRelation}
e^{i\chi}:= 
\frac1{4m_1m_2}
\int\!\frac{\d^Dq}{(2\pi)^{D-2}}\, \delta(q\cdot v_{1})\, \delta(q\cdot v_{2}) \, e^{iq\cdot b}\,
\braket{\phi_1\phi_2|S|\phi_1\phi_2}\,.
\ee
where $b=b_2-b_1$ and $q=p_{1}'-p_{1}=p_{2}-p_{2}'$ is the momentum transfer
from particle 1 to 2 ($p_{i}$ momenta ingoing and $p_{i}'$ momenta outgoing). 
 
 An immediate corollary of \eqn{eq:AmpinG12} and \eqn{centralrel} 
 and a central result of our work is
then the simple relation (holding in the classical limit)
\begin{equation}\label{eq:eikonalId}
\boxed{\mathcal{Z}_{\text{WQFT}}=  e^{i\chi}\, ,}
\end{equation}
i.e.~the free energy of the WQFT is to be identified with the eikonal phase. This is a rather direct link
between the worldline theory and the QFT S-matrix.\footnote{Note that there is a factor $\frac1{2m_i}$ for each worldline,
which comes from inserting $\sigma_i=\frac1{2m_i}\tau_i$ as described above.}
We shall evaluate the eikonal phase to 2PM in \Sec{sec:eikonal} and
establish a relationship to the classical impulse $\Delta p_i^\mu=q^{\mu}$
and scattering angle $\theta$.

\subsection{Time-symmetric vs.~retarded propagators}

Let us finally comment on the use of time-symmetric propagators above versus
retarded (or advanced) ones in the WQFT.
The retarded (or advanced) worldline propagators on an infinite worldline read
\be
\label{eq:raprop}
\Delta_{r/a}(\sigma)= \frac{|\sigma|}{2} \pm \frac{\sigma}{2}\, .
\ee
We claim that switching between these propagators simply amounts to performing shifts
in the background parameters $b^{\mu}$ and $v^{\mu}$. This is best seen in
a classical setting where one seeks to solve an inhomogeneous second-order ordinary differential
equation for $x^{\mu}(\sigma)$. Writing the solution as
$x^{\mu}(\sigma)= b^{\mu}+v^{\mu}\sigma + z^{\mu}(\sigma)$, the
$b^\mu$ and $v^\mu$ terms represent a solution to the homogeneous (force-free) equation, whereas
the perturbatively constructed $z^{\mu}$ is a specific solution to the inhomogeneous solution.

The choice of propagator is equivalent to picking a specific inhomogeneous solution.
Hence all choices for worldline propagators are valid and physically equivalent,
but the meaning of the background constants changes.
To emphasize this we will denote them as follows:
for a retarded propagator $b^\mu$ and $v^\mu$
describe the initial worldline trajectory ($\sigma\to-\infty$),
for an advanced propagator $b'^\mu$ and $v'^\mu$ give
the final worldline ($\sigma\to+\infty$),
and for a time-symmetric propagator $\hat{b}^\mu$ and $\hat{v}^\mu$
an in-between state ($\sigma=0$).
So we identify $p^\mu=mv^\mu$, $p'^\mu=mv'^\mu$ and
$\hat{p}^\mu=\frac12(p^\mu+p'^\mu)=\hat{m}\hat{v}^\mu$
as the ingoing, outgoing, and average momenta respectively,
where $\hat{m}^2=\frac{m^2}2(1+v\cdot v')$ is chosen to ensure $\hat{v}^2=1$.
One may directly compute the shifts in $b^\mu$ and $v^\mu$
for transitions between the propagators from their definitions
in \eqns{eq:feynprop}{eq:raprop}.

The choice of Feynman vs.~retarded propagators is also
meaningful for the gravitons, but in a different way.
The real part of Feynman propagators (as one uses when calculating scattering amplitudes)
yield time symmetric Green's functions,
which is consistent with purely conservative scattering.
For a classically radiating system one should instead use retarded propagators.
This will affect observables like the impulse $\Delta p_i^\mu$,
which after integration will have a different form.
This important subtlety was recently discussed in the context
of the 3PM deflection in \rcite{Damour:2020tta}, resolving a tension with the high-energy limit \cite{Bern:2019crd, Damour:2019lcq}.
It was argued earlier that, from an amplitudes perspective, this tension would be resolved by including the full soft region \cite{DiVecchia:2020ymx}.

\section{WQFT Feynman rules}
\label{sec:feynmanRules}

In the previous section we saw a clear link between
gravitational S-matrices and expectation values
of operators evaluated in the WQFT.
These involve path integrals over not only the gravitational field $h_{\mu\nu}$,
but also the deflection $z^\mu$ and ghosts $\ga^\mu$, $\gb^\mu$, $\gc^\mu$.
In this section we develop a set of Feynman rules
which allow us to calculate these expectation values directly.
By taking a diagrammatic approach we invite
comparisons with the diagrams used to describe scattering amplitudes.

We treat the gravitational field $h_{\mu\nu}(x)$
and deflection $z^\mu(\tau)$ on an equal footing.
As we are not interested in quantum corrections we will work at tree level,
so we can ignore the ghosts.
The graviton is most naturally described in momentum space;
the deflection in \emph{energy} space (or frequency, using $E=\hbar\omega$):
\begin{align}\label{eq:fourier}
h_{\mu\nu}(x)=\int_ke^{-ik\cdot x}h_{\mu\nu}(k)\,,&&
z^\mu(\tau)=\int_\omega e^{-i\omega\tau}z^\mu(\omega)\,,&&
\end{align}
where we have introduced the shorthands
\begin{align}
\int_k:=\int\!\frac{\d^4k}{(2\pi)^4}\,, &&
\int_\omega:=\int\frac{\d\omega}{2\pi}\,.
\end{align}
From this point onwards we specialize to $D=4$.
We will also absorb factors of $(2\pi)$ into the $\delta$-functions:
\begin{align}
\dd(k):=(2\pi)^4\delta^{(4)}(k)\,, &&
\dd(\omega):=(2\pi)\delta(\omega)\,.
\end{align}
The Einstein-Hilbert action~\eqref{eq:einsteinHilbert} being integrated
over all positions $x$ implies the usual momentum conservation at
those interaction vertices;
vertices arising from $S_{\rm pm}$ in 
 \eqref{eq:Spm3} instead conserve the energy $\omega$.

First consider the Einstein-Hilbert action.
The Feynman rules arising from here are the usual ones involving only
the graviton $h_{\mu\nu}$, with propagator
\begin{equation}\label{eq:hPropagator}
\begin{tikzpicture}[baseline={(current bounding box.center)}]
\coordinate (x) at (-.5,0);
\coordinate (y) at (1.5,0);
\draw [photon] (x) -- (y) node [midway, below] {$k$};
\draw [fill] (x) circle (.08) node [above] {$h_{\mu\nu}(x)$};
\draw [fill] (y) circle (.08) node [above] {$h_{\rho\sigma}(y)$};
\end{tikzpicture}=iP_{\mu\nu;\rho\sigma}\int_k
\frac{e^{-ik\cdot(x-y)}}{k^2}\,,
\end{equation}
where $P_{\mu\nu;\rho\sigma}=\eta_{\mu(\rho}\eta_{\sigma)\nu}-
\frac12\eta_{\mu\nu}\eta_{\rho\sigma}$.
We are flexible about the $i\eps$ prescription:
either write the denominator as $k^2+i\eps$,
making it a time-symmetric Feynman propagator,
or $(k^0\pm i\eps)^2-{\mathbf k}^2$,
making it retarded/advanced.
In the retarded case the poles in $k^0$ occur
at $k^0=\pm\sqrt{\mathbf k^2}-i\eps$:
as both are below the real axis the
integration contour must be closed in the lower-half plane.
So the integral is non-zero only when $x^0>y^0$, thus ensuring causality.

Next we consider the worldline action $S_{\rm pm}$ given in \eqn{eq:Spm3}:
\begin{equation}
S_{\rm pm}=-\frac{m}2\int_{-\infty}^\infty\d\tau
\big(g_{\mu\nu}\dot{x}^\mu\dot{x}^\nu+1\big)\,.
\end{equation}
For now ignoring the parts containing $h_{\mu\nu}$,
we expand $\dot{x}^\mu(\tau)=v^\mu+\dot{z}^\mu(\tau)$ to obtain
\begin{equation}
\left.S_{\rm pm}\right|_{h_{\mu\nu}=0}=
-\int_{-\infty}^\infty\d\tau\left(m+m\,\eta_{\mu\nu}v^\mu\dot{z}^\nu+
\frac{m}2\eta_{\mu\nu}\dot{z}^\mu\dot{z}^\nu\right)\,,
\end{equation}
having used $\eta_{\mu\nu}v^\mu v^\nu=1$.
Both the first term (a constant) and the second term (a boundary term)
we can ignore; the third gives us our propagator for $z^\mu$:
\begin{align}\label{eq:zPropagator}
\begin{tikzpicture}[baseline={(current bounding box.center)}]
\coordinate (in) at (-1,0);
\coordinate (out) at (2,0);
\coordinate (x) at (-.3,0);
\coordinate (y) at (1.3,0);
\draw [zUndirected] (x) -- (y) node [midway, below] {$\omega$};
\draw [dotted] (in) -- (x);
\draw [dotted] (y) -- (out);
\draw [fill] (x) circle (.08) node [above] {$z^\mu(\tau_1)$};
\draw [fill] (y) circle (.08) node [above] {$z^\nu(\tau_2)$};
\end{tikzpicture}&=-i\frac{\eta^{\mu\nu}}m\int_\omega
\frac{e^{-i\omega(\tau_1-\tau_2)}}{(\omega\pm i\eps)^2}=
\frac{i\eta^{\mu\nu}}{2m}\left(|\tau_1-\tau_2|\pm(\tau_1-\tau_2)\right)\,.
\end{align}
These are the retarded/advanced versions of the propagator,
which are non-zero when $\tau_1>\tau_2$ or $\tau_1<\tau_2$ respectively.
Using $\sigma=\frac{\tau}{2m}$ we see a precise match
for the same propagator given earlier \eqref{eq:genericProps}.
We define the time-symmetric propagator as simply the averaged combination
of the retarded/advanced propagators.\footnote{
	For the graviton such an averaging procedure would produce only the real part of its Feynman propagator. The missing imaginary part corresponds to dissipation; however, on the worldline there is no dissipation and hence no imaginary part.
}
As we explained in \Sec{sec:toEikonal},
the correct interpretation of $b^\mu$ and $v^\mu$ is sensitive
to the choice of worldline propagator.

Finally we proceed to consider worldline interactions,
all of which involve the gravitational field $h_{\mu\nu}$.
As $S_{\rm pm}$ depends on the gravitational field only through
$g_{\mu\nu}=\eta_{\mu\nu}+m_{\rm Pl}^{-1}h_{\mu\nu}$
(and not the inverse metric $g^{\mu\nu}$)
this conveniently ensures that all such vertices are linear in $h_{\mu\nu}$.
We extract the $\tau$ dependence from $h_{\mu\nu}$
when it is evaluated on the worldline of a black hole:
\begin{align}
\begin{aligned}
h_{\mu\nu}(x(\tau))&=
\int_ke^{ik\cdot(b+v\tau+z(\tau))}h_{\mu\nu}(-k)
=\sum_{n=0}^\infty\frac{i^n}{n!}\!\int_ke^{ik\cdot(b+v\tau)}
\left(k\cdot z(\tau)\right)^nh_{\mu\nu}(-k)\\
&=\sum_{n=0}^\infty\frac{i^n}{n!}\!\int_{k,\omega_1,\ldots,\omega_n}
e^{ik\cdot b}e^{i(k\cdot v+\sum_{i=1}^n\omega_i)\tau}
\left(\prod_{i=1}^nk\cdot z(-\omega_i)\right)h_{\mu\nu}(-k)\,.
\end{aligned}
\end{align}
The product on $z^\mu(-\omega_i)$ produces a tower of vertices
which are fed into the interacting part of the action
$S_{\rm pm}^{\rm int}=S_{\rm pm}-\left.S_{\rm pm}\right|_{h_{\mu\nu}=0}$:
\begin{align}
\begin{aligned}
S_{\rm pm}^{\rm int}&=
-\frac{m}{2m_{\rm Pl}}\int_{-\infty}^\infty\d\tau\,
h_{\mu\nu}(x(\tau))\dot{x}^\mu(\tau)\dot{x}^\nu(\tau)\\
&=
-\frac{m}{2m_{\rm Pl}}\int_{-\infty}^\infty\d\tau\,
h_{\mu\nu}(x(\tau))
\big(v^\mu v^\nu+2v^{(\mu}\dot{z}^{\nu)}(\tau)+\dot{z}^{\mu}(\tau)\dot{z}^{\nu}(\tau)\big)\,.
\end{aligned}
\end{align}
We obtain
\begin{align}\label{eq:allOrderWL}
\begin{aligned}
S_{\rm pm}^{\rm int}=
-\frac{m}{m_{\rm Pl}}\sum_{n=0}^\infty\frac{i^n}{n!}
\int_{k,\omega_1,\ldots,\omega_n}
e^{ik\cdot b}\dd\bigg(k\cdot v+\sum_{i=1}^n\omega_i\bigg)
h_{\mu\nu}(-k)\left(\prod_{i=1}^nz^{\rho_i}(-\omega_i)\right)\!\times&\\
\left(\frac12\left(\prod_{i=1}^nk_{\rho_i}\!\right)\!v^\mu v^\nu+
\sum_{i=1}^n\omega_i\left(\prod_{j\neq i}^nk_{\rho_j}\right)\!
v^{(\mu}\delta^{\nu)}_{\rho_i}+
\sum_{i<j}^n\omega_i\omega_j\left(\prod_{l\neq i,j}^nk_{\rho_l}\right)
\delta^{(\mu}_{\rho_i}\delta^{\nu)}_{\rho_j}\right),&
\end{aligned}
\end{align}
having integrated over $\tau$ to extract the energy-conserving $\delta$-function.
When $n=0$ only the first term in the second line is included;
when $n=1$ only the first two terms.

Let us see how the Feynman rules are read off using
some explicit examples.
At zeroth order in $z^\mu$:
\begin{align}
\left.S_{\rm pm}^{{\rm int}}\right|_{z^0}=-\frac{m}{2m_{\rm Pl}}
\int_ke^{ik\cdot b}\dd(k\cdot v)h_{\mu\nu}(-k)v^\mu v^\nu\,.
\end{align}
This term gives rise to the stress-energy tensor
$T^{\mu\nu}(k)=me^{ik\cdot b}\dd(k\cdot v)v^\mu v^\nu$
(see \eg \rcite{Guevara:2019fsj})
which we interpret as a classical source for $h_{\mu\nu}$.
The Feynman rule is
\begin{align}\label{eq:vertexH}
\begin{tikzpicture}[baseline={(current bounding box.center)}]
\coordinate (in) at (-1,0);
\coordinate (out) at (1,0);
\coordinate (x) at (0,0);
\node (k) at (0,-1.3) {$h_{\mu\nu}(k)$};
\draw [dotted] (in) -- (x);
\draw [dotted] (x) -- (out);
\draw [graviton] (x) -- (k);
\draw [fill] (x) circle (.08);
\end{tikzpicture}=
-i\frac{m}{2m_{\rm Pl}}e^{ik\cdot b}\dd(k\cdot v)v^\mu v^\nu\,,
\end{align}
with $k$ outgoing.
It is a tadpole: the dotted line represents the worldline,
and is intended only as a visual aid.
The linear terms in $z^\mu$ are
\begin{align}
\left.S_{\rm pm}^{{\rm int}}\right|_{z}=
-i\frac{m}{2m_{\rm Pl}}\int_{k,\omega}e^{ik\cdot b}\dd(k\cdot v+\omega)
h_{\mu\nu}(-k)z^\rho(-\omega)
\left(2\omega v^{(\mu}\delta^{\nu)}_\rho+v^\mu v^\nu k_\rho\right)\,,
\end{align}
from which we read off the two-point vertex:
\begin{align}
\begin{tikzpicture}[baseline={(current bounding box.center)}]
\coordinate (in) at (-1,0);
\coordinate (out) at (1,0);
\coordinate (x) at (0,0);
\node (k) at (0,-1.3) {$h_{\mu\nu}(k)$};
\draw (out) node [right] {$z^\rho(\omega)$};
\draw [dotted] (in) -- (x);
\draw [zUndirected] (x) -- (out);
\draw [graviton] (x) -- (k);
\draw [fill] (x) circle (.08);
\end{tikzpicture}=
\frac{m}{2m_{\rm Pl}}e^{ik\cdot b}\dd(k\cdot v+\omega)
\left(2\omega v^{(\mu}\delta^{\nu)}_\rho+v^\mu v^\nu k_\rho\right)\,.
\end{align}
The energy $\omega$ is also taken as outgoing.
Finally, to quadratic order in $z^\mu$:
\begin{align}
\begin{aligned}
\left.S_{\rm pm}^{{\rm int}}\right|_{z^2}=
\frac{m}{2m_{\rm Pl}}\int_{k,\omega_1,\omega_2}
e^{ik\cdot b}\dd(k\cdot v+\omega_1+\omega_2)
h_{\mu\nu}(-k)z^{\rho_1}(-\omega_1)z^{\rho_2}(-\omega_2)\times&\\
\left(\frac12k_{\rho_1} k_{\rho_2}v^\mu v^\nu+
\omega_1k_{\rho_2} v^{(\mu}\delta^{\nu)}_{\rho_1}+
\omega_2k_{\rho_1} v^{(\mu}\delta^{\nu)}_{\rho_2}+
\omega_1\omega_2\delta^{(\mu}_{\rho_1}\delta^{\nu)}_{\rho_2}\right)\,.&
\end{aligned}
\end{align}
The associated trivalent Feynman vertex is
\begin{align}\label{eq:zzH}
\begin{tikzpicture}[baseline={(current bounding box.center)}]
\coordinate (in) at (-1,0);
\coordinate (out1) at (1,0);
\coordinate (out2) at (1,0.5);
\coordinate (x) at (0,0);
\node (k) at (0,-1.3) {$h_{\mu\nu}(k)$};
\draw (out1) node [right] {$z^{\rho_1}(\omega_1)$};
\draw (out2) node [right] {$z^{\rho_2}(\omega_2)$};
\draw [dotted] (in) -- (x);
\draw [zUndirected] (x) -- (out1);
\draw [zUndirected] (x) to[out=40,in=180] (out2);
\draw [graviton] (x) -- (k);
\draw [fill] (x) circle (.08);
\end{tikzpicture}=
i\frac{m}{m_{\rm Pl}}e^{ik\cdot b}\dd(k\cdot v+\omega_1+\omega_2)\times
\qquad\qquad\qquad\qquad\qquad
\end{align}
\vspace{-3.5em}
\begin{align}
\qquad\qquad\qquad\qquad\qquad\quad
\left(\frac12k_{\rho_1} k_{\rho_2}v^\mu v^\nu+
\omega_1k_{\rho_2} v^{(\mu}\delta^{\nu)}_{\rho_1}+
\omega_2k_{\rho_1} v^{(\mu}\delta^{\nu)}_{\rho_2}+
\omega_1\omega_2\delta^{(\mu}_{\rho_1}\delta^{\nu)}_{\rho_2}\right).\nn
\end{align}
While of course the second worldline fluctuation
still travels on the worldline,
we draw it above to distinguish it from its partner.

Given that an $n$-graviton vertex carries an overall $m_{\rm Pl}^{2-n}$,
it might seem odd that each of these $z$-vertices
carries only a single power of $m_{\rm Pl}^{-1}$.
To rectify this we might try rescaling $z^\mu\to m_{\rm Pl}^{-1}z^\mu$,
similar to how we write $g_{\mu\nu}=\eta_{\mu\nu}+m_{\rm Pl}^{-1}h_{\mu\nu}$
for the graviton.
However, we find this operation to be undesirable as it also
rescales the propagator~\eqref{eq:zPropagator} to carry an overall $m_{\rm Pl}^2$.
As we shall see,
despite the higher-point vertices carrying the same overall power of $m_{\rm Pl}$,
their appearance at low orders in the PM expansion is ruled
out by the combinatorics of which diagrams we can draw.

The three vertices given above will be sufficient for
all of the calculations done in this paper.
However, using \eqn{eq:allOrderWL} we can easily generalize
to an $n$th order vertex:
\begin{align}
&V^{{\rm WL},\mu\nu}_{\rho_1\cdots \rho_n}(k;\omega_1,\cdots,\omega_n)=
i^{n-1}\frac{m}{m_{\rm Pl}}
e^{ik\cdot b}\dd\bigg(k\cdot v+\sum_{i=1}^n\omega_i\bigg)\times\\
&\qquad\qquad
\left(\frac12\left(\prod_{i=1}^nk_{\rho_i}\!\right)\!v^\mu v^\nu+
\sum_{i=1}^n\omega_i\left(\prod_{j\neq i}^nk_{\rho_j}\right)\!
v^{(\mu}\delta^{\nu)}_{\rho_i}+
\sum_{i<j}^n\omega_i\omega_j\left(\prod_{l\neq i,j}^nk_{\rho_l}\right)
\delta^{(\mu}_{\rho_i}\delta^{\nu)}_{\rho_j}\right).\nn
\end{align}
An intriguing property of this vertex is that,
should we set the energy on one of the external $z^\mu$ lines to zero,
the resulting expression can also be obtained as a derivative of its
lower-point cousin with respect to the impact factor:
\begin{align}\label{eq:eikonalmaster}
V^{{\rm WL},\mu\nu}_{\rho_1\ldots\rho_{n+1}}(k;\omega_1,\ldots,\omega_n,0)=
\frac{\partial}{\partial b^{\rho_{n+1}}}
V^{{\rm WL},\mu\nu}_{\rho_1\ldots\rho_n}(k;\omega_1,\ldots,\omega_n)\,.
\end{align}
This will be important when we return to the
eikonal phase in \Sec{sec:eikonal}.

\section{Radiation}
\label{sec:radiation}

Having set up the worldline Feynman rules
we begin with our first application:
calculating the radiation far away from a source.
For simplicity, let us first consider a single black hole.
We calculate $k^2\braket{h^{\mu\nu}(k)}_{\rm WQFT}$ for $k^2=0$,
where the expectation value of an operator in the WQFT
was defined in \eqn{eq:expValues}.
This requires us to draw diagrams with a single outgoing graviton line,
which is equivalent to solving Einstein's equation for $h_{\mu\nu}(x)$.
For a single black hole 
\begin{align}
-ik^2\braket{h^{\mu\nu}(k)}_{\rm WQFT}=
-i\frac{m}{2m_{\rm Pl}}e^{ik\cdot b}\,\dd(k\cdot v)v^\mu v^\nu\,,
\end{align}
which is simply the tadpole in \eqn{eq:vertexH}.\footnote{
	Contracting with $P_{\mu\nu;\rho\sigma}$ on an outgoing on-shell graviton line is unnecessary due to the tracelessness of the polarization tensor.
}
In other words, $h_{\mu\nu}$ is directly sourced by the stress-energy tensor.

Like in \Sec{sec:momSpaceRep}, we compare this with the
three-point interaction vertex between two complex
scalars $\varphi$ and a graviton $h_{\mu\nu}$:
\begin{align}\label{eq:ssHVertex}
\begin{aligned}
\begin{tikzpicture}[baseline={(current bounding box.center)}]
\coordinate (in) at (-1,0);
\coordinate (out) at (1,0);
\coordinate (x) at (0,0);
\node (k) at (1,-1) {$k$};
\draw (in) node [left] {$p$};
\draw (out) node [right] {$p'$};
\draw [fill] (x) circle (.08);
\draw [cscalar] (in) -- (x);
\draw [cscalar] (x) -- (out);
\draw [graviton] (x) -- (k);
\end{tikzpicture}&=-
i\kappa\left[p^{(\mu}p'^{\nu)}-
\frac{\eta^{\mu\nu}}2\left(p\cdot p'-m^2\right)\right]\\
&=-i\kappa\left[\hat{p}^\mu\hat{p}^\nu+\frac14\left(\eta^{\mu\nu}k^2-k^\mu k^\nu\right)\right]\,.
\end{aligned}
\end{align}
In the second equality we have inserted
$p=\hat{p}+\frac{k}2$, $p'=\hat{p}-\frac{k}2$,
and using the mass-shell conditions $p^2=p'^2=m^2$ we find that
$\hat{p}^2=m^2-\frac{k^2}4$ and $\hat{p}\cdot k=0$.
In the classical limit we write $k^\mu=\hbar\bar{k}^\mu$ and send $\hbar\to0$,
so we can discard the $k$-dependent terms.
Finally inserting $\hat{p}^\mu=mv^\mu$,\footnote{
	Strictly speaking, we should write $\hat{p}^\mu=\hat{m}\hat{v}^\mu$
	to represent the average momentum.
	However, at this low PM order the difference is inconsequential.
} we see that
\be
k^2 \braket{h^{\mu\nu}(k)}_{\rm WQFT} = 
\frac{i}{2m}e^{ik\cdot b}\dd(k\cdot v)\lim_{\hbar\to 0}
{\cal M}^{\mu\nu}_{\rm GR}(p,k)\,.
\ee
The graviton from a single black hole is therefore identified with
the three-point amplitude (with the polarization tensor $\eps_{\mu\nu}$
stripped away).

This identity follows naturally from our discussion in \Secs{sec:2}{sec:3}.
The exponential and $\delta$-function factors come from
the central relationship in \eqn{centralrel};
the factor of $\frac1{2m}$ from replacing $\sigma=\frac{\tau}{2m}$.
The interpretation of a single black hole radiating a graviton
as a massive three-point amplitude has been widely studied elsewhere,
including for higher spins \cite{Guevara:2018wpp,Guevara:2019fsj}.
The non-spinning black hole is associated with the Schwarzschild solution;
a spinning black hole with the Kerr solution~\cite{Arkani-Hamed:2019ymq}.
The corresponding double copies are closely related to the so-called
Kerr-Schild double copy~\cite{Monteiro:2014cda,Monteiro:2015bna,Luna:2015paa}.

\subsection{Two-body radiation (2PM)}
\label{sect:5.1}

Let us now examine the radiation emitted from the inelastic scattering
of a pair of non-spinning black holes at leading order (2PM).
To begin with consider the five-point scattering amplitude:
\begin{align}\label{eq:5ptAmplitude}
\epsilon_{\mu\nu}{\cal M}^{\mu\nu}_{\rm GR}(p_{i},p'_i,k)&=
\begin{tikzpicture}[baseline={(current bounding box.center)}]
\node (inA) at (-1,1) {$p_1$};
\node (outA) at (1,1) {$p'_1$};
\node (inB) at (-1,-1) {$p_2$};
\node (outB) at (1,-1) {$p'_2$};
\node (k) at (1.3,0) {$k$};
\coordinate (c) at (0,0);
\draw [cscalar] (inA) -- (c);
\draw [cscalar2] (c) -- (outA);
\draw [cscalar] (inB) -- (c);
\draw [cscalar2] (c) -- (outB);
\draw [graviton] (c) -- (k);
\draw [fill] (c) circle (.3);
\end{tikzpicture}\\
&=
\begin{tikzpicture}[scale=.6,baseline={(current bounding box.center)}]
\coordinate (inA) at (-1,1);
\coordinate (outA) at (1,1);
\coordinate (inB) at (-1,-1);
\coordinate (outB) at (1,-1);
\coordinate (xA) at (0,1);
\coordinate (xB) at (0,-1);
\coordinate (c) at (0,0);
\coordinate (k) at (1,0);
\draw (k) node [right] {};
\draw [cscalar] (inA) -- (xA);
\draw [cscalar] (xA) -- (outA);
\draw [cscalar] (inB) -- (xB);
\draw [cscalar] (xB) -- (outB);
\draw [graviton] (c) -- (k);
\draw [photon] (xA) -- (c);
\draw [photon] (xB) -- (c);
\end{tikzpicture}
+\left(
\begin{tikzpicture}[scale=.6,baseline={(current bounding box.center)}]
\coordinate (inA) at (-1,1);
\coordinate (outA) at (1,1);
\coordinate (inB) at (-1,-1);
\coordinate (outB) at (1,-1);
\coordinate (xA) at (0,1);
\coordinate (xB) at (0,-1);
\coordinate (k) at (1,0);
\draw (k) node [right] {};
\draw [cscalar] (inA) -- (xA);
\draw [cscalar] (xA) -- (outA);
\draw [cscalar] (inB) -- (xB);
\draw [cscalar] (xB) -- (outB);
\draw [graviton] (xA) -- (k);
\draw [photon] (xA) -- (xB);
\end{tikzpicture}
+
\begin{tikzpicture}[scale=.6,baseline={(current bounding box.center)}]
\coordinate (inA) at (-1,1);
\coordinate (outA) at (2,1);
\coordinate (inB) at (-1,-1);
\coordinate (outB) at (2,-1);
\coordinate (xA) at (0,1);
\coordinate (yA) at (1,1);
\coordinate (xB) at (0,-1);
\coordinate (k) at (2,0);
\draw (k) node [right] {};
\draw [cscalar] (inA) -- (xA);
\draw [cscalar] (yA) -- (outA);
\draw [cscalar] (inB) -- (xB);
\draw [cscalar] (xB) -- (outB);
\draw [cscalar] (xA) -- (yA);
\draw [graviton] (yA) -- (k);
\draw [photon] (xA) -- (xB);
\end{tikzpicture}
+
\begin{tikzpicture}[scale=.6,baseline={(current bounding box.center)}]
\coordinate (inA) at (-1,1);
\coordinate (outA) at (2,1);
\coordinate (inB) at (-1,-1);
\coordinate (outB) at (2,-1);
\coordinate (xA) at (0,1);
\coordinate (yA) at (1,1);
\coordinate (yB) at (1,-1);
\coordinate (k) at (2,0);
\draw (k) node [right] {};
\draw [cscalar] (inA) -- (xA);
\draw [cscalar] (xA) -- (yA);
\draw [cscalar] (yA) -- (outA);
\draw [cscalar] (inB) -- (yB);
\draw [cscalar] (yB) -- (outB);
\draw [graviton] (xA) to[out=-50,in=180] (k);
\draw [photon] (yA) -- (yB);
\end{tikzpicture}
+(1\leftrightarrow2)\right),\nn
\end{align}
where $\epsilon_{\mu\nu}$ is the polarization of
the emitted graviton with momentum $k$, and $p'_i=p_i-q_i$.
The on-shell conditions $(\hat{p}_i\pm\frac{q_i}2)^2=m_i^2$
imply $\hat{p}_i\cdot q_i=0$;
momentum conservation gives $k=q_1+q_2$.
Inserting the established relation \eqref{centralrel} between an
$n$-graviton form factor and a WQFT correlator into
a generic $\phi_{1}\, \phi_{2}\to \phi_{1}\, \phi_{2}(+h)$ scattering amplitude
\`a la \eqn{eq:AmpinG12} yields a direct link to the WQFT:
\be
\label{eq:LinkA5}
k^2 \Bigl \langle h^{\mu\nu}(k)\Bigr \rangle_{\text{WQFT}} = 
\frac{i}{4m_1 m_2}\int_{q_1,q_2}
\hat{\mu}_{1,2}(k)\lim_{\hbar\to 0}{\cal M}^{\mu\nu}_{\rm GR}(p_{i},p'_i,k)\,.
\ee
We have introduced the integral measure emerging from \eqn{centralrel}:
\begin{equation}\label{intmeasureGold}
\hat{\mu}_{1,2}(k)=
e^{i(q_1\cdot\hat{b}_1+q_2\cdot\hat{b}_2)}
\dd(q_1\cdot\hat{v}_1)\dd(q_2\cdot\hat{v}_2)
\dd(k-q_1-q_2)\,.
\end{equation}
This connection was already established in \rcites{Goldberger:2016iau,Luna:2017dtq};
here we show how individual diagrams may be identified between the two methods.
There are three diagrams in the WQFT,
all with a single outgoing graviton line.

The first diagram contributing to $k^2\braket{h^{\mu\nu}(k)}_{\rm WQFT}$ is
\begin{align}
\begin{tikzpicture}[baseline={(current bounding box.center)}]
\coordinate (inA) at (-1,1);
\coordinate (outA) at (1,1);
\coordinate (inB) at (-1,-1);
\coordinate (outB) at (1,-1);
\coordinate (xA) at (0,1);
\coordinate (xB) at (0,-1);
\coordinate (c) at (0,0);
\coordinate (k) at (1,0);
\draw (k) node [right] {$^{\mu,\nu}$};
\draw [fill] (c) circle (.04);
\draw [dotted] (inA) -- (xA);
\draw [dotted] (xA) -- (outA);
\draw [dotted] (inB) -- (xB);
\draw [dotted] (xB) -- (outB);
\draw [graviton] (c) -- (k) node [midway, above] {$k$};
\draw [photon] (xA) -- (c) node [midway, left] {$q_1\!\downarrow$};
\draw [photon] (xB) -- (c) node [midway, left] {$q_2\!\uparrow$};
\draw (inA) node [left] {$1$};
\draw (inB) node [left] {$2$};
\draw [fill] (xA) circle (.08);
\draw [fill] (xB) circle (.08);
\end{tikzpicture}\!=-\frac{m_1m_2}{8m_{\rm Pl}^3}
\int_{q_1,q_2}\hat{\mu}_{1,2}(k)
V_3^{(\mu\nu)(\rho\sigma)(\lambda\tau)}
\frac{P_{\rho\sigma;\alpha\beta}}{q_1^2}
\frac{P_{\lambda\tau;\gamma\delta}}{q_2^2}
\hat{v}^\alpha_1\hat{v}^\beta_1\hat{v}^\gamma_2\hat{v}^\delta_2\,,
\end{align}
where
$(i/2)m_{\rm Pl}^{-1}V_3^{(\mu\nu)(\rho\sigma)(\lambda\tau)}(k,-q_1,-q_2)$
is the three-graviton vertex.
The delta functions in the measure $\hat{\mu}_{1,2}(k)$
are picked up from the vertices;
we integrate over the intermediate momenta $q_i$.
The counterpart to this diagram in ${\cal M}^{\mu\nu}_{\rm GR}$ is 
the first diagram in \eqn{eq:5ptAmplitude},
so we simply re-interpret the worldlines as scalars.
Showing that the two expressions match in the $\hbar\to0$ limit is trivial:
the graviton propagators and three-graviton vertex are the same in either case,
and we have already shown in \eqn{eq:ssHVertex} that when $\hbar\to0$
the scalar-scalar-graviton vertex maps onto the stress-energy tensor.

A more interesting comparison is with this diagram:
\begin{align}
&\begin{tikzpicture}[baseline={(current bounding box.center)}]
\coordinate (inA) at (-1,1);
\coordinate (outA) at (2,1);
\coordinate (inB) at (-1,-1);
\coordinate (outB) at (2,-1);
\coordinate (xA) at (0,1);
\coordinate (yA) at (1,1);
\coordinate (xB) at (0,-1);
\coordinate (k) at (2,0);
\draw (k) node [right] {$^{\mu,\nu}$};
\draw [dotted] (inA) -- (xA);
\draw [dotted] (yA) -- (outA);
\draw [dotted] (inB) -- (xB);
\draw [dotted] (xB) -- (outB);
\draw [zUndirected] (xA) -- (yA) node [midway, below] {$\overrightarrow{\omega}$};
\draw [fill] (xA) circle (.08);
\draw [fill] (yA) circle (.08);
\draw [fill] (xB) circle (.08);
\draw [graviton] (yA) -- (k) node [midway, below] {$k$};
\draw [photon] (xA) -- (xB) node [midway, left] {$q_2\!\uparrow$};
\draw (inA) node [left] {$1$};
\draw (inB) node [left] {$2$};
\draw [fill] (xA) circle (.08);
\draw [fill] (xB) circle (.08);
\end{tikzpicture}\\&\qquad=-\frac{m_1m_2}{8m_{\rm Pl}^3}\!
\int_{q_1,q_2}\hat{\mu}_{1,2}(k)
\frac{(2\omega \hat{v}_1^{(\mu}\delta^{\nu)}_\rho-\hat{v}_1^\mu \hat{v}_1^\nu k_\rho)
(2\omega \hat{v}_1^{(\sigma}\eta^{\lambda)\rho}-\hat{v}_1^\sigma \hat{v}_1^\lambda q_2^\rho)}
{\omega^2}
\frac{P_{\sigma\lambda;\alpha\beta}}{q_2^2}\hat{v}_2^\alpha \hat{v}_2^\beta\,,\nn
\end{align}
where $\omega=\hat{v}_1\cdot k=\hat{v}_1\cdot q_2$ from the $\delta$-function constraints.
We have have massaged the integral measure into $\hat{\mu}_{1,2}(k)$
by performing the trivial $\omega$ integration:
\begin{align}
\int_{\omega,q_2}\!e^{i(\hat{b}_1\cdot(k-q_2)+\hat{b}_2\cdot q_2)}
\dd(\hat{v}_2\cdot q_2)\dd(\omega-\hat{v}_1\cdot q_2)
\dd(\omega-\hat{v}_1\cdot k)=
\int_{q_1,q_2}\hat{\mu}_{1,2}(k)\,.
\end{align}
This expression arises from the classical limit of \emph{three}
diagrams in ${\cal M}^{\mu\nu}_{\rm GR}$,
again drawn in \eqn{eq:5ptAmplitude}.
We can intuitively see where the $1/\omega^2$ factor
comes from by studying the classical limit of the scalar propagators:
\begin{subequations}\label{eq:propDemo}
\begin{align}
\frac{i}{(\hat{p}_1+\frac{q_1}2+q_2)^2-m_1^2}&=
\frac{i}{2\hat{p}_1\cdot q_2+q_2\cdot k}=
\frac{i}{2\hat{p}_1\cdot q_2}\left(1-\frac{q_2\cdot k}{2\hat{p}_1\cdot q_2}+\cdots\right)\,,\\
\frac{i}{(\hat{p}_1-\frac{q_1}2-q_2)^2-m_1^2}&=
\frac{-i}{2\hat{p}_1\cdot q_2-q_2\cdot k}=
\frac{-i}{2\hat{p}_1\cdot q_2}\left(1+\frac{q_2\cdot k}{2\hat{p}_1\cdot q_2}+\cdots\right)\,,
\end{align}
\end{subequations}
having used $(\hat{p}_1\pm\frac{q_1}2)^2=m_1^2$.
The leading terms cancel when added;
the sub-leading terms give rise to the desired $1/\omega^2$ propagator.

The third diagram is related to the previous one by symmetry.
Adding up the contributions and dropping unnecessary terms we get\footnote{
	In the full metric $g_{\mu\nu}=\eta_{\mu\nu}+m_{\rm Pl}^{-1}h_{\mu\nu}$
	the radiation occurs at $\cO(m_{\rm Pl}^{-4})=\cO(G^2)$, \ie 2PM.
}
\begin{align}
k^2\braket{h^{\mu\nu}(k)}_{\rm WQFT}&=
\frac{m_1m_2}{4m_{\rm Pl}^3}\int_{q_1,q_2}\hat{\mu}_{1,2}(k)\times\\
&\qquad\left[
\frac{\hat{P}_{12}^{(\mu}\big(\hat{P}_{12}^{\nu)}+\hat{\gamma}\hat{Q}_{12}^{\nu)}\big)}{q_1^2q_2^2}+
\frac{2\hat{\gamma}^2-1}8\left(\frac{\hat{Q}_{12}^\mu \hat{Q}_{12}^\nu}{q_1^2q_2^2}-
\frac{\hat{P}_{12}^\mu \hat{P}_{12}^\nu}{(k\cdot \hat{v}_1)^2(k\cdot \hat{v}_2)^2}\right)
\right]\,,\nn
\end{align}
where $\hat{\gamma}=\hat{v}_1\cdot \hat{v}_2$ and
we have recycled some notation from \rcite{Luna:2017dtq}:
\begin{subequations}
\begin{align}
\hat{P}_{12}^\mu&=k\cdot \hat{v}_1\hat{v}_2^\mu-k\cdot \hat{v}_2\hat{v}_1^\mu\,,\\
\hat{Q}_{12}^\mu&=(q_1-q_2)^\mu-
\frac{q_1^2}{k\cdot \hat{v}_1}\hat{v}_1^\mu+\frac{q_2^2}{k\cdot \hat{v}_2}\hat{v}_2^\mu\,.
\end{align}
\end{subequations}
These vectors satisfy $\hat{P}_{12}\cdot k=\hat{Q}_{12}\cdot k=0$,
which makes gauge invariance manifest.

As $\cM^{\mu\nu}_{\rm GR}$ consists of Feynman propagators
(both for internal gravitons and scalars)
using the established link to compute
$k^2\langle h^{\mu\nu}(k)\rangle_{\text{WQFT}}$
gives rise to Feynman-type graviton propagators
and time-symmetric $1/\omega^2$ worldline propagators
in the classical $\hbar\to0$ limit.
In the latter case, one can re-examine \eqn{eq:propDemo}
and carefully track the $i\eps$'s through
the calculation to show that the result is an average of the 
advanced/retarded propagators given in \eqn{eq:zPropagator}.
This is consistent with our use of $\hat{b}^\mu_i$ and $\hat{v}^\mu_i$ ---
as we discussed at the end of \Sec{sec:toEikonal},
the choice of propagators corresponds to picking a specific inhomogeneous
solution to the equations of motion.

In a genuine physical setting one might also wish to
describe the radiation in terms of $b_i^\mu$ and $v_i^\mu$,
corresponding to the initial trajectories of the black holes.
This would require the use of retarded propagators
for the worldline fluctuations in the above calculation,
which should always point towards the outgoing graviton
and thus provide a clear flow of causality.
In the WQFT one could simply adopt these propagators from the start;
if using an amplitude and taking the classical limit one should take
care and change the $i\eps$ prescription before integration.

While for the worldlines the integration constants $b_i^\mu$,
$v_i^\mu$ mediate between different propagator choices,
this possibility is not available for the gravitons:
a retarded propagator is demanded by the physical setup.
Strictly speaking, the expectation value
$k^2\langle h^{\mu\nu}(k)\rangle_{\text{WQFT}}$
as defined above in terms of a path integral \eqref{eq:expValues}
leads to Feynman graviton propagators\footnote{This is rooted in the fact that this expectation value uses a state fixed by boundary conditions (``in-out'') instead of initial conditions (``in-in'') --- see \rcites{Galley:2009px,Galley:2012hx} for a discussion in the worldline EFT context. See also ref.~\cite{Kosower:2018adc} for a derivation of radiation reaction effects from amplitudes, which involves terms quadratic in the Feynman-propagator based amplitudes.} ---
but of course the $i\eps$ prescription can also be adapted
for the gravitons before integration by identifying the flow of causality.

\subsection{Three-body radiation (3PM)}

\begin{figure}
	\centering
	\begin{tikzpicture}[baseline={(current bounding box.center)}]
	\coordinate (inA) at (-1,1);
	\coordinate (outA) at (2,1);
	\coordinate (inB) at (-1,-1);
	\coordinate (outB) at (2,-1);
	\coordinate (xxA) at (-.2,1);
	\coordinate (xA) at (.6,1);
	\coordinate (yA) at (1.4,1);
	\coordinate (yB) at (1.4,-1);
	\coordinate (C) at (1.4,0);
	\coordinate (k) at (2.5,0);
	\draw [dotted] (inA) -- (xA);
	\draw [zUndirected] (xA) -- (yA) node [midway, above] {$\stackrel{\displaystyle\omega}{\rightarrow}$};
	\draw [dotted] (yA) -- (outA);
	\draw [dotted] (inB) -- (outB);
	\draw [photon] (yA) -- (C);
	\draw [photon] (C) -- (yB);
	\draw [photon] (xxA) to[out=-80,in=-100] (xA);
	\draw [graviton] (C) -- (k);
	\draw [fill] (xA) circle (.08);
	\draw [fill] (xxA) circle (.08);
	\draw [fill] (yA) circle (.08);
	\draw [fill] (yB) circle (.08);
	\draw [fill] (C) circle (.04);
	\end{tikzpicture}
	\caption{An example of a self-energy diagram,
		which does not appear in a three-body calculation at the same PM order.
		On support of the $\delta$-function constraints we have $\omega=0$,
		which gives rise to a singularity in the $1/\omega^2$ propagator.}
	\label{fig:selfEnergy}
\end{figure}
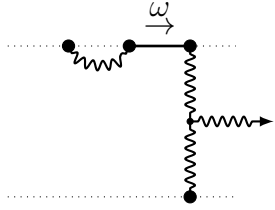

At 3PM we find it convenient to first study the three-body problem,
and the specialization to the two-body problem should just be a special 
case.\footnote{The two-body waveform at this order was previously discussed in the context of
soft theorems from an amplitudes perspective
\cite{Saha:2019tub,Sahoo:2020ryf}
(see also \rcite{Sahoo:2018lxl}).}
A three-body starting point was also
used in \rcite{Shen:2018ebu} to study two-body
radiation in dilaton gravity,
and proved helpful when considering the double copy.
It allows us to identify additional symmetries of the diagrams,
and avoid self-energy graphs which would otherwise give singularities in the two-body problem.
An example of such an unwanted graph is given in \Fig{fig:selfEnergy}.

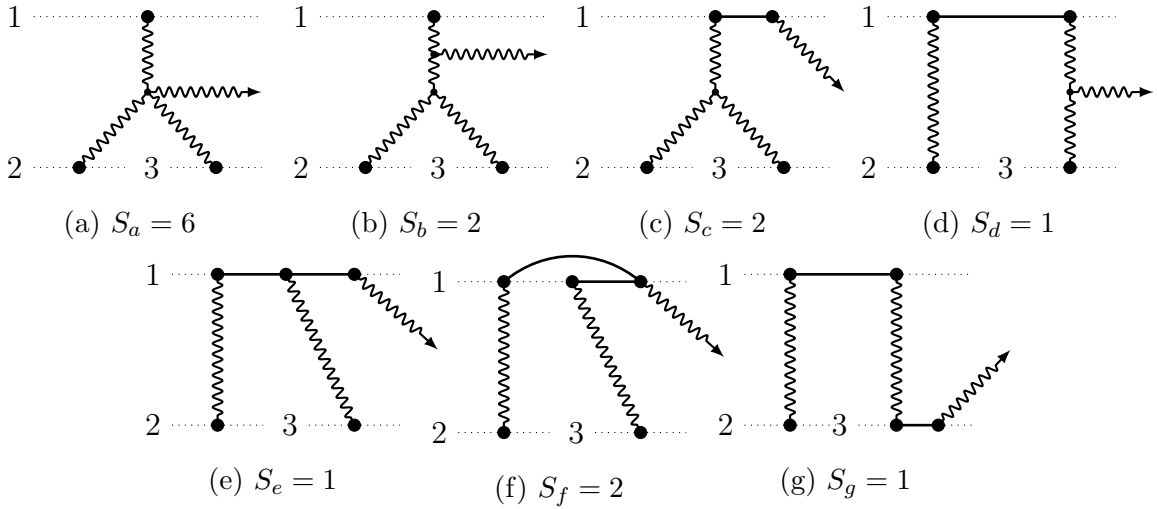
\begin{figure}[t]
  \centering
  \begin{subfigure}{0.24\textwidth}
    \centering
    \begin{tikzpicture}[baseline={(current bounding box.center)}]
	\coordinate (inA) at (-1,1);
	\coordinate (outA) at (2,1);
	\coordinate (inB) at (-1,-1);
	\coordinate (outB) at (2,-1);
	\coordinate (C) at (.5,0);
	\coordinate (A) at (.5,1);
	\coordinate (xB) at (-.4,-1);
	\coordinate (yB) at (1.4,-1);
	\coordinate (xxB) at (.2,-1);
	\coordinate (yyB) at (.8,-1);
	\coordinate (k) at (2,0);
	\draw [dotted] (inA) -- (A);
	\draw [dotted] (A) -- (outA);
	\draw [dotted] (inB) -- (xB);
	\draw [dotted] (xB) -- (xxB);
	\draw [dotted] (yyB) -- (yB);
	\draw [dotted] (yB) -- (outB);
	\draw [fill] (C) circle (.04);
	\draw [fill] (A) circle (.08);
	\draw [fill] (xB) circle (.08);
	\draw [fill] (yB) circle (.08);
	\draw [photon] (C) -- (A);
	\draw [photon] (xB) -- (C);
	\draw [photon] (yB) -- (C);
	\draw [graviton] (C) -- (k);
	\draw (inA) node [left] {$1$};
	\draw (inB) node [left] {$2$};
	\draw (yyB) node [left] {$3$};
	\end{tikzpicture}
	\caption{$S_a=6$}
  \end{subfigure}
  \begin{subfigure}{0.24\textwidth}
    \centering
    \begin{tikzpicture}[baseline={(current bounding box.center)}]
	\coordinate (inA) at (-1,1);
	\coordinate (outA) at (2,1);
	\coordinate (inB) at (-1,-1);
	\coordinate (outB) at (2,-1);
	\coordinate (C) at (.5,0);
	\coordinate (D) at (.5,.5);
	\coordinate (A) at (.5,1);
	\coordinate (xB) at (-.4,-1);
	\coordinate (yB) at (1.4,-1);
	\coordinate (xxB) at (.2,-1);
	\coordinate (yyB) at (.8,-1);
	\coordinate (k) at (2,.5);
	\draw [dotted] (inA) -- (A);
	\draw [dotted] (A) -- (outA);
	\draw [dotted] (inB) -- (xB);
	\draw [dotted] (xB) -- (xxB);
	\draw [dotted] (yyB) -- (yB);
	\draw [dotted] (yB) -- (outB);
	\draw [fill] (C) circle (.04);
	\draw [fill] (D) circle (.04);
	\draw [fill] (A) circle (.08);
	\draw [fill] (xB) circle (.08);
	\draw [fill] (yB) circle (.08);
	\draw [photon] (C) -- (A);
	\draw [photon] (xB) -- (C);
	\draw [photon] (yB) -- (C);
	\draw [graviton] (D) -- (k);
	\draw (inA) node [left] {$1$};
	\draw (inB) node [left] {$2$};
	\draw (yyB) node [left] {$3$};
	\end{tikzpicture}
	\caption{$S_b=2$}
  \end{subfigure}
  \begin{subfigure}{0.24\textwidth}
    \centering
    \begin{tikzpicture}[baseline={(current bounding box.center)}]
	\coordinate (inA) at (-1,1);
	\coordinate (outA) at (2,1);
	\coordinate (inB) at (-1,-1);
	\coordinate (outB) at (2,-1);
	\coordinate (C) at (.5,0);
	\coordinate (xA) at (.5,1);
	\coordinate (yA) at (1.25,1);
	\coordinate (xB) at (-.4,-1);
	\coordinate (yB) at (1.4,-1);
	\coordinate (xxB) at (.2,-1);
	\coordinate (yyB) at (.8,-1);
	\coordinate (k) at (2.2,0);
	\draw [dotted] (inA) -- (xA);
	\draw [zUndirected] (xA) -- (yA);
	\draw [dotted] (yA) -- (outA);
	\draw [dotted] (inB) -- (xB);
	\draw [dotted] (xB) -- (xxB);
	\draw [dotted] (yyB) -- (yB);
	\draw [dotted] (yB) -- (outB);
	\draw [fill] (C) circle (.04);
	\draw [fill] (xA) circle (.08);
	\draw [fill] (yA) circle (.08);
	\draw [fill] (xB) circle (.08);
	\draw [fill] (yB) circle (.08);
	\draw [photon] (C) -- (xA);
	\draw [photon] (xB) -- (C);
	\draw [photon] (yB) -- (C);
	\draw [graviton] (yA) -- (k);
	\draw (inA) node [left] {$1$};
	\draw (inB) node [left] {$2$};
	\draw (yyB) node [left] {$3$};
	\end{tikzpicture}
	\caption{$S_c=2$}
  \end{subfigure}
  \begin{subfigure}{0.24\textwidth}
    \centering
    \begin{tikzpicture}[baseline={(current bounding box.center)}]
	\coordinate (inA) at (-1,1);
	\coordinate (outA) at (2,1);
	\coordinate (inB) at (-1,-1);
	\coordinate (outB) at (2,-1);
	\coordinate (xA) at (-.4,1);
	\coordinate (yA) at (1.4,1);
	\coordinate (xB) at (-.4,-1);
	\coordinate (yB) at (1.4,-1);
	\coordinate (xxB) at (.2,-1);
	\coordinate (yyB) at (.8,-1);
	\coordinate (C) at (1.4,0);
	\coordinate (k) at (2.5,0);
	\draw [dotted] (inA) -- (xA);
	\draw [zUndirected] (xA) -- (yA);
	\draw [dotted] (yA) -- (outA);
	\draw [dotted] (inB) -- (xB);
	\draw [dotted] (xB) -- (xxB);
	\draw [dotted] (yyB) -- (yB);
	\draw [dotted] (yB) -- (outB);
	\draw [fill] (xA) circle (.08);
	\draw [fill] (xB) circle (.08);
	\draw [fill] (yA) circle (.08);
	\draw [fill] (yB) circle (.08);
	\draw [fill] (C) circle (.04);
	\draw [photon] (xA) -- (xB);
	\draw [photon] (yA) -- (C);
	\draw [photon] (C) -- (yB);
	\draw [graviton] (C) -- (k);
	\draw (inA) node [left] {$1$};
	\draw (inB) node [left] {$2$};
	\draw (yyB) node [left] {$3$};
	\end{tikzpicture}
	\caption{$S_d=1$}
  \end{subfigure}
  \begin{subfigure}{0.24\textwidth}
    \centering
    \begin{tikzpicture}[baseline={(current bounding box.center)}]
	\coordinate (inA) at (-1,1);
	\coordinate (outA) at (2,1);
	\coordinate (inB) at (-1,-1);
	\coordinate (outB) at (2,-1);
	\coordinate (xA) at (-.4,1);
	\coordinate (yA) at (.5,1);
	\coordinate (zA) at (1.4,1);
	\coordinate (xB) at (-.4,-1);
	\coordinate (yB) at (1.4,-1);
	\coordinate (xxB) at (.2,-1);
	\coordinate (yyB) at (.8,-1);
	\coordinate (k) at (2.5,0);
	\draw [dotted] (inA) -- (xA);
	\draw [zUndirected] (xA) -- (yA);
	\draw [zUndirected] (yA) -- (zA);
	\draw [dotted] (zA) -- (outA);
	\draw [dotted] (inB) -- (xB);
	\draw [dotted] (xB) -- (xxB);
	\draw [dotted] (yyB) -- (yB);
	\draw [dotted] (yB) -- (outB);
	\draw [fill] (xA) circle (.08);
	\draw [fill] (xB) circle (.08);
	\draw [fill] (yA) circle (.08);
	\draw [fill] (yB) circle (.08);
	\draw [fill] (zA) circle (.08);
	\draw [photon] (xA) -- (xB);
	\draw [photon] (yA) -- (yB);
	\draw [graviton] (zA) -- (k);
	\draw (inA) node [left] {$1$};
	\draw (inB) node [left] {$2$};
	\draw (yyB) node [left] {$3$};
	\end{tikzpicture}
	\caption{$S_e=1$}
  \end{subfigure}
  \begin{subfigure}{0.24\textwidth}
    \centering
    \begin{tikzpicture}[baseline={(current bounding box.center)}]
	\coordinate (inA) at (-1,1);
	\coordinate (outA) at (2,1);
	\coordinate (inB) at (-1,-1);
	\coordinate (outB) at (2,-1);
	\coordinate (xA) at (-.4,1);
	\coordinate (yA) at (.5,1);
	\coordinate (zA) at (1.4,1);
	\coordinate (xB) at (-.4,-1);
	\coordinate (yB) at (1.4,-1);
	\coordinate (xxB) at (.2,-1);
	\coordinate (yyB) at (.8,-1);
	\coordinate (k) at (2.5,0);
	\draw [dotted] (inA) -- (xA);
	\draw [dotted] (xA) -- (yA);
	\draw [zUndirected] (xA) to[out=40,in=140] (zA);
	\draw [zUndirected] (yA) -- (zA);
	\draw [dotted] (zA) -- (outA);
	\draw [dotted] (inB) -- (xB);
	\draw [dotted] (xB) -- (xxB);
	\draw [dotted] (yyB) -- (yB);
	\draw [dotted] (yB) -- (outB);
	\draw [fill] (xA) circle (.08);
	\draw [fill] (xB) circle (.08);
	\draw [fill] (yA) circle (.08);
	\draw [fill] (yB) circle (.08);
	\draw [fill] (zA) circle (.08);
	\draw [photon] (xA) -- (xB);
	\draw [photon] (yA) -- (yB);
	\draw [graviton] (zA) -- (k);
	\draw (inA) node [left] {$1$};
	\draw (inB) node [left] {$2$};
	\draw (yyB) node [left] {$3$};
	\end{tikzpicture}
	\caption{$S_f=2$}
  \end{subfigure}
  \begin{subfigure}{0.24\textwidth}
    \centering
    \begin{tikzpicture}[baseline={(current bounding box.center)}]
	\coordinate (inA) at (-1,1);
	\coordinate (outA) at (2,1);
	\coordinate (inB) at (-1,-1);
	\coordinate (outB) at (2,-1);
	\coordinate (xA) at (-.4,1);
	\coordinate (yA) at (1,1);
	\coordinate (xB) at (-.4,-1);
	\coordinate (yB) at (1,-1);
	\coordinate (zB) at (1.55,-1);
	\coordinate (xxB) at (.2,-1);
	\coordinate (yyB) at (.5,-1);
	\coordinate (k) at (2.5,0);
	\draw [dotted] (inA) -- (xA);
	\draw [zUndirected] (xA) -- (yA);
	\draw [dotted] (yA) -- (outA);
	\draw [dotted] (inB) -- (xB);
	\draw [dotted] (xB) -- (xxB);
	\draw [dotted] (yyB) -- (yB);
	\draw [zUndirected] (yB) -- (zB);
	\draw [dotted] (zB) -- (outB);
	\draw [fill] (xA) circle (.08);
	\draw [fill] (xB) circle (.08);
	\draw [fill] (yA) circle (.08);
	\draw [fill] (yB) circle (.08);
	\draw [fill] (zB) circle (.08);
	\draw [photon] (xA) -- (xB);
	\draw [photon] (yA) -- (yB);
	\draw [graviton] (zB) -- (k);
	\draw (inA) node [left] {$1$};
	\draw (inB) node [left] {$2$};
	\draw (yyB) node [left] {$3$};
	\end{tikzpicture}
	\caption{$S_g=1$}
  \end{subfigure}
  \caption{\small The seven diagrams contributing to the radiation 
  	$k^2\braket{h^{\mu\nu}(k)}_{\rm WQFT}$ at order 3PM,
  	and their respective symmetry factors.
  	All seven graphs represent tree-level contributions,
  	the worldlines being drawn only as a visual aid.
  	The outgoing momentum from each worldline is $q_i^\mu$.
  }
  \label{fig:subradiationGraphs}
\end{figure}

The radiation is fully described by the seven diagrams in
\Fig{fig:subradiationGraphs}:
\begin{equation}
k^2\braket{h^{\mu\nu}(k)}_{\rm WQFT}=\frac{m_1m_2m_3}{m_{\rm Pl}^5}
\sum_{S_3}\sum_{i\in\{a,\ldots,g\}}\int_{q_1,q_2,q_3}
\mu_{1,2,3}(k)\frac1S_i\frac{N_i^{\mu\nu}}{D_i}\,,
\end{equation}
where we now use retarded propagators both for the gravitons and worldline
(but omit the propagator on the outgoing line with momentum $k^\mu$).
As the diagrams must connect with all three worldlines,
self-energy diagrams of the kind in \Fig{fig:selfEnergy} are avoided.
We sum over $3!$ permutations of the worldlines,
swapping $q^\mu_i$, $b^\mu_i$ and $v^\mu_i$ in each case.
By design these permutations preserve the integral measure emerging from \eqn{centralrel}:
\begin{equation}
\mu_{1,2,3}(k)=
e^{i(q_1\cdot b_1+q_2\cdot b_2+q_3\cdot b_3)}
\dd(v_1\cdot q_1)\dd(v_2\cdot q_2)\dd(v_3\cdot q_3)
\dd(k-q_1-q_2-q_3)\,,
\end{equation}
which (after an appropriate rearrangement) is the same for all seven diagrams.
Each symmetry factor $S_i$ corrects for an ``overcount'' in the sum ---
for example, diagram (a) is invariant under all 6 permutations,
so dividing by $S_a=6$ accounts for this.
The propagator factors are
\begin{align}
\begin{aligned}
D_a&=q_1^2q_2^2q_3^2\,, &
D_b&=q_1^2q_2^2q_3^2q_{23}^2\,, \\
D_c&=q_2^2q_3^2q_{23}^2(\omega_1+i\eps)^2\,, &
D_d&=q_2^2q_3^2q_{12}^2(\omega_{12}+i\eps)^2\,, \\
D_e&=q_2^2q_3^2(\omega_1+i\eps)^2(\omega_{12}+i\eps)^2\,, &
D_f&=q_2^2q_3^2(\omega_{12}+i\eps)^2(\omega_{13}+i\eps)^2\,, \\
D_g&=q_2^2 q_{12}^2(\omega_{12}+i\eps)^2(\omega_3+i\eps)^2\,,
\end{aligned}
\end{align}
where $ q_{ij}=q_i+ q_j$ and we have introduced new variables:
\begin{align}
\omega_i=v_i\cdot k\,, &&
\omega_{ij}=v_i\cdot q_j\,.
\end{align}
On support of $\mu_{1,2,3}(k)$,
$\omega_i=\sum_{j=1}^3\omega_{ij}$ and $\omega_{ii}=0$ (no sum on $i$).
To confirm this result we have checked the off-shell ($k^2\neq0$) Ward identity
\begin{align}
k_\mu\braket{h^{\mu\nu}(k)}_{\rm WQFT}=0\,,
\end{align}
which holds already at the integrand level.

Our final result for $k^2\braket{h^{\mu\nu}(k)}_{\rm WQFT}$
is presented in an ancillary file
attached to the \texttt{arXiv} submission of this paper,
with expressions given for each of the seven numerators $N_i^{\mu\nu}$
in \Fig{fig:subradiationGraphs}.
In a separate \texttt{Mathematica} file we also explicitly
demonstrate the off-shell Ward identity;\footnote{
	Our demonstration file relies on the tensor computer algebra package
	\texttt{xAct} \cite{DBLP:journals/corr/abs-0803-0862}.
}
the on-shell graviton can easily be obtained by setting $k^2=0$.
We claim that the same integrand
(with $b_i^\mu\to\hat{b}_i^\mu$, $v_i^\mu\to\hat{v}_i^\mu$)
can also be obtained from a seven-point scalar-graviton
amplitude with three pairs of distinctly flavored scalars:
\begin{equation}\label{eq:LinkA7}
k^2 \Bigl \langle h^{\mu\nu}(k)\Bigr \rangle_{\text{WQFT}} = 
\frac{i}{8m_1 m_2m_3}\int_{q_1,q_2,q_3}
\hat{\mu}_{1,2,3}(k)\lim_{\hbar\to 0}{\cal M}^{\mu\nu}_{\rm GR}(p_{i},p'_i,k)\,,
\end{equation}
but checking this explicitly we save for future work.

\section{Deflections}
\label{sec:deflections}

Let us now switch to a purely conservative setting.
We compute the impulse on a single black hole in a binary scattering,
which classically can be expressed as
\begin{align}\label{eq:impulseFirst}
\Delta p_i^\mu=m_i\Delta\dot{x}_i^\mu=m_i\int\!\d\dot{x}_i^\mu(\tau_i)=
m_i\int_{-\infty}^\infty\d\tau_i\frac{\d^2z_i^\mu(\tau_i)}{\d\tau_i^2}\,.
\end{align}
Even though this is a total derivative,
in the present context it does not integrate to zero.
In the WQFT (where $z_i^\mu(\tau_i)$ is promoted to an operator)
our task is therefore to calculate the expectation value
\begin{align}\label{eq:impulseOpdef}
\Delta p_i^\mu=
m_i\int_{-\infty}^\infty\d\tau_i
\Bigl\langle
\frac{\d^2z_i^\mu(\tau_i)}{\d\tau_i^2} \Bigr\rangle_{\rm WQFT}\,.
\end{align}
Inserting the Fourier space definition of $z^\mu$
\eqref{eq:fourier} the impulse becomes
\begin{align}\label{eq:impulsefinal}
\Delta p_i^\mu=
m_i\int_\omega(-i\omega)^2 \WQFTbraket{z_i^\mu(\omega) \dd(\omega)}=
\left.-m_i \omega^2 \WQFTbraket{z_i^\mu(\omega)} \right|_{\omega=0} \,.
\end{align}
Hence the impulse follows from  drawing tree-level graphs with a cut external $z_i^\mu$ line,
multiplied by a factor of $i$ for the correct normalization.
This is analogous to how we computed $k^2\braket{h^{\mu\nu}(k)}_{\rm WQFT}$
with $k^2=0$ in \Sec{sec:radiation}.
By using retarded worldline propagators we ensure a
flow of causality towards the outgoing line;
time-symmetric Feynman propagators for the gravitons imply
a purely elastic scattering of the black holes.
To include radiative effects one could instead use retarded graviton propagators,
but in the integrals that follow we shall assume the former.

As a demonstration,
we shall now compute the conservative deflection $\Delta p_1^\mu$ up to order 2PM,
specifically reproducing the integrands by K\"alin and Porto~\cite{Kalin:2020mvi}
whose integrated result matches earlier
work by \eg Westpfahl \cite{Westpfahl:1985tsl}.
Our method does not require the determination of an
effective action for the black holes.

\subsection{Leading order (1PM)}

At leading order $\Delta p_1^\mu$ is described by a single diagram:
\begin{align}\label{eq:leadingDefDiagram}
\begin{tikzpicture}[baseline={(current bounding box.center)}]
\coordinate (inA) at (-1,1);
\coordinate (outA) at (1,1);
\coordinate (inB) at (-1,-1);
\coordinate (outB) at (1,-1);
\coordinate (xA) at (0,1);
\coordinate (xB) at (0,-1);
\draw [dotted] (inA) -- (xA);
\draw [dotted] (inB) -- (xB);
\draw [dotted] (xB) -- (outB);
\draw [zUndirected] (xA) -- (outA) node [midway, below] {$\quad\omega=0$};
\draw [fill] (xA) circle (.08);
\draw [fill] (xB) circle (.08);
\draw [photon] (xA) -- (xB) node [midway, left] {$q\!\uparrow$};
\draw (inA) node [left] {$1$};
\draw (inB) node [left] {$2$};
\end{tikzpicture}=i\frac{m_1 m_2}{4m_{\rm Pl}^2}\!
\int_q e^{iq\cdot b}\dd(q\cdot v_1)\dd(q\cdot v_2)
(-v_1^\nu v_1^\rho q^\mu)
\frac{P_{\nu\rho;\sigma\lambda}}{q^2}v_2^\sigma v_2^\lambda\,,
\end{align}
where $b^\mu=b^\mu_2-b^\mu_1$ ($b^2<0$).
Cleaning this up we deduce that
\begin{align}\label{eq:deflection}
\Delta p_1^\mu&=\frac{m_1m_2}{8m_{\rm Pl}^2}(2\gamma^2-1)
\frac{\partial}{\partial b_{1,\mu}}
\int_q\frac{\dd(q\cdot v_1)\dd(q\cdot v_2)}{q^2}e^{iq\cdot b}\,,
\end{align}
where $\gamma=v_1\cdot v_2>1$.
This matches eq.~(4.9) in \rcite{Kalin:2020mvi}.

The integral above can be performed in a variety of ways
(see also \eg \rcite{DiVecchia:2019kta});
to maintain covariance we find it convenient to decompose
$q=q_\parallel+q_\perp$,
where $q_\parallel=P_\parallel q$ is parallel and
$q_\perp=P_\perp q$ is perpendicular
to the plane described by the two-form $P = v_1 \wedge v_2$.\footnote{
	We thank Gregor K\"alin and Rafael Porto for sharing details of this method with us.
}
The projectors $P_\parallel$ and $P_\perp$ are
\begin{subequations}
\begin{align}
P_\parallel^{\mu\nu} &= \frac{P^{\mu\rho} P_\rho{}^\nu}{|P|^2}
=-\frac1{\gamma^2-1} \left[ v_1^\mu v_1^\nu - 2 \gamma v_1^{(\mu} v_2^{\nu)} + v_2^\mu v_2^\nu \right] \,, \\
P_\perp^{\mu\nu} &= \eta^{\mu\nu} - P_\parallel^{\mu\nu} \,,
\end{align}
\end{subequations}
where $|P|^2 \equiv - \frac{1}{2} P_{\mu\nu}P^{\mu\nu} = \gamma^2 - 1$.
$P_{\parallel\mu\nu}$ is the induced metric of the parallel plane,
so we adopt the notation $\eta_\parallel^{\mu\nu} \equiv P_\parallel^{\mu\nu}$;
the corresponding volume form is $\epsilon_{\parallel\mu\nu} = -\epsilon_{\parallel\nu\mu}$.
It holds that $\det_\parallel \eta_\parallel = -1$ and $\epsilon_\parallel^{\mu\nu} \epsilon_\parallel^{\alpha\beta} = - \eta_\parallel^{\mu\alpha} \eta_\parallel^{\nu\beta} + \eta_\parallel^{\mu\beta} \eta_\parallel^{\nu\alpha}$.
The Dirac deltas impose $q_\parallel\cdot v_1=q_\parallel\cdot v_2=0$,
and upon eliminating them we get an additional factor from the Jacobian determinant:
\begin{equation}
\dd(\underbrace{v_1\cdot q_\parallel}_x)
\dd(\underbrace{v_2\cdot q_\parallel}_y) =
\underbrace{\left|\frac{\partial(x,y)}{\partial q_\parallel}\right|^{-1}}_{\det^{-1}_\parallel(v_1, v_2) = |P|^{-1}}
\dd\big(q_\parallel^\mu\big)\,,
\end{equation}
where $\det_\parallel(v_1,v_2)=\frac12\epsilon_\parallel^{\mu\nu}P_{\mu\nu}$
and $(\epsilon_\parallel^{\mu\nu}P_{\mu\nu})^2=-2P_{\mu\nu}P^{\mu\nu}=4|P|^2$.
Therefore
\begin{align}
\int_q\frac{\dd(q\cdot v_1)\dd(q\cdot v_2)}{q^2}e^{iq\cdot b}=
\frac1{|P|}\int_{q_\perp}\frac{e^{iq_\perp\cdot b}}{q_\perp^2}=
\frac{\log|b|}{2\pi|P|} + \text{const} \,,
\end{align}
and when plugged into \eqn{eq:deflection} this yields the final result:\footnote{
	The overall sign is consistent with an attractive gravitational force:
	$\Delta p_1^\mu$ aligns with $b^\mu=b_2^\mu-b_1^\mu$,
	which points from the first to the second black hole.
}
\begin{align}\label{eq:1PMdeflection}
\Delta p_1^\mu=2Gm_1m_2\frac{(2\gamma^2-1)}{\sqrt{\gamma^2-1}}\frac{b^\mu}{|b^2|}\,,
\end{align}
where $m_{\rm Pl}^{-1}=\sqrt{32\pi G}$. 
This result can also be found in (for example) \rcites{Bini:2018ywr,Guevara:2019fsj}.

\subsection{Sub-leading order (2PM)}

At $\cO(G^2)$ there are four contributing diagrams,
with two each proportional to $m_1m_2^2$ and $m_1^2m_2$.
As they carry different integral measures we treat them separately.
In the first category:
\begin{align}\label{eq:2PMdeflection1}
\!\!\!\!\!\!\!\!\!\!\!\!\!\!\!\!\!\!\!\!\!\!\!\!
\begin{tikzpicture}[baseline={(current bounding box.center)}]
\coordinate (inA) at (-1,1);
\coordinate (outA) at (2,1);
\coordinate (inB) at (-1,-1);
\coordinate (outB) at (2,-1);
\coordinate (xA) at (0,1);
\coordinate (yA) at (1,1);
\coordinate (xB) at (0,-1);
\coordinate (yB) at (1,-1);
\draw [dotted] (inA) -- (xA);
\draw [zUndirected] (xA) -- (yA) node [midway, above] {$\stackrel{\displaystyle\widetilde{\omega}}{\rightarrow}$};
\draw [zUndirected] (yA) -- (outA) node [midway, above] {$\omega\!=\!0$};
\draw [dotted] (inB) -- (xB);
\draw [dotted] (xB) -- (yB);
\draw [dotted] (yB) -- (outB);
\draw [fill] (xA) circle (.08);
\draw [fill] (yA) circle (.08);
\draw [fill] (xB) circle (.08);
\draw [fill] (yB) circle (.08);
\draw [photon] (xA) -- (xB) node [midway, left] {$k\!\uparrow$};
\draw [photon] (yA) -- (yB) node [midway, right] {$\downarrow\!k\!-\!q$};
\draw (inA) node [left] {$1$};
\draw (inB) node [left] {$2$};
\end{tikzpicture}&=i\frac{m_1m_2^2}{64m_{\rm Pl}^4}
\int_{k,q}
\frac{(q^\mu-k^\mu)\dd(q\cdot v_1)\dd(q\cdot v_2)\dd(k\cdot v_2)}
{k^2(k-q)^2(k\cdot v_1+i\eps)^2}e^{iq\cdot b}\times
\end{align}
\vspace{-3.5em}
\begin{align}
\qquad\qquad\qquad\qquad\qquad\qquad
\left[(2\gamma^2-1)^2k\cdot(k-q)+8\gamma^2(k\cdot v_1)^2\right]\,.\nn
\end{align}
The integral measure is
$\int_{k,q,\tilde{\omega}}\dd(k\cdot v_2)\dd(\tilde{\omega}-k\cdot v_1)
\dd(\tilde{\omega}-(k-q)\cdot v_1)\dd((q-k)\cdot v_2)$
in its initial form;
$\tilde{\omega}$ integration yields $\widetilde{\omega}=k\cdot v_1$
and leaves the three remaining $\delta$-functions in \eqn{eq:2PMdeflection1}.
This diagram matches eq.~(4.14) of ref.~\cite{Kalin:2020mvi},
up to terms that vanish upon integration
(those that do not contribute to long-range interactions).
The other diagram with the same integral measure is
\begin{align}
\!\!\!\!\!\!\!\!\!\!\!\!\!\!\!\!\!\!\!\!\!\!\!\!
\begin{tikzpicture}[baseline={(current bounding box.center)}]
\coordinate (inA) at (-1,1);
\coordinate (outA) at (2,1);
\coordinate (inB) at (-1,-1);
\coordinate (outB) at (2,-1);
\coordinate (C) at (.5,0);
\coordinate (A) at (.5,1);
\coordinate (xB) at (-.4,-1);
\coordinate (yB) at (1.4,-1);
\draw [dotted] (inA) -- (A);
\draw [zUndirected] (A) -- (outA) node [midway, above] {$\omega\!=\!0$};
\draw [dotted] (inB) -- (xB);
\draw [dotted] (xB) -- (yB);
\draw [dotted] (yB) -- (outB);
\draw [fill] (C) circle (.08);
\draw [fill] (A) circle (.08);
\draw [fill] (xB) circle (.08);
\draw [fill] (yB) circle (.08);
\draw [photon] (C) -- (A) node [midway, left] {$q\!\uparrow$};
\draw [photon] (xB) -- (C) node [midway, left] {$k\!\nearrow$};
\draw [photon] (yB) -- (C) node [midway, right] {$\searrow\!k\!-\!q$};
\draw (inA) node [left] {$1$};
\draw (inB) node [left] {$2$};
\end{tikzpicture}&=-i\frac{m_1m_2^2}{16m_{\rm Pl}^4}
\int_{k,q}
\frac{q^\mu\dd(q\cdot v_1)\dd(q\cdot v_2)\dd(k\cdot v_2)}
{k^2q^2(k-q)^2}e^{iq\cdot b}\times
\end{align}
\vspace{-3.5em}
\begin{align}
\qquad\qquad\qquad\qquad\qquad\qquad
\left[\gamma^2q^2+(k\cdot v_1)^2+(2\gamma^2-1)k\cdot(k-q)\right]\,.\nn
\end{align}
This agrees with eq.~(4.15) of ref.~\cite{Kalin:2020mvi}
(up to a symmetry factor of $1/2$).
In the second category we have
\begin{align}
\!\!\!\!\!\!\!\!\!\!\!\!\!\!\!\!\!\!\!\!\!\!\!\!
\begin{tikzpicture}[baseline={(current bounding box.center)}]
\coordinate (inA) at (-1,1);
\coordinate (outA) at (2,1);
\coordinate (inB) at (-1,-1);
\coordinate (outB) at (2,-1);
\coordinate (xA) at (0,1);
\coordinate (yA) at (1,1);
\coordinate (xB) at (0,-1);
\coordinate (yB) at (1,-1);
\draw [dotted] (inA) -- (xA);
\draw [zUndirected] (xB) -- (yB) node [midway, above] {$\stackrel{\displaystyle\widetilde{\omega}}{\rightarrow}$};
\draw [zUndirected] (yA) -- (outA) node [midway, above] {$\omega\!=\!0$};
\draw [dotted] (inB) -- (xB);
\draw [dotted] (xA) -- (yA);
\draw [dotted] (yB) -- (outB);
\draw [fill] (xA) circle (.08);
\draw [fill] (yA) circle (.08);
\draw [fill] (xB) circle (.08);
\draw [fill] (yB) circle (.08);
\draw [photon] (xA) -- (xB) node [midway, left] {$k\!\uparrow$};
\draw [photon] (yA) -- (yB) node [midway, right] {$\downarrow\!k\!-\!q$};
\draw (inA) node [left] {$1$};
\draw (inB) node [left] {$2$};
\end{tikzpicture}&=i\frac{m_1^2m_2}{64m_{\rm Pl}^4}
\int_{k,q}
\frac{(q^\mu-k^\mu)\dd(q\cdot v_1)\dd(q\cdot v_2)\dd(k\cdot v_1)}
{k^2(k-q)^2(k\cdot v_2-i\eps)^2}e^{iq\cdot b}\times
\end{align}
\vspace{-3.5em}
\begin{align}
\qquad\qquad\qquad\qquad\qquad\qquad
\left[(2\gamma^2-1)^2k\cdot(k-q)+8\gamma^2(k\cdot v_2)^2\right]\,,\nn
\end{align}
which (except for the outgoing $\omega=0$ line) is
related to \eqref{eq:2PMdeflection1} by symmetry;
the $\delta$-function constraint yields $\tilde{\omega}=-k\cdot v_2$.
Finally,
\begin{align}
\!\!\!\!\!\!\!\!\!\!\!\!\!\!\!\!\!\!\!\!\!\!\!\!
\begin{tikzpicture}[baseline={(current bounding box.center)}]
\coordinate (inA) at (-1,1);
\coordinate (outA) at (2,1);
\coordinate (inB) at (-1,-1);
\coordinate (outB) at (2,-1);
\coordinate (xA) at (-.4,1);
\coordinate (yA) at (1.4,1);
\coordinate (C) at (.5,0);
\coordinate (B) at (.5,-1);
\draw [dotted] (inA) -- (xA);
\draw [dotted] (xA) -- (yA);
\draw [zUndirected] (yA) -- (outA) node [midway, above] {$\omega\!=\!0$};
\draw [dotted] (inB) -- (B);
\draw [dotted] (B) -- (outB);
\draw [fill] (C) circle (.08);
\draw [fill] (xA) circle (.08);
\draw [fill] (B) circle (.08);
\draw [fill] (yA) circle (.08);
\draw [photon] (C) -- (B) node [midway, left] {$q\!\uparrow$};
\draw [photon] (xA) -- (C) node [midway, left] {$k\!\nwarrow$};
\draw [photon] (yA) -- (C) node [midway, right] {$\swarrow\!k\!-\!q$};
\draw (inA) node [left] {$1$};
\draw (inB) node [left] {$2$};
\end{tikzpicture}&=i\frac{m_1^2m_2}{16m_{\rm Pl}^4}
\int_{k,q}
\frac{(k^\mu-q^\mu)\dd(q\cdot v_1)\dd(q\cdot v_2)\dd(k\cdot v_2)}
{k^2q^2(k-q)^2}e^{iq\cdot b}\times
\end{align}
\vspace{-3.5em}
\begin{align}
\qquad\qquad\qquad\qquad\qquad\qquad
\left[\gamma^2q^2+(k\cdot v_2)^2+(2\gamma^2-1)k\cdot(k-q)\right]\,.\nn
\end{align}
Not included are diagrams involving self-interactions of
gravitons on a single worldline,
which also do not contribute to the final integrated result.

Taken together, these four diagrams make up the 2PM deflection.
As the integration was already discussed at length in \rcite{Kalin:2020mvi}
we will not reiterate the details;
instead we simply present the final result for the conservative impulse at this order:
\begin{align}\label{eq:2PMdeflection}
\begin{aligned}
\Delta p_1^\mu&=
\frac{Gm_1m_2b^\mu}{|b^2|}
\left(\frac{2(2\gamma^2-1)}{\sqrt{\gamma^2-1}}+\frac{3\pi}4
\frac{(5\gamma^2-1)}{\sqrt{\gamma^2-1}}\frac{G(m_1+m_2)}{|b|}\right)\\
&\qquad-
2G^2m_1m_2\frac{(2\gamma^2-1)^2}{(\gamma^2-1)^2|b^2|}
\left((\gamma m_1+m_2)v_1^\mu-(\gamma m_2+m_1)v_2^\mu\right)\,.
\end{aligned}
\end{align}
This includes the 1PM result already given in \eqn{eq:1PMdeflection},
and agrees with Westpfahl's result~\cite{Westpfahl:1985tsl}.
It also satisfies
\begin{equation}
p_1^2=(p_1+\Delta p_1)^2
\end{equation}
up to terms $\cO(G^3)$, using $b\cdot v_i=0$.
This is consistent with our use of retarded worldline propagators:
$p_i^\mu=m_iv_{i}^\mu$ are the \emph{incoming} momenta.

Should we flip the sign on $i\eps$ throughout our calculation above,
\ie use advanced instead of retarded worldline propagators,
then the result \eqref{eq:2PMdeflection} in terms of
${b_i'}^\mu$ and ${v_i'}^\mu$ is identical except
with the signs on ${v_i'}^\mu$ reversed.
This is consistent with the impulse instead obeying ${p'_1}^2=(p'_1-\Delta p_1)^2$,
where ${p'_i}^\mu=m_i{v'_i}^\mu$ are the \emph{outgoing} momenta.
Similarly, if we use time-symmetric worldline propagators
(an average of advanced/retarded)
then the terms proportional to $\hat{v}_i^\mu$ vanish altogether ---
more on this in the next section.
The impulse obeys
$(\hat{p}_1+\frac{\Delta p_1}2)^2=(\hat{p}_1-\frac{\Delta p_1}2)^2$,
\ie $\hat{p}_1\cdot\Delta p_1=0$.

\section{Eikonal phase}\label{sec:eikonal}

Having now computed the deflection $\Delta p_1^\mu$ up to 2PM order,
let us finally explain its connection to scattering amplitudes.
Unlike with the emitted graviton $k^2\braket{h^{\mu\nu}(k)}$ computed
in \Sec{sec:radiation}, it is not immediately obvious how to obtain
$\Delta p_i^\mu$ from an amplitudes perspective.
The reason is simple: unlike $h_{\mu\nu}(x)$,
the worldline fluctuations $z_i^\mu(\tau)$ do not live in the amplitudes;
instead we have the scalars $\phi_i(x)$.
So the trick we used in \eqn{eq:LinkA5} to integrate out the scalars
in a five-point amplitude,
leaving behind the expectation $\braket{h^{\mu\nu}(k)}$, does not work.
Instead we need to make use of the four-point scalar amplitude
$\mathcal{M}_{\phi_1\phi_2\to \phi_1\phi_2}$.

From an amplitudes perspective the eikonal phase of \eqn{eq:eikonalRelation} is a very useful scalar quantity, as
it captures the impulse and other classical observables.  Writing the S-matrix in terms of a scattering amplitude \eqn{eq:eikonalRelation} gives rise to
\begin{align}\label{eq:eikonalA}
\begin{aligned}
  e^{i \chi} &= 1 + \frac{i}{4m_1m_2}\int_q e^{i q \cdot b} \dd(v_1 \cdot q) \dd(v_2 \cdot q) \lim_{\hbar\to 0}\mathcal{M}_{\phi_{1}\, \phi_{2}\to \phi_{1}\, \phi_{2}}(q) \\
  &= 1 + i\int_{q_\perp} e^{i q_\perp \cdot b} \frac{\mathcal{M}_{\phi_{1}\, \phi_{2}\to \phi_{1}\, \phi_{2}}(q)}{4 m_1 m_2 \sqrt{\gamma^2 - 1}} \,,
\end{aligned}
\end{align}
where $\chi$ is the eikonal phase and $q$ is the transferred momentum.
It was demonstrated in refs.~\cite{Bjerrum-Bohr:2018xdl,Bern:2020buy} (see also \rcite{Kalin:2020mvi}) that the 2PM deflection in the center-of-mass system
is obtained via
\begin{equation}\label{eq:dpeikonal}
  \Delta p_\perp = \frac{\partial \chi}{\partial b} \,,
\end{equation}
where $p_\perp = P_\perp p_2 = - P_\perp p_1$.
A similar relation holds for the scattering angle \cite{Bjerrum-Bohr:2018xdl} ---
see also eq.~(4.5) in \rcite{Damour:2019lcq}.
It is suggestive that such a relation can be extended to higher orders and changes in spins \cite{Maybee:2019jus,Bern:2020buy}.
That is, the eikonal phase $\chi$ appears to be a generator
for all observables of conservative classical scattering.
It can also be related to bound orbits via analytic continuation \cite{Antonelli:2020}, 
analogously to the investigation in \rcites{Kalin:2019rwq,Kalin:2019inp}.

From our WQFT perspective we have shown in \Sec{sec:toEikonal}
that the classical part of the eikonal phase $\chi$ is given by the free 
energy of the WQFT at tree level (integrating out the $z^{\mu}$ and $h_{\mu\nu}$ fields).
We use Feynman propagators for the gravitons
(which also occur in the QFT S-matrix)
and time-symmetric propagators for the worldlines.
So $\hat{b}_{i}^\mu$ and $\hat{v}_{i}^\mu$ are identified with the
average of the incoming/outgoing momenta $\hat{p}_i^\mu$.
Recalling \eqns{ZWQFTdef}{eq:eikonalId}, the eikonal phase is then
\begin{align}\label{eq:defeikonal}
  e^{i \chi (\hat{b}_{i},\hat{v}_{i})} &= Z_{\text{WQFT}}\\
  &= \text{const} \times \int \cD[h_{\mu\nu}, z_1^\mu, z_2^\mu] \, 
  \exp\Bigl[i\Big(S_{\rm EH}+S_{\rm gf}-\sum_{i=1}^{n}\int_{-\infty}^{\infty}\!\d\tau_{i}\, \frac{m_{i}}{2}
g_{\mu\nu} \xdot_{i}^{\mu}\xdot_{i}^{\nu} \Big) \Bigr ]\,,\nn
\end{align}
where we have dropped the Lee-Yang ghost contributions
which are irrelevant for the classical limit.
Instead of \eqn{eq:dpeikonal} we will demonstrate that
\begin{equation}\label{eq:dpeikonal2}
  \Delta p_{1,\mu} =  i \frac{\partial}{\partial\hat{b}_{1}^{\mu}} 
 \log Z_{\text{WQFT}} = -\frac{\partial \chi}{\partial\hat{b}_1^\mu}
\end{equation}
holds in our formalism to all orders as a consequence of \eqn{eq:eikonalmaster}.
This should satisfy
$(\hat{p}_1+\frac{\Delta p_1}2)^2=(\hat{p}_1-\frac{\Delta p_1}2)^2$,
\ie $\hat{p}_1\cdot\Delta p_1=0$.
Note how connected amplitudes get mapped into WQFT diagrams in $\mathcal{Z}_{\text{WQFT}}$ that are disconnected in general, and finally exponentiate into connected WQFT diagrams in $\chi$.
This manifests the exponentiation of amplitudes in the classical limit.

Let us now prove \eqn{eq:dpeikonal2}.
On the one hand, from \eqn{eq:impulsefinal} we have that 
\begin{multline}
  \Delta p_1^\sigma =
\left.-m_1 \omega^2 \WQFTbraket{z_1^\mu(\omega)} \right|_{\omega=0} \\
  = - m_1  \Bigg \langle  \omega^2 z_1^\sigma(\omega) \exp \Bigg [ \overbrace{\int \sum_{n=0}^\infty \frac{1}{n!} V^{\text{WL},\mu\nu}_{1,\rho_1\dots\rho_n}(k; \omega_1,\dots, \omega_n) h_{\mu\nu}(k) z_1^{\rho_1}(\omega_1) \dots z_1^{\rho_n}}^{=:(\text{WL}_1)} \\ + (\text{WL}_2) + (\text{GR}) \Bigg]  \Bigg \rangle_\text{free WQFT} \Bigg|_{\omega=0} ,
\end{multline}
with $\langle \ldots\rangle_{\text{free WQFT}}$ denoting the vacuum expectation
value of the non-interacting theory and (GR) the graviton vertices.
This expression involves only connected diagrams, or Wick contractions.
We perform all Wick contractions involving the \emph{external} $z_1^\sigma(\omega)$,
noting that contractions with an $n$-vertex
on the worldline give an overall factor of $n$.
Thereafter we set $\omega=0$, essentially canceling the external propagator:
\begin{multline}
  \Delta p_1^\sigma
  = i \Bigg \langle  \Bigg( \int \sum_{n=1}^\infty \frac{1}{(n-1)!} V^{\text{WL},\mu\nu}_{1,\rho_1\dots\rho_n}(k; \omega_1,\dots, \omega_{n-1}, \omega) h_{\mu\nu} z_1^{\rho_1} \dots z_1^{\rho_{n-1}} \eta^{\sigma\rho_n} \Bigg) \\
   e^{(\text{WL}_1) + (\text{WL}_2) + (\text{GR})}  \Bigg \rangle_{\text{free WQFT}}\, 
   \Bigg|_{\omega=0} \, .
\end{multline}
On the other hand, again considering connected contractions only (leaving the $\log$ implicit), it holds that
\begin{align}
  &\Delta p_1^\sigma = i \frac{\partial i \chi}{\partial b_{1,\sigma}}
  = i \bigg \langle  \frac{\partial}{\partial b_{1,\sigma}} e^{(\text{WL}_1) + (\text{WL}_2) + (\text{GR})}  \bigg \rangle_{\text{free WQFT}} \\
  &= i \Bigg \langle  \Bigg( \int \sum_{n=0}^\infty \frac{1}{n!} \underbrace{\frac{\partial}{\partial b_{1,\sigma}} V^{\text{WL},\mu\nu}_{1,\rho_1\dots\rho_n}(k; \omega_1,\dots, \omega_{n})}_{V^{\text{WL},\mu\nu}_{1,\rho_1\dots\rho_{n+1}}(k; \omega_1,\dots, \omega_n,0) \eta^{\sigma\rho_{n+1}}} h_{\mu\nu} z_1^{\rho_1} \dots z_1^{\rho_n} \Bigg)  e^{(\text{WL}_1) + (\text{WL}_2) + (\text{GR})} \Bigg \rangle_{\text{free WQFT}} , \nonumber
\end{align}
making crucial use of \eqn{eq:eikonalmaster}.
This is the same as the preceding equation when shifting $n$ and using $\omega=0$, thus showing eq.~\eqref{eq:dpeikonal2}.

Let us now work out the eikonal to 2PM order.
The corresponding diagrams in the eikonal phase are
\begin{align}\label{eq:eikonal2PM}
i \chi = \begin{tikzpicture}[baseline={(current bounding box.center)}]
\coordinate (inA) at (-1,1);
\coordinate (outA) at (1,1);
\coordinate (inB) at (-1,-1);
\coordinate (outB) at (1,-1);
\coordinate (xA) at (0,1);
\coordinate (xB) at (0,-1);
\draw [dotted] (inA) -- (xA);
\draw [dotted] (inB) -- (xB);
\draw [dotted] (xB) -- (outB);
\draw [dotted] (xA) -- (outA);
\draw [fill] (xA) circle (.08);
\draw [fill] (xB) circle (.08);
\draw [photon] (xA) -- (xB);
\end{tikzpicture}\quad+
\begin{tikzpicture}[baseline={(current bounding box.center)}]
\coordinate (inA) at (-1,1);
\coordinate (outA) at (2,1);
\coordinate (inB) at (-1,-1);
\coordinate (outB) at (2,-1);
\coordinate (xA) at (0,1);
\coordinate (yA) at (1,1);
\coordinate (xB) at (0,-1);
\coordinate (yB) at (1,-1);
\draw [dotted] (inA) -- (xA);
\draw [zParticleF] (xA) -- (yA);
\draw [dotted] (yA) -- (outA);
\draw [dotted] (inB) -- (xB);
\draw [dotted] (xB) -- (yB);
\draw [dotted] (yB) -- (outB);
\draw [fill] (xA) circle (.08);
\draw [fill] (yA) circle (.08);
\draw [fill] (xB) circle (.08);
\draw [fill] (yB) circle (.08);
\draw [photon] (xA) -- (xB);
\draw [photon] (yA) -- (yB);
\end{tikzpicture}\quad+
\begin{tikzpicture}[baseline={(current bounding box.center)}]
\coordinate (inA) at (-1,1);
\coordinate (outA) at (2,1);
\coordinate (inB) at (-1,-1);
\coordinate (outB) at (2,-1);
\coordinate (C) at (.5,0);
\coordinate (A) at (.5,1);
\coordinate (xB) at (-.4,-1);
\coordinate (yB) at (1.4,-1);
\draw [dotted] (inA) -- (A);
\draw [dotted] (A) -- (outA);
\draw [dotted] (inB) -- (xB);
\draw [dotted] (xB) -- (yB);
\draw [dotted] (yB) -- (outB);
\draw [fill] (C) circle (.08);
\draw [fill] (A) circle (.08);
\draw [fill] (xB) circle (.08);
\draw [fill] (yB) circle (.08);
\draw [photon] (C) -- (A);
\draw [photon] (xB) -- (C);
\draw [photon] (yB) -- (C);
\end{tikzpicture} + \mathcal{O}(G^3)\,,
\end{align}
where mirror diagrams are left implicit.
Assembling the contributions in \eqref{eq:eikonal2PM} and performing
the integrals one finds
\begin{equation}
\chi=
Gm_1m_2
\left(-\frac{2(2\hat{\gamma}^2-1)}{\sqrt{\hat{\gamma}^2-1}}\log|\hat{b}|+\frac{3\pi}4
\frac{(5\hat{\gamma}^2-1)}{\sqrt{\hat{\gamma}^2-1}}\frac{G(m_1+m_2)}{|\hat{b}|}\right)+\cO(G^3)\,,
\end{equation}
in agreement with \rcite{Bjerrum-Bohr:2018xdl}.
Taking the derivative with respect to $\hat{b}_{1}^\mu$ we obtain the impulse:
\begin{equation}\label{eq:2PMdeflectionB}
\Delta p_1^\mu=
\frac{Gm_1m_2\hat{b}^\mu}{|\hat{b}^2|}
\left(\frac{2(2\hat{\gamma}^2-1)}{\sqrt{\hat{\gamma}^2-1}}+\frac{3\pi}4
\frac{(5\hat{\gamma}^2-1)}{\sqrt{\hat{\gamma}^2-1}}\frac{G(m_1+m_2)}{|\hat{b}|}\right)
+\cO(G^3)\,.
\end{equation}
As it depends on $\hat{b}_{i}^\mu$ and $\hat{v}_{i}^\mu$,
this expression is different from \eqn{eq:2PMdeflection} derived earlier
(which depended on $b_{i}^\mu$ and $v_{i}^\mu$).
It satisfies $\hat{p}_1\cdot\Delta p_1=0$ as expected.

One may naturally ask whether,
having obtained the conservative deflection in \eqn{eq:2PMdeflectionB}
in terms of $\hat{b}^\mu$ and $\hat{v}_{i}^\mu$,
one could subsequently extract the (arguably more useful)
result in \eqn{eq:2PMdeflection} involving
involving $b^\mu$ and $v_{i}^\mu$.
At this PM order we make certain simplifying assumptions:
$\hat{v}^\mu_{i}=v^\mu_{i}+\cO(G)$ implies $\hat{\gamma}=\gamma+\cO(G^2)$.
Also, $\hat{b}^\mu=b^\mu+x_1v_1^\mu+x_2v_2^\mu$ where $x_i\sim\cO(G)$
and $b\cdot v_i=0$ implies $|\hat{b}|=|b|+\cO(G^2)$.
So there is no need to distinguish between different
versions of $|b|$ and $\gamma$ at 2PM.
To reproduce the result in \eqn{eq:2PMdeflection} we simply need to find
the terms in the $v_{i}^\mu$ directions:
it was demonstrated in \rcite{Kalin:2020fhe} that
by demanding $p_1^2=(p_1+\Delta p_1)^2$ (to all orders in $G$)
the missing terms are reproducible by iteration from lower orders.

From the deflection $\Delta p_1^\mu$
we can also find the scattering angle $\theta$
(see \eg \rcites{Bjerrum-Bohr:2018xdl,Parra-Martinez:2020dzs,Vines:2018gqi}).
In the center-of-mass (COM) frame $p_1=(E_1,\mathbf{p})$,
$p_2=(E_2,-\mathbf{p})$ and
\begin{equation}
|\Delta\mathbf{p}|=2|\mathbf{p}|\sin\!\left(\frac\theta2\right)\,.
\end{equation}
In the COM frame one can also deduce that
\begin{equation}
|\mathbf{p}|=\frac{m_1m_2\sqrt{\gamma^2-1}}{\sqrt{m_1^2+m_2^2+2m_1m_2\gamma}}\,.
\end{equation}
The total angular momentum is given by
$J=|\mathbf{b}\times\mathbf{p}|=|\mathbf{b}||\mathbf{p}|$.
Putting the pieces together, we find that
\begin{equation}
\sin\!\left(\frac\theta2\right)=
\frac{Gm_1m_2}{J}
\left(\frac{(2\gamma^2-1)}{\sqrt{\gamma^2-1}}+\frac{3\pi}8
\frac{Gm_1m_2(m_1+m_2)(5\gamma^2-1)}{J\sqrt{m_1^2+m_2^2+2m_1m_2\gamma}}\right)
+\cO(G^3)
\end{equation}
fully describes the scattering angle to order 2PM.

\section{Discussion}\label{sec:discussion}

In this paper we have examined the link between scattering amplitudes
and observables in a worldline quantum field theory (WQFT).
The link is manifested by a worldline path integral representation
of the graviton-dressed scalar propagator,
which can be inserted into a formal definition of the S-matrix
in terms of time-ordered correlators ---
formally integrating out the scalars on external lines.
By taking the classical $\hbar\to0$ limit we can interpret the results as
expectation values of operators in the WQFT.
Performing LSZ reduction on the time-ordered correlators,
\ie cutting the propagators on their external lines,
corresponds with allowing the worldlines of each black hole
to span an infinite proper time domain $\tau\in[-\infty,\infty]$.
We also derived the crucial relationship in the classical $\hbar\to0$ limit:
\begin{equation}
\mathcal{Z}_{\text{WQFT}}= e^{i\chi}\,,
\end{equation}
\ie the free energy of the WQFT corresponds precisely with
the eikonal phase of a $2\to2$ scalar S-matrix.

Path integrals in the WQFT involve not only the graviton $h_{\mu\nu}(x)$
but also the deflection of each black hole $z_i^\mu(\tau_i)$,
where $x_i^\mu(\tau_i)=b_i^\mu+\tau_i v_i^\mu+z_i^\mu(\tau_i)$
is the full unbound trajectory.
We therefore developed a set of Feynman rules to compute expectation values
of WQFT operators directly in Fourier space.
For the graviton $h_{\mu\nu}(k)$ this of course means momentum space;
for the deflection $z_i^\mu(\omega)$ it instead means \emph{energy space}.
Feynman vertices arising from the purely-gravitational Einstein-Hilbert action
conserve four-momentum as usual;
vertices arising from a worldline conserve only the energy along that worldline.
So even though in these classical calculations we remain at tree level,
we see ``loop integrals'' arising due to
the lack of momentum conservation at worldline vertices.
These are precisely the integrals one would obtain when working to higher orders
in $G$ from an amplitudes perspective.

Of particular significance was our choice of $i\eps$
prescription for the propagators, being either retarded or time-symmetric.
For the worldline, using retarded propagators identifies the background parameters
$b_i^\mu$ and $v_i^\mu$ with the incoming momenta $p_i^\mu=m_iv_i^\mu$
at $\tau_i=-\infty$;
time-symmetric propagators identify $\hat{b}_i^\mu$ and $\hat{v}_i^\mu$
with the intermediate momenta $\hat{p}_i^\mu$ at $\tau_i=0$.
For the gravitons, using Feynman propagators implies a time-reversal symmetric dynamics,
hence we are dealing with a purely conservative scattering scenario;
retarded propagators are applicable to a radiating system
and incorporate radiation-reaction effects.

For the scattering of two non-spinning black holes
we considered two 2PM observables:
the radiation, $k^2\left.\braket{h_{\mu\nu}(k)}\right|_{k^2=0}$,
and the impulse on one of the black holes,
$\Delta p_i^\mu=\left.-m_i\omega^2z_i^\mu(\omega)\right|_{\omega=0}$,
the latter reproducing recent results by K\"{a}lin and Porto~\cite{Kalin:2020mvi}.
We also computed the 3PM three-body radiation,
from which 3PM two-body radiation should be obtainable as a special case.
We drew tree-level graphs with a single outgoing line ---
either an outgoing graviton line or
an outgoing deflection mode $z^\mu$ ---
and cut the energy/momentum on that line.

The connection with amplitudes is straightforward for radiation.
As was observed in \rcite{Luna:2017dtq},
at 2PM order $k^2\left.\braket{h_{\mu\nu}(k)}\right|_{k^2=0}$ is
straightforwardly obtained from a five-point amplitude with two pairs
of distinctly flavored external scalars by integrating
over internal momenta (with an appropriate integral measure).
This formula we derived by integrating out the scalars,
leaving the emitted graviton $h_{\mu\nu}(k)$ unaffected.
For the deflection $\Delta p_i^\mu$ there is no clear amplitudes analog;
however, from the eikonal phase $\chi$ we derived $\Delta p_i^\mu$ up to 2PM
by differentiating with respect to the impact parameter $b_i^\mu$ ---
a relationship which we showed extends to higher PM orders.

Setting aside the link to amplitudes,
the WQFT offers a number of advantages
over other methods for obtaining post-Minkowskian (PM) integrands:
\begin{enumerate}
	\item One has the benefits of a fully diagrammatic approach without needing to take the classical $\hbar\to0$ limit.
	\item Generating tree-level graphs is simple to achieve algorithmically --- for instance using Berends-Giele recursion~\cite{Berends:1987me}.
	\item There is no need to obtain an effective action by integrating out the gravitons.
	\item The $i\eps$ prescriptions are flexible: with different choices corresponding to retarded or time-symmetric worldline propagators one can identify the background parameters $b_i^\mu$ and $v_i^\mu$ with either the incoming or intermediate momenta. Similarly, for the graviton one can incorporate radiation by using retarded propagators.
\end{enumerate}
Our approach complements ongoing work in the PM regime on tackling the integrals
required to compute different gravitational observables
(see \eg \rcite{Parra-Martinez:2020dzs}).

There are numerous opportunities for follow-up work.
Of course, we would like to compute observables to higher PM orders:
the eikonal phase $\chi$ and deflection $\Delta p_i^\mu$
at order 4PM are obvious targets.
We also believe that spin can be incorporated in a natural way,
by including classical spin vectors $S_i^\mu$ for each black hole
with their own propagators and worldline vertices. In fact, the WQFTs of supersymmetric spinning particles already exist \cite{Bastianelli:2019xhi,Bonezzi:2020jjq}.
It will be interesting to see how this relates to ongoing work on amplitudes in
higher-spin field theories (see \eg \rcite{Bern:2020buy}).
Finally, as the link to amplitudes is now readily apparent,
it may be worth revisiting the double copy to see how it translates
into this new formalism --- hopefully clarifying the observed breakdown in \rcite{Plefka:2019hmz}.

\begin{acknowledgments}
The authors would like to thank Roberto Bonezzi, Gregor K\"alin, Rafael Porto
and Christian Schubert
for interesting and helpful discussions.
GM's research is funded by the Deutsche Forschungsgemeinschaft (DFG, German Research Foundation) --- Projektnummer 417533893/GRK2575 ``Rethinking Quantum Field Theory'', the Swedish Research Council
under grant 621-2014-5722,
the Knut and Alice Wallenberg Foundation under grant KAW 2018.0116,
and the Ragnar S\"{o}derberg Foundation (Swedish Foundations' Starting Grant).
\end{acknowledgments}

\appendix

\section{Derivation of the momentum space propagator}
\label{app:propDeriv}

In this Appendix we further elaborate on the derivation of a momentum space
representation of the massive scalar propagator coupled to gravitons,
\eqn{eq:Fromaggio}.
The basic ingredient for us to evaluate is
\be
\label{thisone}
\Bigl \langle{\prod_{l=1}^{N}\int_{0}^s\!\d \tau_{l}e^{ik_{l}\cdot(x+\dx\, \frac{\tau_{l}}{s})} \epsilon^{(l)\, \mu\nu}
\tilde D^{(l)}_{\mu\nu}(x,x',\{ k_{l}\})} \Bigr \rangle\,,
\ee
with
\be
\label{eq:Dtildemn}
\tilde D^{(l)}_{\mu\nu}(x,x',\{ k_{l}\})= \Bigl [\Bigl (\frac{\dx^{\mu}}{s}+{\dot q}^{\mu}(\tau_{l})\Bigr )\Bigl
(\frac{\dx^{\nu}}{s}+{\dot q}^{\nu}(\tau_{l})\Bigr ) + \ga^{\mu}(\tau_{l})\ga^{\nu}(\tau_{l})+ \gb^{\mu}(\tau_{l})\gc^{\nu}(\tau_{l}) \Bigr ] e^{ik_{l}\cdot q(\tau_{l})} \, .
\ee
We also note that
\be
Z= \int D[q]  e^{-i\int_{0}^s\!\d \tau \frac{1}{4}{\dot q}^{2}} = \frac{1}{(-4\pi i s)^{D/2}}\,,
\ee
which is the free partition function.

To evaluate the correlation function we use a number of tricks.
Firstly, we introduce a scalar function $F$
of the polarization vectors $\eps$, $\alpha$, $\beta$, $\gamma$,
the latter two of which are anti-commuting:
\be
F(\epsilon, \alpha, \beta, \gamma) :=
\Vev{\exp\Bigl [  \sum_{l=1}^{N} \epsilon_{l}\cdot \dot q (\tau_{l}) +  \alpha_{l}\cdot \ga (\tau_{l}) +
\beta_{l}\cdot \gb (\tau_{l}) +\gamma_{l}\cdot \gc (\tau_{l}) + ik_{l}\cdot q (\tau_{l}) \Bigr ] }\, .
\ee
Then we may write
\be
\eqref{thisone}=
\prod_{l=1}^{N}\int_{0}^s\!\d\tau_{l}
\Bigl [ ( \sfrac{\dx_{\mu}}{s}+\partial_{\epsilon_{l}^{\mu}})   ( \sfrac{\dx_{\nu}}{s}+\partial_{\epsilon_{l}^{\nu}})   
+ \partial_{\alpha_{l}^{\mu}}\partial_{\alpha_{l}^{\nu}}+ \partial_{\beta_{l}^{\mu}}
\partial_{\gamma_{\alpha_{l}^{\nu}}}
\Bigr ]\, F(\epsilon, \alpha, \beta, \gamma) \Bigr | _{\alpha=\beta=\gamma=\epsilon=0} \,.\nn
\ee
To compute $F(\epsilon, \alpha, \beta, \gamma) $ we use the fact that for operators $\mathcal {O}$ linear in quantum fields one has the free-field correlation function relation
\be
\label{eq:Fevaled}
\vev{e^{\mathcal{O}}} = \frac{1}{(-4\pi si)^{D/2}}\, \exp [\sfrac{1}{2} \vev{\mathcal O \mathcal O} ]\, .
\ee
Hence we find
\be
F(\epsilon, \alpha, \beta, \gamma) = \frac{1}{(-4\pi si)^{D/2}}\, \exp [\sfrac{1}{2} \vev{\sum_{l,l'=1}^{N}\mathcal O_{l} \mathcal O_{l'}} ]\, ,
\ee
with $\mathcal O_{l}= \epsilon_{l}\cdot \dot q (\tau_{l}) +  \alpha_{l}\cdot \ga (\tau_{l}) +
\beta_{l}\cdot \gb (\tau_{l}) +\gamma_{l}\cdot \gc (\tau_{l}) + ik_{l}\cdot q (\tau_{l}) $.
We then compute using \eqn{eq:DeltaProps}
\begin{align}
\begin{aligned}
i\vev{\mathcal{O}_{l}\mathcal{O}_{l'}}= & 2\delta(\tau_{l}-\tau_{l'})\, [\epsilon_{l}\cdot \epsilon_{l'} + \alpha_{l}\cdot \alpha_{l'}
-2 \gamma_{l}\cdot \beta_{l'}- 2\gamma_{l'}\cdot \beta_{l}]  \\
& -\frac{2}{s} [  ik_{l}\tau_{l} + \epsilon_{l}]\cdot  [ ik_{l'}\tau_{l'} + \epsilon_{l'}]
+ ik_{l}\cdot  [  ik_{l'}\tau_{l'} + \epsilon_{l'}]+ ik_{l'}\cdot  [  ik_{l}\tau_{l} + \epsilon_{l}] \\
& -i (\epsilon_{l}\cdot k_{l'}-\epsilon_{l'}\cdot k_{l})\, \text{sign}(\tau_{l}-\tau_{l'}) + k_{l}\cdot k_{l'}|\tau_{l}-\tau_{l'}|
\, .
\end{aligned}
\end{align}
For this it is helpful to note the derivatives of the
propagator given in \eqn{eq:DeltaProps}:
\begin{align}
\begin{aligned}
\partial_{\tau}\Delta(\tau,\tau') &= 
	\sfrac{1}{2}\text{sign}(\tau-\tau') + \frac{\tau'}{s} -\frac{1}{2}\,, \\
\partial_{\tau'}\Delta(\tau,\tau') &= 
	-\sfrac{1}{2}\text{sign}(\tau-\tau') + \frac{\tau}{s} -\frac{1}{2}\,, \\
\partial_{\tau}\partial_{\tau'}\Delta(\tau,\tau') &= 
	\frac{1}{s} - \delta(\tau-\tau')\,,  \\
\partial_{\tau}^{2}\Delta(\tau,\tau') &= \delta(\tau-\tau') \, .
\end{aligned}
\end{align}
The second trick lies in promoting also the $\dx^{\mu}/s$ terms in \eqn{eq:Dtildemn} into the exponent of the evaluated $F(\epsilon, \alpha, \beta, \gamma)$
by manually adding $\sum_{l=1}^{N}\epsilon_{l}\cdot \frac{\Delta x}{s}$
to the exponent on the right-hand side of \eqn{eq:Fevaled}.
Then we perform the space-time integrals over $x$ and $x'$,
giving a total momentum-conserving delta function and a Gaussian integral.

\bibliographystyle{JHEP}
\bibliography{paper}

\providecommand{\href}[2]{#2}\begingroup\raggedright\begin{thebibliography}{100}

\bibitem{Chandrasekhar:1979qb}
S.~Chandrasekhar, \emph{{Beauty and the Quest for Beauty in Scinece}},  in
  \emph{{International Symposium in Honor of Robert R. Wilson}}, 1, 1979.

\bibitem{Dixon:2011xs}
L.~J. Dixon, \emph{{Scattering amplitudes: the most perfect microscopic
  structures in the universe}},
  \href{https://doi.org/10.1088/1751-8113/44/45/454001}{\emph{J. Phys. A}
  {\bfseries 44} (2011) 454001}
  [\href{https://arxiv.org/abs/1105.0771}{{\ttfamily 1105.0771}}].

\bibitem{Abbott:2016blz}
{\scshape LIGO Scientific, Virgo} collaboration, B.~Abbott et~al.,
  \emph{{Observation of Gravitational Waves from a Binary Black Hole Merger}},
  \href{https://doi.org/10.1103/PhysRevLett.116.061102}{\emph{Phys. Rev. Lett.}
  {\bfseries 116} (2016) 061102}
  [\href{https://arxiv.org/abs/1602.03837}{{\ttfamily 1602.03837}}].

\bibitem{TheLIGOScientific:2017qsa}
{\scshape LIGO Scientific, Virgo} collaboration, B.~Abbott et~al.,
  \emph{{GW170817: Observation of Gravitational Waves from a Binary Neutron
  Star Inspiral}},
  \href{https://doi.org/10.1103/PhysRevLett.119.161101}{\emph{Phys. Rev. Lett.}
  {\bfseries 119} (2017) 161101}
  [\href{https://arxiv.org/abs/1710.05832}{{\ttfamily 1710.05832}}].

\bibitem{LIGOScientific:2018mvr}
{\scshape LIGO Scientific, Virgo} collaboration, B.~Abbott et~al.,
  \emph{{GWTC-1: A Gravitational-Wave Transient Catalog of Compact Binary
  Mergers Observed by LIGO and Virgo during the First and Second Observing
  Runs}}, \href{https://doi.org/10.1103/PhysRevX.9.031040}{\emph{Phys. Rev. X}
  {\bfseries 9} (2019) 031040}
  [\href{https://arxiv.org/abs/1811.12907}{{\ttfamily 1811.12907}}].

\bibitem{Purrer:2019jcp}
M.~P\"urrer and C.-J. Haster, \emph{{Gravitational waveform accuracy
  requirements for future ground-based detectors}},
  \href{https://doi.org/10.1103/PhysRevResearch.2.023151}{\emph{Phys. Rev.
  Res.} {\bfseries 2} (2020) 023151}
  [\href{https://arxiv.org/abs/1912.10055}{{\ttfamily 1912.10055}}].

\bibitem{Blanchet:2013haa}
L.~Blanchet, \emph{{Gravitational Radiation from Post-Newtonian Sources and
  Inspiralling Compact Binaries}},
  \href{https://doi.org/10.12942/lrr-2014-2}{\emph{Living Rev. Rel.} {\bfseries
  17} (2014) 2} [\href{https://arxiv.org/abs/1310.1528}{{\ttfamily
  1310.1528}}].

\bibitem{Schafer:2018kuf}
G.~Sch\"afer and P.~Jaranowski, \emph{{Hamiltonian formulation of general
  relativity and post-Newtonian dynamics of compact binaries}},
  \href{https://doi.org/10.1007/s41114-018-0016-5}{\emph{Living Rev. Rel.}
  {\bfseries 21} (2018) 7} [\href{https://arxiv.org/abs/1805.07240}{{\ttfamily
  1805.07240}}].

\bibitem{Futamase:2007zz}
T.~Futamase and Y.~Itoh, \emph{{The post-Newtonian approximation for
  relativistic compact binaries}},
  \href{https://doi.org/10.12942/lrr-2007-2}{\emph{Living Rev. Rel.} {\bfseries
  10} (2007) 2}.

\bibitem{Pati:2000vt}
M.~E. Pati and C.~M. Will, \emph{{PostNewtonian gravitational radiation and
  equations of motion via direct integration of the relaxed Einstein equations.
  1. Foundations}},
  \href{https://doi.org/10.1103/PhysRevD.62.124015}{\emph{Phys. Rev. D}
  {\bfseries 62} (2000) 124015}
  [\href{https://arxiv.org/abs/gr-qc/0007087}{{\ttfamily gr-qc/0007087}}].

\bibitem{Bel:1981be}
L.~Bel, T.~Damour, N.~Deruelle, J.~Ibanez and J.~Martin,
  \emph{{Poincar\'e-invariant gravitational field and equations of motion of
  two pointlike objects: The postlinear approximation of general relativity}},
  \href{https://doi.org/10.1007/BF00756073}{\emph{Gen. Rel. Grav.} {\bfseries
  13} (1981) 963}.

\bibitem{Westpfahl:1985tsl}
K.~Westpfahl, \emph{{High-Speed Scattering of Charged and Uncharged Particles
  in General Relativity}},
  \href{https://doi.org/10.1002/prop.2190330802}{\emph{Fortsch. Phys.}
  {\bfseries 33} (1985) 417}.

\bibitem{Goldberger:2004jt}
W.~D. Goldberger and I.~Z. Rothstein, \emph{{An Effective field theory of
  gravity for extended objects}},
  \href{https://doi.org/10.1103/PhysRevD.73.104029}{\emph{Phys. Rev. D}
  {\bfseries 73} (2006) 104029}
  [\href{https://arxiv.org/abs/hep-th/0409156}{{\ttfamily hep-th/0409156}}].

\bibitem{Goldberger:2006bd}
W.~D. Goldberger and I.~Z. Rothstein, \emph{{Towers of Gravitational
  Theories}}, \href{https://doi.org/10.1142/S0218271806009698}{\emph{Gen. Rel.
  Grav.} {\bfseries 38} (2006) 1537}
  [\href{https://arxiv.org/abs/hep-th/0605238}{{\ttfamily hep-th/0605238}}].

\bibitem{Goldberger:2009qd}
W.~D. Goldberger and A.~Ross, \emph{{Gravitational radiative corrections from
  effective field theory}},
  \href{https://doi.org/10.1103/PhysRevD.81.124015}{\emph{Phys. Rev. D}
  {\bfseries 81} (2010) 124015}
  [\href{https://arxiv.org/abs/0912.4254}{{\ttfamily 0912.4254}}].

\bibitem{Kol:2007bc}
B.~Kol and M.~Smolkin, \emph{{Non-Relativistic Gravitation: From Newton to
  Einstein and Back}},
  \href{https://doi.org/10.1088/0264-9381/25/14/145011}{\emph{Class. Quant.
  Grav.} {\bfseries 25} (2008) 145011}
  [\href{https://arxiv.org/abs/0712.4116}{{\ttfamily 0712.4116}}].

\bibitem{Goldberger:2007hy}
W.~D. Goldberger, \emph{{Les Houches lectures on effective field theories and
  gravitational radiation}},  in \emph{{Les Houches Summer School - Session 86:
  Particle Physics and Cosmology: The Fabric of Spacetime}}, 1, 2007,
  \href{https://arxiv.org/abs/hep-ph/0701129}{{\ttfamily hep-ph/0701129}}.

\bibitem{Foffa:2013qca}
S.~Foffa and R.~Sturani, \emph{{Effective field theory methods to model compact
  binaries}}, \href{https://doi.org/10.1088/0264-9381/31/4/043001}{\emph{Class.
  Quant. Grav.} {\bfseries 31} (2014) 043001}
  [\href{https://arxiv.org/abs/1309.3474}{{\ttfamily 1309.3474}}].

\bibitem{Rothstein:2014sra}
I.~Z. Rothstein, \emph{{Progress in effective field theory approach to the
  binary inspiral problem}},
  \href{https://doi.org/10.1007/s10714-014-1726-y}{\emph{Gen. Rel. Grav.}
  {\bfseries 46} (2014) 1726}.

\bibitem{Porto:2016pyg}
R.~A. Porto, \emph{{The effective field theorist\textquoteright{}s approach to
  gravitational dynamics}},
  \href{https://doi.org/10.1016/j.physrep.2016.04.003}{\emph{Phys. Rept.}
  {\bfseries 633} (2016) 1} [\href{https://arxiv.org/abs/1601.04914}{{\ttfamily
  1601.04914}}].

\bibitem{Levi:2018nxp}
M.~Levi, \emph{{Effective Field Theories of Post-Newtonian Gravity: A
  comprehensive review}},
  \href{https://doi.org/10.1088/1361-6633/ab12bc}{\emph{Rept. Prog. Phys.}
  {\bfseries 83} (2020) 075901}
  [\href{https://arxiv.org/abs/1807.01699}{{\ttfamily 1807.01699}}].

\bibitem{Damour:2014jta}
T.~Damour, P.~Jaranowski and G.~Sch\"afer, \emph{{Nonlocal-in-time action for
  the fourth post-Newtonian conservative dynamics of two-body systems}},
  \href{https://doi.org/10.1103/PhysRevD.89.064058}{\emph{Phys. Rev. D}
  {\bfseries 89} (2014) 064058}
  [\href{https://arxiv.org/abs/1401.4548}{{\ttfamily 1401.4548}}].

\bibitem{Damour:2016abl}
T.~Damour, P.~Jaranowski and G.~Sch\"afer, \emph{{Conservative dynamics of
  two-body systems at the fourth post-Newtonian approximation of general
  relativity}}, \href{https://doi.org/10.1103/PhysRevD.93.084014}{\emph{Phys.
  Rev. D} {\bfseries 93} (2016) 084014}
  [\href{https://arxiv.org/abs/1601.01283}{{\ttfamily 1601.01283}}].

\bibitem{Bernard:2016wrg}
L.~Bernard, L.~Blanchet, A.~Boh\'e, G.~Faye and S.~Marsat, \emph{{Energy and
  periastron advance of compact binaries on circular orbits at the fourth
  post-Newtonian order}},
  \href{https://doi.org/10.1103/PhysRevD.95.044026}{\emph{Phys. Rev. D}
  {\bfseries 95} (2017) 044026}
  [\href{https://arxiv.org/abs/1610.07934}{{\ttfamily 1610.07934}}].

\bibitem{Foffa:2016rgu}
S.~Foffa, P.~Mastrolia, R.~Sturani and C.~Sturm, \emph{{Effective field theory
  approach to the gravitational two-body dynamics, at fourth post-Newtonian
  order and quintic in the Newton constant}},
  \href{https://doi.org/10.1103/PhysRevD.95.104009}{\emph{Phys. Rev. D}
  {\bfseries 95} (2017) 104009}
  [\href{https://arxiv.org/abs/1612.00482}{{\ttfamily 1612.00482}}].

\bibitem{Damour:2017ced}
T.~Damour and P.~Jaranowski, \emph{{Four-loop static contribution to the
  gravitational interaction potential of two point masses}},
  \href{https://doi.org/10.1103/PhysRevD.95.084005}{\emph{Phys. Rev. D}
  {\bfseries 95} (2017) 084005}
  [\href{https://arxiv.org/abs/1701.02645}{{\ttfamily 1701.02645}}].

\bibitem{Foffa:2019rdf}
S.~Foffa and R.~Sturani, \emph{{Conservative dynamics of binary systems to
  fourth Post-Newtonian order in the EFT approach I: Regularized Lagrangian}},
  \href{https://doi.org/10.1103/PhysRevD.100.024047}{\emph{Phys. Rev. D}
  {\bfseries 100} (2019) 024047}
  [\href{https://arxiv.org/abs/1903.05113}{{\ttfamily 1903.05113}}].

\bibitem{Foffa:2019yfl}
S.~Foffa, R.~A. Porto, I.~Rothstein and R.~Sturani, \emph{{Conservative
  dynamics of binary systems to fourth Post-Newtonian order in the EFT approach
  II: Renormalized Lagrangian}},
  \href{https://doi.org/10.1103/PhysRevD.100.024048}{\emph{Phys. Rev. D}
  {\bfseries 100} (2019) 024048}
  [\href{https://arxiv.org/abs/1903.05118}{{\ttfamily 1903.05118}}].

\bibitem{Blumlein:2020pog}
J.~Bl\"umlein, A.~Maier, P.~Marquard and G.~Sch\"afer, \emph{{Fourth
  post-Newtonian Hamiltonian dynamics of two-body systems from an effective
  field theory approach}},
  \href{https://doi.org/10.1016/j.nuclphysb.2020.115041}{\emph{Nucl. Phys. B}
  {\bfseries 955} (2020) 115041}
  [\href{https://arxiv.org/abs/2003.01692}{{\ttfamily 2003.01692}}].

\bibitem{Porto:2017dgs}
R.~A. Porto and I.~Z. Rothstein, \emph{{Apparent ambiguities in the
  post-Newtonian expansion for binary systems}},
  \href{https://doi.org/10.1103/PhysRevD.96.024062}{\emph{Phys. Rev. D}
  {\bfseries 96} (2017) 024062}
  [\href{https://arxiv.org/abs/1703.06433}{{\ttfamily 1703.06433}}].

\bibitem{Marchand:2017pir}
T.~Marchand, L.~Bernard, L.~Blanchet and G.~Faye, \emph{{Ambiguity-Free
  Completion of the Equations of Motion of Compact Binary Systems at the Fourth
  Post-Newtonian Order}},
  \href{https://doi.org/10.1103/PhysRevD.97.044023}{\emph{Phys. Rev. D}
  {\bfseries 97} (2018) 044023}
  [\href{https://arxiv.org/abs/1707.09289}{{\ttfamily 1707.09289}}].

\bibitem{Galley:2015kus}
C.~R. Galley, A.~K. Leibovich, R.~A. Porto and A.~Ross, \emph{{Tail effect in
  gravitational radiation reaction: Time nonlocality and renormalization group
  evolution}}, \href{https://doi.org/10.1103/PhysRevD.93.124010}{\emph{Phys.
  Rev. D} {\bfseries 93} (2016) 124010}
  [\href{https://arxiv.org/abs/1511.07379}{{\ttfamily 1511.07379}}].

\bibitem{Foffa:2019hrb}
S.~Foffa, P.~Mastrolia, R.~Sturani, C.~Sturm and W.~J. Torres~Bobadilla,
  \emph{{Static two-body potential at fifth post-Newtonian order}},
  \href{https://doi.org/10.1103/PhysRevLett.122.241605}{\emph{Phys. Rev. Lett.}
  {\bfseries 122} (2019) 241605}
  [\href{https://arxiv.org/abs/1902.10571}{{\ttfamily 1902.10571}}].

\bibitem{Blumlein:2019zku}
J.~Bl\"umlein, A.~Maier and P.~Marquard, \emph{{Five-Loop Static Contribution
  to the Gravitational Interaction Potential of Two Point Masses}},
  \href{https://doi.org/10.1016/j.physletb.2019.135100}{\emph{Phys. Lett. B}
  {\bfseries 800} (2020) 135100}
  [\href{https://arxiv.org/abs/1902.11180}{{\ttfamily 1902.11180}}].

\bibitem{Bini:2019nra}
D.~Bini, T.~Damour and A.~Geralico, \emph{{Novel approach to binary dynamics:
  application to the fifth post-Newtonian level}},
  \href{https://doi.org/10.1103/PhysRevLett.123.231104}{\emph{Phys. Rev. Lett.}
  {\bfseries 123} (2019) 231104}
  [\href{https://arxiv.org/abs/1909.02375}{{\ttfamily 1909.02375}}].

\bibitem{Blumlein:2020pyo}
J.~Bl\"umlein, A.~Maier, P.~Marquard and G.~Sch\"afer, \emph{{The fifth-order
  post-Newtonian Hamiltonian dynamics of two-body systems from an effective
  field theory approach: potential contributions}},
  \href{https://arxiv.org/abs/2010.13672}{{\ttfamily 2010.13672}}.

\bibitem{Blumlein:2020znm}
J.~Bl\"umlein, A.~Maier, P.~Marquard and G.~Sch\"afer, \emph{{Testing binary
  dynamics in gravity at the sixth post-Newtonian level}},
  \href{https://doi.org/10.1016/j.physletb.2020.135496}{\emph{Phys. Lett. B}
  {\bfseries 807} (2020) 135496}
  [\href{https://arxiv.org/abs/2003.07145}{{\ttfamily 2003.07145}}].

\bibitem{Cheung:2020gyp}
C.~Cheung and M.~P. Solon, \emph{{Classical gravitational scattering at $
  \mathcal{O} $(G$^{3}$) from Feynman diagrams}},
  \href{https://doi.org/10.1007/JHEP06(2020)144}{\emph{JHEP} {\bfseries 06}
  (2020) 144} [\href{https://arxiv.org/abs/2003.08351}{{\ttfamily
  2003.08351}}].

\bibitem{Bini:2020nsb}
D.~Bini, T.~Damour and A.~Geralico, \emph{{Sixth post-Newtonian local-in-time
  dynamics of binary systems}},
  \href{https://doi.org/10.1103/PhysRevD.102.024061}{\emph{Phys. Rev. D}
  {\bfseries 102} (2020) 024061}
  [\href{https://arxiv.org/abs/2004.05407}{{\ttfamily 2004.05407}}].

\bibitem{Bini:2020wpo}
D.~Bini, T.~Damour and A.~Geralico, \emph{{Binary dynamics at the fifth and
  fifth-and-a-half post-Newtonian orders}},
  \href{https://doi.org/10.1103/PhysRevD.102.024062}{\emph{Phys. Rev. D}
  {\bfseries 102} (2020) 024062}
  [\href{https://arxiv.org/abs/2003.11891}{{\ttfamily 2003.11891}}].

\bibitem{Bini:2020hmy}
D.~Bini, T.~Damour and A.~Geralico, \emph{{Sixth post-Newtonian
  nonlocal-in-time dynamics of binary systems}},
  \href{https://doi.org/10.1103/PhysRevD.102.084047}{\emph{Phys. Rev. D}
  {\bfseries 102} (2020) 084047}
  [\href{https://arxiv.org/abs/2007.11239}{{\ttfamily 2007.11239}}].

\bibitem{Bini:2020uiq}
D.~Bini, T.~Damour, A.~Geralico, S.~Laporta and P.~Mastrolia,
  \emph{{Gravitational dynamics at $O(G^6)$: perturbative gravitational
  scattering meets experimental mathematics}},
  \href{https://arxiv.org/abs/2008.09389}{{\ttfamily 2008.09389}}.

\bibitem{Blanchet:2001aw}
L.~Blanchet, B.~R. Iyer and B.~Joguet, \emph{{Gravitational waves from
  inspiralling compact binaries: Energy flux to third postNewtonian order}},
  \href{https://doi.org/10.1103/PhysRevD.65.064005}{\emph{Phys. Rev. D}
  {\bfseries 65} (2002) 064005}
  [\href{https://arxiv.org/abs/gr-qc/0105098}{{\ttfamily gr-qc/0105098}}].

\bibitem{Blanchet:2004ek}
L.~Blanchet, T.~Damour, G.~Esposito-Farese and B.~R. Iyer, \emph{{Gravitational
  radiation from inspiralling compact binaries completed at the third
  post-Newtonian order}},
  \href{https://doi.org/10.1103/PhysRevLett.93.091101}{\emph{Phys. Rev. Lett.}
  {\bfseries 93} (2004) 091101}
  [\href{https://arxiv.org/abs/gr-qc/0406012}{{\ttfamily gr-qc/0406012}}].

\bibitem{Blanchet:2008je}
L.~Blanchet, G.~Faye, B.~R. Iyer and S.~Sinha, \emph{{The Third post-Newtonian
  gravitational wave polarisations and associated spherical harmonic modes for
  inspiralling compact binaries in quasi-circular orbits}},
  \href{https://doi.org/10.1088/0264-9381/25/16/165003}{\emph{Class. Quant.
  Grav.} {\bfseries 25} (2008) 165003}
  [\href{https://arxiv.org/abs/0802.1249}{{\ttfamily 0802.1249}}].

\bibitem{Leibovich:2019cxo}
A.~K. Leibovich, N.~T. Maia, I.~Z. Rothstein and Z.~Yang, \emph{{Second
  post-Newtonian order radiative dynamics of inspiralling compact binaries in
  the Effective Field Theory approach}},
  \href{https://doi.org/10.1103/PhysRevD.101.084058}{\emph{Phys. Rev. D}
  {\bfseries 101} (2020) 084058}
  [\href{https://arxiv.org/abs/1912.12546}{{\ttfamily 1912.12546}}].

\bibitem{Porto:2005ac}
R.~A. Porto, \emph{{Post-Newtonian corrections to the motion of spinning bodies
  in NRGR}}, \href{https://doi.org/10.1103/PhysRevD.73.104031}{\emph{Phys. Rev.
  D} {\bfseries 73} (2006) 104031}
  [\href{https://arxiv.org/abs/gr-qc/0511061}{{\ttfamily gr-qc/0511061}}].

\bibitem{Levi:2015msa}
M.~Levi and J.~Steinhoff, \emph{{Spinning gravitating objects in the effective
  field theory in the post-Newtonian scheme}},
  \href{https://doi.org/10.1007/JHEP09(2015)219}{\emph{JHEP} {\bfseries 09}
  (2015) 219} [\href{https://arxiv.org/abs/1501.04956}{{\ttfamily
  1501.04956}}].

\bibitem{Levi:2020kvb}
M.~Levi, A.~J. Mcleod and M.~Von~Hippel, \emph{{N$^3$LO gravitational
  spin-orbit coupling at order $G^4$}},
  \href{https://arxiv.org/abs/2003.02827}{{\ttfamily 2003.02827}}.

\bibitem{Antonelli:2020aeb}
A.~Antonelli, C.~Kavanagh, M.~Khalil, J.~Steinhoff and J.~Vines,
  \emph{{Gravitational spin-orbit coupling through third-subleading
  post-Newtonian order: from first-order self-force to arbitrary mass ratios}},
  \href{https://doi.org/10.1103/PhysRevLett.125.011103}{\emph{Phys. Rev. Lett.}
  {\bfseries 125} (2020) 011103}
  [\href{https://arxiv.org/abs/2003.11391}{{\ttfamily 2003.11391}}].

\bibitem{Levi:2016ofk}
M.~Levi and J.~Steinhoff, \emph{{Complete conservative dynamics for
  inspiralling compact binaries with spins at fourth post-Newtonian order}},
  \href{https://arxiv.org/abs/1607.04252}{{\ttfamily 1607.04252}}.

\bibitem{Porto:2006bt}
R.~A. Porto and I.~Z. Rothstein, \emph{{The Hyperfine Einstein-Infeld-Hoffmann
  potential}}, \href{https://doi.org/10.1103/PhysRevLett.97.021101}{\emph{Phys.
  Rev. Lett.} {\bfseries 97} (2006) 021101}
  [\href{https://arxiv.org/abs/gr-qc/0604099}{{\ttfamily gr-qc/0604099}}].

\bibitem{Porto:2008tb}
R.~A. Porto and I.~Z. Rothstein, \emph{{Spin(1)Spin(2) Effects in the Motion of
  Inspiralling Compact Binaries at Third Order in the Post-Newtonian
  Expansion}}, \href{https://doi.org/10.1103/PhysRevD.78.044012}{\emph{Phys.
  Rev. D} {\bfseries 78} (2008) 044012}
  [\href{https://arxiv.org/abs/0802.0720}{{\ttfamily 0802.0720}}].

\bibitem{Levi:2008nh}
M.~Levi, \emph{{Next to Leading Order gravitational Spin1-Spin2 coupling with
  Kaluza-Klein reduction}},
  \href{https://doi.org/10.1103/PhysRevD.82.064029}{\emph{Phys. Rev. D}
  {\bfseries 82} (2010) 064029}
  [\href{https://arxiv.org/abs/0802.1508}{{\ttfamily 0802.1508}}].

\bibitem{Porto:2008jj}
R.~A. Porto and I.~Z. Rothstein, \emph{{Next to Leading Order Spin(1)Spin(1)
  Effects in the Motion of Inspiralling Compact Binaries}},
  \href{https://doi.org/10.1103/PhysRevD.78.044013}{\emph{Phys. Rev. D}
  {\bfseries 78} (2008) 044013}
  [\href{https://arxiv.org/abs/0804.0260}{{\ttfamily 0804.0260}}].

\bibitem{Porto:2010tr}
R.~A. Porto, \emph{{Next to leading order spin-orbit effects in the motion of
  inspiralling compact binaries}},
  \href{https://doi.org/10.1088/0264-9381/27/20/205001}{\emph{Class. Quant.
  Grav.} {\bfseries 27} (2010) 205001}
  [\href{https://arxiv.org/abs/1005.5730}{{\ttfamily 1005.5730}}].

\bibitem{Perrodin:2010dy}
D.~L. Perrodin, \emph{{Subleading Spin-Orbit Correction to the Newtonian
  Potential in Effective Field Theory Formalism}},  in \emph{{12th Marcel
  Grossmann Meeting on General Relativity}}, pp.~725--727, 5, 2010,
  \href{https://arxiv.org/abs/1005.0634}{{\ttfamily 1005.0634}},
  \href{https://doi.org/10.1142/9789814374552_0041}{DOI}.

\bibitem{Levi:2010zu}
M.~Levi, \emph{{Next to Leading Order gravitational Spin-Orbit coupling in an
  Effective Field Theory approach}},
  \href{https://doi.org/10.1103/PhysRevD.82.104004}{\emph{Phys. Rev. D}
  {\bfseries 82} (2010) 104004}
  [\href{https://arxiv.org/abs/1006.4139}{{\ttfamily 1006.4139}}].

\bibitem{Mishra:2016whh}
C.~K. Mishra, A.~Kela, K.~Arun and G.~Faye, \emph{{Ready-to-use post-Newtonian
  gravitational waveforms for binary black holes with nonprecessing spins: An
  update}}, \href{https://doi.org/10.1103/PhysRevD.93.084054}{\emph{Phys. Rev.
  D} {\bfseries 93} (2016) 084054}
  [\href{https://arxiv.org/abs/1601.05588}{{\ttfamily 1601.05588}}].

\bibitem{Buonanno:2012rv}
A.~Buonanno, G.~Faye and T.~Hinderer, \emph{{Spin effects on gravitational
  waves from inspiraling compact binaries at second post-Newtonian order}},
  \href{https://doi.org/10.1103/PhysRevD.87.044009}{\emph{Phys. Rev. D}
  {\bfseries 87} (2013) 044009}
  [\href{https://arxiv.org/abs/1209.6349}{{\ttfamily 1209.6349}}].

\bibitem{Porto:2010zg}
R.~A. Porto, A.~Ross and I.~Z. Rothstein, \emph{{Spin induced multipole moments
  for the gravitational wave flux from binary inspirals to third Post-Newtonian
  order}}, \href{https://doi.org/10.1088/1475-7516/2011/03/009}{\emph{JCAP}
  {\bfseries 03} (2011) 009} [\href{https://arxiv.org/abs/1007.1312}{{\ttfamily
  1007.1312}}].

\bibitem{Porto:2012as}
R.~A. Porto, A.~Ross and I.~Z. Rothstein, \emph{{Spin induced multipole moments
  for the gravitational wave amplitude from binary inspirals to 2.5
  Post-Newtonian order}},
  \href{https://doi.org/10.1088/1475-7516/2012/09/028}{\emph{JCAP} {\bfseries
  09} (2012) 028} [\href{https://arxiv.org/abs/1203.2962}{{\ttfamily
  1203.2962}}].

\bibitem{Maia:2017gxn}
N.~T. Maia, C.~R. Galley, A.~K. Leibovich and R.~A. Porto, \emph{{Radiation
  reaction for spinning bodies in effective field theory I: Spin-orbit
  effects}}, \href{https://doi.org/10.1103/PhysRevD.96.084064}{\emph{Phys. Rev.
  D} {\bfseries 96} (2017) 084064}
  [\href{https://arxiv.org/abs/1705.07934}{{\ttfamily 1705.07934}}].

\bibitem{Maia:2017yok}
N.~T. Maia, C.~R. Galley, A.~K. Leibovich and R.~A. Porto, \emph{{Radiation
  reaction for spinning bodies in effective field theory II: Spin-spin
  effects}}, \href{https://doi.org/10.1103/PhysRevD.96.084065}{\emph{Phys. Rev.
  D} {\bfseries 96} (2017) 084065}
  [\href{https://arxiv.org/abs/1705.07938}{{\ttfamily 1705.07938}}].

\bibitem{Kalin:2020mvi}
G.~K\"alin and R.~A. Porto, \emph{{Post-Minkowskian Effective Field Theory for
  Conservative Binary Dynamics}},
  \href{https://doi.org/10.1007/JHEP11(2020)106}{\emph{JHEP} {\bfseries 11}
  (2020) 106} [\href{https://arxiv.org/abs/2006.01184}{{\ttfamily
  2006.01184}}].

\bibitem{Bini:2020flp}
D.~Bini, T.~Damour and A.~Geralico, \emph{{Scattering of tidally interacting
  bodies in post-Minkowskian gravity}},
  \href{https://doi.org/10.1103/PhysRevD.101.044039}{\emph{Phys. Rev. D}
  {\bfseries 101} (2020) 044039}
  [\href{https://arxiv.org/abs/2001.00352}{{\ttfamily 2001.00352}}].

\bibitem{Cheung:2020sdj}
C.~Cheung and M.~P. Solon, \emph{{Tidal Effects in the Post-Minkowskian
  Expansion}},
  \href{https://doi.org/10.1103/PhysRevLett.125.191601}{\emph{Phys. Rev. Lett.}
  {\bfseries 125} (2020) 191601}
  [\href{https://arxiv.org/abs/2006.06665}{{\ttfamily 2006.06665}}].

\bibitem{Kalin:2020lmz}
G.~K\"alin, Z.~Liu and R.~A. Porto, \emph{{Conservative Tidal Effects in
  Compact Binary Systems to Next-to-Leading Post-Minkowskian Order}},
  \href{https://doi.org/10.1103/PhysRevD.102.124025}{\emph{Phys. Rev. D}
  {\bfseries 102} (2020) 124025}
  [\href{https://arxiv.org/abs/2008.06047}{{\ttfamily 2008.06047}}].

\bibitem{Haddad:2020que}
K.~Haddad and A.~Helset, \emph{{Tidal effects in quantum field theory}},
  \href{https://doi.org/10.1007/JHEP12(2020)024}{\emph{JHEP} {\bfseries 12}
  (2020) 024} [\href{https://arxiv.org/abs/2008.04920}{{\ttfamily
  2008.04920}}].

\bibitem{Kalin:2020fhe}
G.~K\"alin, Z.~Liu and R.~A. Porto, \emph{{Conservative Dynamics of Binary
  Systems to Third Post-Minkowskian Order from the Effective Field Theory
  Approach}}, \href{https://doi.org/10.1103/PhysRevLett.125.261103}{\emph{Phys.
  Rev. Lett.} {\bfseries 125} (2020) 261103}
  [\href{https://arxiv.org/abs/2007.04977}{{\ttfamily 2007.04977}}].

\bibitem{Ledvinka:2008tk}
T.~Ledvinka, G.~Schaefer and J.~Bicak, \emph{{Relativistic Closed-Form
  Hamiltonian for Many-Body Gravitating Systems in the Post-Minkowskian
  Approximation}},
  \href{https://doi.org/10.1103/PhysRevLett.100.251101}{\emph{Phys. Rev. Lett.}
  {\bfseries 100} (2008) 251101}
  [\href{https://arxiv.org/abs/0807.0214}{{\ttfamily 0807.0214}}].

\bibitem{Damour:2016gwp}
T.~Damour, \emph{{Gravitational scattering, post-Minkowskian approximation and
  Effective One-Body theory}},
  \href{https://doi.org/10.1103/PhysRevD.94.104015}{\emph{Phys. Rev. D}
  {\bfseries 94} (2016) 104015}
  [\href{https://arxiv.org/abs/1609.00354}{{\ttfamily 1609.00354}}].

\bibitem{Blanchet:2018yvb}
L.~Blanchet and A.~S. Fokas, \emph{{Equations of motion of self-gravitating
  $N$-body systems in the first post-Minkowskian approximation}},
  \href{https://doi.org/10.1103/PhysRevD.98.084005}{\emph{Phys. Rev. D}
  {\bfseries 98} (2018) 084005}
  [\href{https://arxiv.org/abs/1806.08347}{{\ttfamily 1806.08347}}].

\bibitem{Kovacs:1977}
S.~J. {Kovacs} and K.~S. {Thorne}, \emph{{The generation of gravitational
  waves. III. Derivation of bremsstrahlung formulae.}},
  \href{https://doi.org/10.1086/155576}{\emph{Astrophysical Journal} {\bfseries
  217} (1977) 252}.

\bibitem{Kovacs:1978}
J.~{Kovacs}, S.~J. and K.~S. {Thorne}, \emph{{The generation of gravitational
  waves. IV. Bremsstrahlung.}},
  \href{https://doi.org/10.1086/156350}{\emph{Astrophysical Journal} {\bfseries
  224} (1978) 62}.

\bibitem{Bini:2017xzy}
D.~Bini and T.~Damour, \emph{{Gravitational spin-orbit coupling in binary
  systems, post-Minkowskian approximation and effective one-body theory}},
  \href{https://doi.org/10.1103/PhysRevD.96.104038}{\emph{Phys. Rev. D}
  {\bfseries 96} (2017) 104038}
  [\href{https://arxiv.org/abs/1709.00590}{{\ttfamily 1709.00590}}].

\bibitem{Vines:2017hyw}
J.~Vines, \emph{{Scattering of two spinning black holes in post-Minkowskian
  gravity, to all orders in spin, and effective-one-body mappings}},
  \href{https://doi.org/10.1088/1361-6382/aaa3a8}{\emph{Class. Quant. Grav.}
  {\bfseries 35} (2018) 084002}
  [\href{https://arxiv.org/abs/1709.06016}{{\ttfamily 1709.06016}}].

\bibitem{Bini:2018ywr}
D.~Bini and T.~Damour, \emph{{Gravitational spin-orbit coupling in binary
  systems at the second post-Minkowskian approximation}},
  \href{https://doi.org/10.1103/PhysRevD.98.044036}{\emph{Phys. Rev. D}
  {\bfseries 98} (2018) 044036}
  [\href{https://arxiv.org/abs/1805.10809}{{\ttfamily 1805.10809}}].

\bibitem{Iwasaki:1971vb}
Y.~Iwasaki, \emph{{Quantum theory of gravitation vs. classical theory. -
  fourth-order potential}},
  \href{https://doi.org/10.1143/PTP.46.1587}{\emph{Prog. Theor. Phys.}
  {\bfseries 46} (1971) 1587}.

\bibitem{Duff:1973zz}
M.~Duff, \emph{{Quantum Tree Graphs and the Schwarzschild Solution}},
  \href{https://doi.org/10.1103/PhysRevD.7.2317}{\emph{Phys. Rev. D} {\bfseries
  7} (1973) 2317}.

\bibitem{Holstein:2004dn}
B.~R. Holstein and J.~F. Donoghue, \emph{{Classical physics and quantum
  loops}}, \href{https://doi.org/10.1103/PhysRevLett.93.201602}{\emph{Phys.
  Rev. Lett.} {\bfseries 93} (2004) 201602}
  [\href{https://arxiv.org/abs/hep-th/0405239}{{\ttfamily hep-th/0405239}}].

\bibitem{Neill:2013wsa}
D.~Neill and I.~Z. Rothstein, \emph{{Classical Space-Times from the S Matrix}},
  \href{https://doi.org/10.1016/j.nuclphysb.2013.09.007}{\emph{Nucl. Phys. B}
  {\bfseries 877} (2013) 177}
  [\href{https://arxiv.org/abs/1304.7263}{{\ttfamily 1304.7263}}].

\bibitem{Bjerrum-Bohr:2013bxa}
N.~Bjerrum-Bohr, J.~F. Donoghue and P.~Vanhove, \emph{{On-shell Techniques and
  Universal Results in Quantum Gravity}},
  \href{https://doi.org/10.1007/JHEP02(2014)111}{\emph{JHEP} {\bfseries 02}
  (2014) 111} [\href{https://arxiv.org/abs/1309.0804}{{\ttfamily 1309.0804}}].

\bibitem{Luna:2017dtq}
A.~Luna, I.~Nicholson, D.~O'Connell and C.~D. White, \emph{{Inelastic Black
  Hole Scattering from Charged Scalar Amplitudes}},
  \href{https://doi.org/10.1007/JHEP03(2018)044}{\emph{JHEP} {\bfseries 03}
  (2018) 044} [\href{https://arxiv.org/abs/1711.03901}{{\ttfamily
  1711.03901}}].

\bibitem{Bjerrum-Bohr:2018xdl}
N.~J. Bjerrum-Bohr, P.~H. Damgaard, G.~Festuccia, L.~Plant\'e and P.~Vanhove,
  \emph{{General Relativity from Scattering Amplitudes}},
  \href{https://doi.org/10.1103/PhysRevLett.121.171601}{\emph{Phys. Rev. Lett.}
  {\bfseries 121} (2018) 171601}
  [\href{https://arxiv.org/abs/1806.04920}{{\ttfamily 1806.04920}}].

\bibitem{Kosower:2018adc}
D.~A. Kosower, B.~Maybee and D.~O'Connell, \emph{{Amplitudes, Observables, and
  Classical Scattering}},
  \href{https://doi.org/10.1007/JHEP02(2019)137}{\emph{JHEP} {\bfseries 02}
  (2019) 137} [\href{https://arxiv.org/abs/1811.10950}{{\ttfamily
  1811.10950}}].

\bibitem{Cheung:2018wkq}
C.~Cheung, I.~Z. Rothstein and M.~P. Solon, \emph{{From Scattering Amplitudes
  to Classical Potentials in the Post-Minkowskian Expansion}},
  \href{https://doi.org/10.1103/PhysRevLett.121.251101}{\emph{Phys. Rev. Lett.}
  {\bfseries 121} (2018) 251101}
  [\href{https://arxiv.org/abs/1808.02489}{{\ttfamily 1808.02489}}].

\bibitem{Cristofoli:2019neg}
A.~Cristofoli, N.~Bjerrum-Bohr, P.~H. Damgaard and P.~Vanhove,
  \emph{{Post-Minkowskian Hamiltonians in general relativity}},
  \href{https://doi.org/10.1103/PhysRevD.100.084040}{\emph{Phys. Rev. D}
  {\bfseries 100} (2019) 084040}
  [\href{https://arxiv.org/abs/1906.01579}{{\ttfamily 1906.01579}}].

\bibitem{Bern:2019nnu}
Z.~Bern, C.~Cheung, R.~Roiban, C.-H. Shen, M.~P. Solon and M.~Zeng,
  \emph{{Scattering Amplitudes and the Conservative Hamiltonian for Binary
  Systems at Third Post-Minkowskian Order}},
  \href{https://doi.org/10.1103/PhysRevLett.122.201603}{\emph{Phys. Rev. Lett.}
  {\bfseries 122} (2019) 201603}
  [\href{https://arxiv.org/abs/1901.04424}{{\ttfamily 1901.04424}}].

\bibitem{Bern:2019crd}
Z.~Bern, C.~Cheung, R.~Roiban, C.-H. Shen, M.~P. Solon and M.~Zeng,
  \emph{{Black Hole Binary Dynamics from the Double Copy and Effective
  Theory}}, \href{https://doi.org/10.1007/JHEP10(2019)206}{\emph{JHEP}
  {\bfseries 10} (2019) 206}
  [\href{https://arxiv.org/abs/1908.01493}{{\ttfamily 1908.01493}}].

\bibitem{Vaidya:2014kza}
V.~Vaidya, \emph{{Gravitational spin Hamiltonians from the S matrix}},
  \href{https://doi.org/10.1103/PhysRevD.91.024017}{\emph{Phys. Rev. D}
  {\bfseries 91} (2015) 024017}
  [\href{https://arxiv.org/abs/1410.5348}{{\ttfamily 1410.5348}}].

\bibitem{Guevara:2018wpp}
A.~Guevara, A.~Ochirov and J.~Vines, \emph{{Scattering of Spinning Black Holes
  from Exponentiated Soft Factors}},
  \href{https://doi.org/10.1007/JHEP09(2019)056}{\emph{JHEP} {\bfseries 09}
  (2019) 056} [\href{https://arxiv.org/abs/1812.06895}{{\ttfamily
  1812.06895}}].

\bibitem{Guevara:2019fsj}
A.~Guevara, A.~Ochirov and J.~Vines, \emph{{Black-hole scattering with general
  spin directions from minimal-coupling amplitudes}},
  \href{https://doi.org/10.1103/PhysRevD.100.104024}{\emph{Phys. Rev. D}
  {\bfseries 100} (2019) 104024}
  [\href{https://arxiv.org/abs/1906.10071}{{\ttfamily 1906.10071}}].

\bibitem{Maybee:2019jus}
B.~Maybee, D.~O'Connell and J.~Vines, \emph{{Observables and amplitudes for
  spinning particles and black holes}},
  \href{https://doi.org/10.1007/JHEP12(2019)156}{\emph{JHEP} {\bfseries 12}
  (2019) 156} [\href{https://arxiv.org/abs/1906.09260}{{\ttfamily
  1906.09260}}].

\bibitem{Bern:2020buy}
Z.~Bern, A.~Luna, R.~Roiban, C.-H. Shen and M.~Zeng, \emph{{Spinning Black Hole
  Binary Dynamics, Scattering Amplitudes and Effective Field Theory}},
  \href{https://arxiv.org/abs/2005.03071}{{\ttfamily 2005.03071}}.

\bibitem{Damgaard:2019lfh}
P.~H. Damgaard, K.~Haddad and A.~Helset, \emph{{Heavy Black Hole Effective
  Theory}}, \href{https://doi.org/10.1007/JHEP11(2019)070}{\emph{JHEP}
  {\bfseries 11} (2019) 070}
  [\href{https://arxiv.org/abs/1908.10308}{{\ttfamily 1908.10308}}].

\bibitem{Aoude:2020onz}
R.~Aoude, K.~Haddad and A.~Helset, \emph{{On-shell heavy particle effective
  theories}}, \href{https://doi.org/10.1007/JHEP05(2020)051}{\emph{JHEP}
  {\bfseries 05} (2020) 051}
  [\href{https://arxiv.org/abs/2001.09164}{{\ttfamily 2001.09164}}].

\bibitem{Damour:2019lcq}
T.~Damour, \emph{{Classical and quantum scattering in post-Minkowskian
  gravity}}, \href{https://doi.org/10.1103/PhysRevD.102.024060}{\emph{Phys.
  Rev. D} {\bfseries 102} (2020) 024060}
  [\href{https://arxiv.org/abs/1912.02139}{{\ttfamily 1912.02139}}].

\bibitem{Damour:2017zjx}
T.~Damour, \emph{{High-energy gravitational scattering and the general
  relativistic two-body problem}},
  \href{https://doi.org/10.1103/PhysRevD.97.044038}{\emph{Phys. Rev. D}
  {\bfseries 97} (2018) 044038}
  [\href{https://arxiv.org/abs/1710.10599}{{\ttfamily 1710.10599}}].

\bibitem{Kalin:2019rwq}
G.~K\"alin and R.~A. Porto, \emph{{From Boundary Data to Bound States}},
  \href{https://doi.org/10.1007/JHEP01(2020)072}{\emph{JHEP} {\bfseries 01}
  (2020) 072} [\href{https://arxiv.org/abs/1910.03008}{{\ttfamily
  1910.03008}}].

\bibitem{Kalin:2019inp}
G.~K\"alin and R.~A. Porto, \emph{{From boundary data to bound states. Part II.
  Scattering angle to dynamical invariants (with twist)}},
  \href{https://doi.org/10.1007/JHEP02(2020)120}{\emph{JHEP} {\bfseries 02}
  (2020) 120} [\href{https://arxiv.org/abs/1911.09130}{{\ttfamily
  1911.09130}}].

\bibitem{Bastianelli:2000pt}
F.~Bastianelli, O.~Corradini and P.~van Nieuwenhuizen, \emph{{Dimensional
  regularization of the path integral in curved space on an infinite time
  interval}}, \href{https://doi.org/10.1016/S0370-2693(00)00978-3}{\emph{Phys.
  Lett. B} {\bfseries 490} (2000) 154}
  [\href{https://arxiv.org/abs/hep-th/0007105}{{\ttfamily hep-th/0007105}}].

\bibitem{Bastianelli:2002fv}
F.~Bastianelli and A.~Zirotti, \emph{{Worldline formalism in a gravitational
  background}},
  \href{https://doi.org/10.1016/S0550-3213(02)00683-1}{\emph{Nucl. Phys. B}
  {\bfseries 642} (2002) 372}
  [\href{https://arxiv.org/abs/hep-th/0205182}{{\ttfamily hep-th/0205182}}].

\bibitem{Fabbrichesi:1993kz}
M.~Fabbrichesi, R.~Pettorino, G.~Veneziano and G.~Vilkovisky, \emph{{Planckian
  energy scattering and surface terms in the gravitational action}},
  \href{https://doi.org/10.1016/0550-3213(94)90361-1}{\emph{Nucl. Phys. B}
  {\bfseries 419} (1994) 147}
  [\href{https://arxiv.org/abs/hep-th/9309037}{{\ttfamily hep-th/9309037}}].

\bibitem{Amati:1988tn}
D.~Amati, M.~Ciafaloni and G.~Veneziano, \emph{{Can Space-Time Be Probed Below
  the String Size?}},
  \href{https://doi.org/10.1016/0370-2693(89)91366-X}{\emph{Phys. Lett. B}
  {\bfseries 216} (1989) 41}.

\bibitem{Amati:1990xe}
D.~Amati, M.~Ciafaloni and G.~Veneziano, \emph{{Higher Order Gravitational
  Deflection and Soft Bremsstrahlung in Planckian Energy Superstring
  Collisions}}, \href{https://doi.org/10.1016/0550-3213(90)90375-N}{\emph{Nucl.
  Phys. B} {\bfseries 347} (1990) 550}.

\bibitem{Feynman:1950ir}
R.~Feynman, \emph{{Mathematical formulation of the quantum theory of
  electromagnetic interaction}},
  \href{https://doi.org/10.1103/PhysRev.80.440}{\emph{Phys. Rev.} {\bfseries
  80} (1950) 440}.

\bibitem{Schwartz:2013pla}
M.~D. Schwartz, \emph{{Quantum Field Theory and the Standard Model}}. Cambridge
  University Press, 3, 2014.

\bibitem{Strassler:1992zr}
M.~J. Strassler, \emph{{Field theory without Feynman diagrams: One loop
  effective actions}},
  \href{https://doi.org/10.1016/0550-3213(92)90098-V}{\emph{Nucl. Phys. B}
  {\bfseries 385} (1992) 145}
  [\href{https://arxiv.org/abs/hep-ph/9205205}{{\ttfamily hep-ph/9205205}}].

\bibitem{Bern:1991aq}
Z.~Bern and D.~A. Kosower, \emph{{The Computation of loop amplitudes in gauge
  theories}}, \href{https://doi.org/10.1016/0550-3213(92)90134-W}{\emph{Nucl.
  Phys. B} {\bfseries 379} (1992) 451}.

\bibitem{Schubert:2001he}
C.~Schubert, \emph{{Perturbative quantum field theory in the string inspired
  formalism}}, \href{https://doi.org/10.1016/S0370-1573(01)00013-8}{\emph{Phys.
  Rept.} {\bfseries 355} (2001) 73}
  [\href{https://arxiv.org/abs/hep-th/0101036}{{\ttfamily hep-th/0101036}}].

\bibitem{Edwards:2019eby}
J.~P. Edwards and C.~Schubert, \emph{{Quantum mechanical path integrals in the
  first quantised approach to quantum field theory}},  12, 2019,
  \href{https://arxiv.org/abs/1912.10004}{{\ttfamily 1912.10004}}.

\bibitem{DeWitt:1957at}
B.~S. DeWitt, \emph{{Dynamical theory in curved spaces. 1. A Review of the
  classical and quantum action principles}},
  \href{https://doi.org/10.1103/RevModPhys.29.377}{\emph{Rev. Mod. Phys.}
  {\bfseries 29} (1957) 377}.

\bibitem{Parker:1979mf}
L.~Parker, \emph{{PATH INTEGRALS FOR A PARTICLE IN CURVED SPACE}},
  \href{https://doi.org/10.1103/PhysRevD.19.438}{\emph{Phys. Rev. D} {\bfseries
  19} (1979) 438}.

\bibitem{Bekenstein:1981xe}
J.~Bekenstein and L.~Parker, \emph{{Path Integral Evaluation of Feynman
  Propagator in Curved Space-time}},
  \href{https://doi.org/10.1103/PhysRevD.23.2850}{\emph{Phys. Rev. D}
  {\bfseries 23} (1981) 2850}.

\bibitem{Bastianelli:1992ct}
F.~Bastianelli and P.~van Nieuwenhuizen, \emph{{Trace anomalies from quantum
  mechanics}}, \href{https://doi.org/10.1016/0550-3213(93)90285-W}{\emph{Nucl.
  Phys. B} {\bfseries 389} (1993) 53}
  [\href{https://arxiv.org/abs/hep-th/9208059}{{\ttfamily hep-th/9208059}}].

\bibitem{Ahmadiniaz:2019ppj}
N.~Ahmadiniaz, F.~M. Balli, O.~Corradini, J.~M. D\'avila and C.~Schubert,
  \emph{{Compton-like scattering of a scalar particle with $N$ photons and one
  graviton}},
  \href{https://doi.org/10.1016/j.nuclphysb.2019.114877}{\emph{Nucl. Phys. B}
  {\bfseries 950} (2020) 114877}
  [\href{https://arxiv.org/abs/1908.03425}{{\ttfamily 1908.03425}}].

\bibitem{Daikouji:1995dz}
K.~Daikouji, M.~Shino and Y.~Sumino, \emph{{Bern-Kosower rule for scalar QED}},
  \href{https://doi.org/10.1103/PhysRevD.53.4598}{\emph{Phys. Rev. D}
  {\bfseries 53} (1996) 4598}
  [\href{https://arxiv.org/abs/hep-ph/9508377}{{\ttfamily hep-ph/9508377}}].

\bibitem{Ahmadiniaz:2015kfq}
N.~Ahmadiniaz, A.~Bashir and C.~Schubert, \emph{{Multiphoton amplitudes and
  generalized Landau-Khalatnikov-Fradkin transformation in scalar QED}},
  \href{https://doi.org/10.1103/PhysRevD.93.045023}{\emph{Phys. Rev. D}
  {\bfseries 93} (2016) 045023}
  [\href{https://arxiv.org/abs/1511.05087}{{\ttfamily 1511.05087}}].

\bibitem{Ahmadiniaz:2015xoa}
N.~Ahmadiniaz, F.~Bastianelli and O.~Corradini, \emph{{Dressed scalar
  propagator in a non-Abelian background from the worldline formalism}},
  \href{https://doi.org/10.1103/PhysRevD.93.025035}{\emph{Phys. Rev. D}
  {\bfseries 93} (2016) 025035}
  [\href{https://arxiv.org/abs/1508.05144}{{\ttfamily 1508.05144}}].

\bibitem{Bonocore:2020xuj}
D.~Bonocore, \emph{{Asymptotic dynamics on the worldline for spinning
  particles}},  \href{https://arxiv.org/abs/2009.07863}{{\ttfamily
  2009.07863}}.

\bibitem{Amati:1987wq}
D.~Amati, M.~Ciafaloni and G.~Veneziano, \emph{{Superstring Collisions at
  Planckian Energies}},
  \href{https://doi.org/10.1016/0370-2693(87)90346-7}{\emph{Phys. Lett. B}
  {\bfseries 197} (1987) 81}.

\bibitem{Damour:2020tta}
T.~Damour, \emph{{Radiative contribution to classical gravitational scattering
  at the third order in $G$}},
  \href{https://doi.org/10.1103/PhysRevD.102.124008}{\emph{Phys. Rev. D}
  {\bfseries 102} (2020) 124008}
  [\href{https://arxiv.org/abs/2010.01641}{{\ttfamily 2010.01641}}].

\bibitem{DiVecchia:2020ymx}
P.~Di~Vecchia, C.~Heissenberg, R.~Russo and G.~Veneziano, \emph{{Universality
  of ultra-relativistic gravitational scattering}},
  \href{https://doi.org/10.1016/j.physletb.2020.135924}{\emph{Phys. Lett. B}
  {\bfseries 811} (2020) 135924}
  [\href{https://arxiv.org/abs/2008.12743}{{\ttfamily 2008.12743}}].

\bibitem{Arkani-Hamed:2019ymq}
N.~Arkani-Hamed, Y.-t. Huang and D.~O'Connell, \emph{{Kerr black holes as
  elementary particles}},
  \href{https://doi.org/10.1007/JHEP01(2020)046}{\emph{JHEP} {\bfseries 01}
  (2020) 046} [\href{https://arxiv.org/abs/1906.10100}{{\ttfamily
  1906.10100}}].

\bibitem{Monteiro:2014cda}
R.~Monteiro, D.~O'Connell and C.~D. White, \emph{{Black holes and the double
  copy}}, \href{https://doi.org/10.1007/JHEP12(2014)056}{\emph{JHEP} {\bfseries
  12} (2014) 056} [\href{https://arxiv.org/abs/1410.0239}{{\ttfamily
  1410.0239}}].

\bibitem{Monteiro:2015bna}
R.~Monteiro, D.~O'Connell and C.~D. White, \emph{{Gravity as a double copy of
  gauge theory: from amplitudes to black holes}},
  \href{https://doi.org/10.1142/S0218271815420080}{\emph{Int. J. Mod. Phys. D}
  {\bfseries 24} (2015) 1542008}.

\bibitem{Luna:2015paa}
A.~Luna, R.~Monteiro, D.~O'Connell and C.~D. White, \emph{{The classical double
  copy for Taub\textendash{}NUT spacetime}},
  \href{https://doi.org/10.1016/j.physletb.2015.09.021}{\emph{Phys. Lett. B}
  {\bfseries 750} (2015) 272}
  [\href{https://arxiv.org/abs/1507.01869}{{\ttfamily 1507.01869}}].

\bibitem{Goldberger:2016iau}
W.~D. Goldberger and A.~K. Ridgway, \emph{{Radiation and the classical double
  copy for color charges}},
  \href{https://doi.org/10.1103/PhysRevD.95.125010}{\emph{Phys. Rev. D}
  {\bfseries 95} (2017) 125010}
  [\href{https://arxiv.org/abs/1611.03493}{{\ttfamily 1611.03493}}].

\bibitem{Galley:2009px}
C.~R. Galley and M.~Tiglio, \emph{{Radiation reaction and gravitational waves
  in the effective field theory approach}},
  \href{https://doi.org/10.1103/PhysRevD.79.124027}{\emph{Phys. Rev. D}
  {\bfseries 79} (2009) 124027}
  [\href{https://arxiv.org/abs/0903.1122}{{\ttfamily 0903.1122}}].

\bibitem{Galley:2012hx}
C.~R. Galley, \emph{{Classical Mechanics of Nonconservative Systems}},
  \href{https://doi.org/10.1103/PhysRevLett.110.174301}{\emph{Phys. Rev. Lett.}
  {\bfseries 110} (2013) 174301}
  [\href{https://arxiv.org/abs/1210.2745}{{\ttfamily 1210.2745}}].

\bibitem{Saha:2019tub}
A.~P. Saha, B.~Sahoo and A.~Sen, \emph{{Proof of the classical soft graviton
  theorem in $D$ = 4}},
  \href{https://doi.org/10.1007/JHEP06(2020)153}{\emph{JHEP} {\bfseries 06}
  (2020) 153} [\href{https://arxiv.org/abs/1912.06413}{{\ttfamily
  1912.06413}}].

\bibitem{Sahoo:2020ryf}
B.~Sahoo, \emph{{Classical Sub-subleading Soft Photon and Soft Graviton
  Theorems in Four Spacetime Dimensions}},
  \href{https://doi.org/10.1007/JHEP12(2020)070}{\emph{JHEP} {\bfseries 12}
  (2020) 070} [\href{https://arxiv.org/abs/2008.04376}{{\ttfamily
  2008.04376}}].

\bibitem{Sahoo:2018lxl}
B.~Sahoo and A.~Sen, \emph{{Classical and Quantum Results on Logarithmic Terms
  in the Soft Theorem in Four Dimensions}},
  \href{https://doi.org/10.1007/JHEP02(2019)086}{\emph{JHEP} {\bfseries 02}
  (2019) 086} [\href{https://arxiv.org/abs/1808.03288}{{\ttfamily
  1808.03288}}].

\bibitem{Shen:2018ebu}
C.-H. Shen, \emph{{Gravitational Radiation from Color-Kinematics Duality}},
  \href{https://doi.org/10.1007/JHEP11(2018)162}{\emph{JHEP} {\bfseries 11}
  (2018) 162} [\href{https://arxiv.org/abs/1806.07388}{{\ttfamily
  1806.07388}}].

\bibitem{DBLP:journals/corr/abs-0803-0862}
J.~M. Mart{\'{\i}}n{-}Garc{\'{\i}}a, \emph{{xPerm: fast index canonicalization
  for tensor computer algebra}}, {\emph{CoRR} {\bfseries abs/0803.0862} (2008)
  } [\href{https://arxiv.org/abs/0803.0862}{{\ttfamily 0803.0862}}].

\bibitem{DiVecchia:2019kta}
P.~Di~Vecchia, S.~G. Naculich, R.~Russo, G.~Veneziano and C.~D. White, \emph{{A
  tale of two exponentiations in $ \mathcal{N} $ = 8 supergravity at subleading
  level}}, \href{https://doi.org/10.1007/JHEP03(2020)173}{\emph{JHEP}
  {\bfseries 03} (2020) 173}
  [\href{https://arxiv.org/abs/1911.11716}{{\ttfamily 1911.11716}}].

\bibitem{Antonelli:2020}
A.~Antonelli, C.~Kavanagh, M.~Khalil, J.~Steinhoff and J.~Vines,
  \emph{{Gravitational spin-orbit and aligned spin$_1$-spin$_2$ couplings
  through third-subleading post-Newtonian orders}},
  \href{https://doi.org/10.1103/PhysRevD.102.124024}{\emph{Phys. Rev. D}
  {\bfseries 102} (2020) 124024}
  [\href{https://arxiv.org/abs/2010.02018}{{\ttfamily 2010.02018}}].

\bibitem{Parra-Martinez:2020dzs}
J.~Parra-Martinez, M.~S. Ruf and M.~Zeng, \emph{{Extremal black hole scattering
  at $\mathcal{O}(G^3)$: graviton dominance, eikonal exponentiation, and
  differential equations}},
  \href{https://doi.org/10.1007/JHEP11(2020)023}{\emph{JHEP} {\bfseries 11}
  (2020) 023} [\href{https://arxiv.org/abs/2005.04236}{{\ttfamily
  2005.04236}}].

\bibitem{Vines:2018gqi}
J.~Vines, J.~Steinhoff and A.~Buonanno, \emph{{Spinning-black-hole scattering
  and the test-black-hole limit at second post-Minkowskian order}},
  \href{https://doi.org/10.1103/PhysRevD.99.064054}{\emph{Phys. Rev. D}
  {\bfseries 99} (2019) 064054}
  [\href{https://arxiv.org/abs/1812.00956}{{\ttfamily 1812.00956}}].

\bibitem{Berends:1987me}
F.~A. Berends and W.~Giele, \emph{{Recursive Calculations for Processes with n
  Gluons}}, \href{https://doi.org/10.1016/0550-3213(88)90442-7}{\emph{Nucl.
  Phys. B} {\bfseries 306} (1988) 759}.

\bibitem{Bastianelli:2019xhi}
F.~Bastianelli, R.~Bonezzi, O.~Corradini and E.~Latini, \emph{{One-loop quantum
  gravity from the $\mathcal N=4$ spinning particle}},
  \href{https://doi.org/10.1007/JHEP11(2019)124}{\emph{JHEP} {\bfseries 11}
  (2019) 124} [\href{https://arxiv.org/abs/1909.05750}{{\ttfamily
  1909.05750}}].

\bibitem{Bonezzi:2020jjq}
R.~Bonezzi, A.~Meyer and I.~Sachs, \emph{{A Worldline Theory for
  Supergravity}}, \href{https://doi.org/10.1007/JHEP06(2020)103}{\emph{JHEP}
  {\bfseries 06} (2020) 103}
  [\href{https://arxiv.org/abs/2004.06129}{{\ttfamily 2004.06129}}].

\bibitem{Plefka:2019hmz}
J.~Plefka, C.~Shi, J.~Steinhoff and T.~Wang, \emph{{Breakdown of the classical
  double copy for the effective action of dilaton-gravity at NNLO}},
  \href{https://doi.org/10.1103/PhysRevD.100.086006}{\emph{Phys. Rev. D}
  {\bfseries 100} (2019) 086006}
  [\href{https://arxiv.org/abs/1906.05875}{{\ttfamily 1906.05875}}].

\end{thebibliography}\endgroup

\end{document}